\documentclass{aa}

\usepackage{amsmath,hyperref,txfonts}
\hypersetup{colorlinks,citecolor=blue,linkcolor=blue,urlcolor=blue}

\newcommand{\ab}[1]{\left|#1\right|}
\newcommand{\av}[1]{\left\langle#1\right\rangle}
\newcommand{\br}[1]{\left[#1\right]}
\newcommand{\cu}[1]{\left\{#1\right\}}
\newcommand{\pa}[1]{\left(#1\right)}
\newcommand{\ed}{\mathop{}\!\mathrm{d}}
\DeclareMathOperator{\arcsinh}{arcsinh}
\DeclareMathOperator{\sign}{sign}

\DeclareMathOperator{\rmsd}{RMSD}

\begin{document}

\title{Photon ring test of the Kerr hypothesis: Variation in the ring shape}

\author{
H.~Paugnat\inst{1}
\and
A.~Lupsasca\inst{2}
\and
F.~H.~Vincent\inst{1}
\and 
M.~Wielgus\inst{3}
}

\institute{
LESIA, Observatoire de Paris, 5 place Jules Janssen, 92195 Meudon, France.
\email{hadrien.paugnat@polytechnique.org}
\and
Princeton Gravity Initiative, Princeton University, Princeton NJ 08544, USA
\and
Max-Planck-Institut f\"ur Radioastronomie, Auf dem H\"ugel 69, D-53121 Bonn, Germany
}

\abstract
{The Event Horizon Telescope (EHT) collaboration recently released horizon-scale images of the supermassive black hole M87*.
These images are consistently described by an optically thin, lensed accretion flow in the Kerr spacetime.
General relativity (GR) predicts that higher-resolution images of such a flow would present thin, ring-shaped features produced by photons on extremely bent orbits.
Recent theoretical work suggests that these ``photon rings'' produce clear interferometric signatures that depend very little on the astrophysical configuration and whose observation could therefore provide a stringent consistency test of the Kerr hypothesis.}
{We wish to understand how the photon rings of a Kerr black hole vary with its surrounding emission.
Gralla, Lupsasca, and Marrone (GLM) found that the shape of high-order photon rings follows a specific functional form that is insensitive to the details of the astrophysical source, and proposed a method for measuring this GR-predicted shape via space-based interferometry.
We wish to assess the robustness of this prediction by checking it for a variety of astrophysical profiles, black hole spins, and observer inclinations.}
{We use the ray tracing code \texttt{Gyoto} to simulate images of thin equatorial disks accreting onto a Kerr black hole.
We extract the shape of the resulting photon rings from their interferometric signatures using a refinement of the method developed by GLM.
We repeat this analysis for hundreds of models with different emission profiles, black hole spins, and observer inclinations.}
{We identify the width of the photon ring and its angular variation as a main obstacle to the method's success.
We qualitatively describe how this width varies with the emission profile, black hole spin, and observer inclination.
At low inclinations, our improved method is robust enough to confirm the shape prediction for a variety of emission profiles; however, the choice of baseline is critical to the method's success.
At high inclinations, we encounter qualitatively new effects that are caused by the ring's non-uniform width and require further refinements to the method.
We also explore how the photon ring shape could constrain black hole spin and inclination.}
{}


\maketitle

\section{Introduction}

The existence of black holes (BH) is a key prediction of general relativity (GR) in the strong-field regime.
More specifically, the theory posits that the spacetime geometry around these compact objects is described by the Kerr metric.
This ``Kerr hypothesis'' underlies a considerable amount of astrophysics and also plays a driving role in theoretical physics, but it has yet to be directly tested.
A number of experiments are now poised to change this.
The recent successes obtained in gravitational-wave and radio interferometry by LIGO-Virgo, GRAVITY, and EHT \citep{LIGO2016,GRAVITY,EHT1} herald the advent of a new era of black hole astronomy, and their future extensions promise to deliver observations that will enable tests of the Kerr hypothesis with unprecedented precision.

In April 2019, the EHT collaboration published the very first images resolving the nucleus of the galaxy M87 \citep{EHT1,EHT2,EHT3,EHT4,EHT5,EHT6}.
Reconstructed from 1.3\,mm very-long-baseline interferometry (VLBI) observations, these images achieved an angular resolution comparable to the expected size of M87*---the supermassive compact object at the center of M87---and revealed a thick, non-uniformly bright ring with a diameter of $\sim\!40\,\mu$as encircling a central brightness deficit.
These observed features are roughly compatible with those appearing in simulated black hole images \citep[e.g.,][]{Luminet1979,James2015,EHT5}, and indeed M87* may be consistently modeled as a Kerr BH surrounded by an accretion flow whose emission originates within a few Schwarzschild radii of the event horizon.
Direct (weakly lensed) images of such a flow present an asymmetric bright ring that is consistent with the 2017 EHT observations \citep[e.g.,][]{Gralla2019,Johnson2020,Chael2021}.

Due to the limited resolution of these first EHT images of M87*, this consistency test of the Kerr hypothesis remains weak; instead, the \citet{EHT6} has thus far assumed the Kerr nature of M87*, so as to constrain its mass from observations.
However, as technology continues to improve, increasingly better images will be obtained, and these should enable correspondingly sharper probes of the spacetime geometry around M87*.
Nonetheless, it is not immediately clear how to best extract relativistic signatures from such probes, which raises the question: how could sharper images of M87* in principle lead to more stringent tests of the Kerr hypothesis?

Addressing this question is made particularly difficult by the inherent asymmetry of the imaging problem.
Thanks to curved-spacetime ray tracing codes such as \texttt{Gyoto} \citep{Vincent2011} and others \citep{Gold2020}, solving the \textit{forward} problem has become a fairly routine task, and determining the observational appearance of a compact object with specified properties (spacetime metric, emission model) is now relatively straightforward.
The \textit{inverse} problem, on the other hand, remains nigh intractable: inferring the nature of an astrophysical source from its visual appearance requires one to disentangle effects of the spacetime geometry from properties of the emission, a problem subject to enormous degeneracy.
A veritable zoo of theoretically proposed objects (besides the Kerr BH) have been shown to also produce simulated images with features compatible with the recent EHT observations,\footnote{A non-exhaustive list includes: exotic compact objects \citep[e.g.,][]{Vincent2016,Herdeiro2021},
wormholes \citep[e.g.,][]{Wielgus2020}, non-singular black holes \citep{Lamy2018}, etc.
}
and many of these alternative models cannot yet be ruled out by observations with current EHT resolution \citep[e.g.,][]{Vincent2021}.

This present blurriness is not the only limitation: even with higher resolution, images that resolve only the direct emission will likely never separate astrophysical and geometrical effects.
(This partly explains the present lack of robust constraints on the spin of M87*.)
As such, these images may always be compatible with models of a Kerr BH surrounded by some (possibly exotic) emission, and hence they could never lead one to definitively conclude a violation of the Kerr hypothesis, as this would require achieving greater confidence in the astrophysics than in GR---this could of course break the degeneracy in principle, but seems very unlikely in practice \citep[e.g.,][]{Gralla2020,Bauer2021}.

Recent work suggests a promising path forward.
The strong gravity of a Kerr BH creates around it a ``photon shell''---a region outside its event horizon in which its gravity is so strong that light rays can become trapped on unstably bound orbits \citep{Bardeen1973,Teo2003}.
As a result, sources in the vicinity of the black hole can produce multiple relativistic images in the observer sky: provided that the emission is optically thin, the primary image (consisting of weakly lensed photons that travel directly to the observer after being emitted) is generically superimposed with a series of mirror images arising from highly bent photons that circumnavigated the photon shell multiple times on their way from source to observer \citep{Johnson2020,GrallaLupsasca2020a}.
If the emission is not spherical (so that it does not fully surround the hole), then these successive images form a discrete sequence of ``photon rings'' that are labeled by photon half-orbit number $n$ and converge to a ``critical curve'' in the observer sky corresponding to the light rays that asymptote to unstably bound orbits in the photon shell.
Crucially, GR predicts that these rings must display a self-similar structure governed by Kerr critical exponents $\gamma$, $\delta$, and $\tau$ that respectively control the successive demagnification, rotation, and time delay of successive images \citep{Johnson2020,GrallaLupsasca2020a}.
This suggests that observing this characteristic lensing pattern may provide a way to extract information about the black hole geometry, and to discriminate between different spacetimes \citep[e.g.,][]{Wielgus2021}.

The photon ring has not yet been conclusively observed, but its GR-predicted substructure has already been numerically confirmed with state-of-the-art (general relativistic magneto-hydrodynamics) GRMHD-simulated images of M87*, in which at least $n=3$ subrings can be distinguished \citep{Johnson2020}.
Although these rings (which are observable) are distinct from the theoretical critical curve (which is not observable), they closely track its shape, which is insensitive to the astrophysical details of the surrounding emission and entirely fixed by the Kerr metric \citep{Bardeen1973,GrallaLupsasca2020c}.
Critical curves have been extensively discussed in the literature and spacetimes other than Kerr are known to produce different shapes \citep[e.g.,][]{Johannsen2010,Johannsen2013, Cunha2018,Okounkova2019,Medeiros2020}.
Thus, measuring the shape of a high-$n$ photon ring of M87* may be a good proxy for probing the shape of its critical curve, and hence its geometry.

This idea was recently made precise by \citet{GLM2020}, who established that, in a certain class of models, the shape of the $n=2$ subring is extremely close to that of the critical curve.
Moreover, they showed that on very long baselines $u\sim300$~G$\lambda$, which could be accessible with space-based\footnote{The Earth's atmosphere limits the frequency, and hence the baseline length, of observations made with Earth-bound elements of the EHT.
As such, a photon ring shape measurement likely requires an interferometer with a space leg.
The ngEHT---the currently envisioned next-generation EHT \citep{Doeleman2019}---is slated to augment its coverage of Earth baselines with additional ground-based stations; this will improve the image fidelity but not resolve narrow features like the photon ring.
Other proposals to deploy a radio telescope to space include \citet{Haworth2019,Pesce2019,Gurvits2021}.
}
extensions of the EHT, this subring presents a clean interferometric signature, as expected from Fourier theory \citep{Johnson2020,Gralla2020}.
They argued that a space mission targeting M87* could resolve the shape of its $n=2$ photon ring with sufficient precision to enable a sharp comparison with the GR prediction: simulations of their experiment achieved sub-percent precision, forecasting a stringent consistency test of the Kerr hypothesis.
We will refer to their proposed procedure as the GLM method.

Of course, this first study is subject to multiple limitations, as it relies on a restricted class of emission models consisting of stationary, axisymmetric disks composed of circular-equatorial orbiters.
More precisely: 1) it only investigated a choice of 3 toy emission profiles, viewed from observer inclinations of 10$^\circ-30^\circ$ (the range of expected relevance to M87*); 2) it considered only geometrically thin disks; 3) it ignored astrophysical fluctuations, which amounts to studying (coherently) time-averaged images.

It is crucial to assess the robustness of the GLM method in greater generality.
The aim of this paper is to extend the GLM analysis to many more configurations; that is, to broaden its scope by performing a parameter survey of emission profiles, as well as BH spins and inclinations, in order to remove the first limitation outlined above.
This work remains confined to the framework of thin, equatorial accretion disks; a parallel study of the impact of disk thickness is ongoing and will soon address the second limitation \citep{Vincent2022}.
In the meantime, it is worth noting that recent work by \citet{Chael2021} suggests that geometrically thin models yield effective approximations to numerically modeled GRMHD accretion flows.

Finally, another objective of this paper is to initiate the study of black-hole-spin estimates based on astrophysics-independent shape measurements of the $n=2$ photon ring.
A first step is taken in this direction by establishing that photon rings arising from astrophysically plausible configurations constitute only a small subset of the full range of theoretically allowed shapes.

The paper is organized as follows.
First, Sec.~\ref{sec:Theory} reviews the theoretical foundations of the GLM method: the critical curve is defined and contrasted with the concept of ``black hole shadow,'' the notion of ``lensing bands'' is introduced, and the expected interferometric signature of the photon ring is described.
Next, Sec.~\ref{sec:Implementation} presents a numerical implementation of the GLM method that improves upon the original technique.
Then, Sec.~\ref{sec:BaselineWindow} describes how the choice of baseline may affect a shape measurement of the photon ring, especially if its thickness varies substantially with angle.
After that, Sec.~\ref{sec:EmissionSurvey} presents the results of a survey over emission profiles while Sec.~\ref{sec:SpinSurvey} examines the influence of BH spin and inclination.
We find that a measurement is always possible at low inclinations, while new complications that may arise at high inclination necessitate further refinements to the method.
We provide further discussion in Sec.~\ref{sec:Discussion}, before ending with our conclusions and perspectives in Sec.~\ref{sec:Conclusion}.

\section{Theoretical overview}
\label{sec:Theory}

\subsection{Shadow, critical curve, and photon rings}
\label{subsec:Shadow}

In the introduction, we asserted that the photon ring generically decomposes into a stack of discrete subrings.
For completeness, here we list the (mild) astrophysical assumptions required for the presence of this substructure; this also provides us with an opportunity to distinguish the photon ring from the ``black hole shadow'' and critical curve, terms deserving of disambiguation.

First, the emission region must be optically thin, so that light may cross it multiple times before being reabsorbed; otherwise, 
highly lensed ($n>0$) images may be obscured \citep[e.g.,][]{Beckwith2005}.
The validity of this assumption depends on both the source and the observing frequency.
As indicated in Fig.~3 of \citet{Johnson2020}, GRMHD predicts it to likely hold for 230\,GHz photons collected from M87* by present-day EHT, and it is even more likely true at the higher 345\,GHz frequency that ngEHT will target; by contrast, the core of Centaurus A is only expected to be optically thin past 1--5\,THz \citep{Janssen2021}.

Second, the emission region must have a gap rather than fully surround the hole; otherwise, the discrete structure visible in the intensity cross-sections displayed in \citet{Johnson2020,Chael2021}, which resemble the layers of a wedding cake, will not be present.
Instead, one will see a continuous cross-section with a mild (logarithmic) divergence near the
the critical curve: the image-plane curve, first derived by \citet{Bardeen1973} (who used the term ``apparent boundary''), corresponding to light rays that asymptote to unstably bound spherical photon orbits in the photon shell.
The resulting image will then display a distinctive feature known as the ``black hole shadow'': a central brightness deficit precisely bounded by the critical curve.
This effect was first highlighted by \citet{Falcke2000} in the context of a BH surrounded by a spherical, radially infalling accretion flow, and later revisited by \citet{Johannsen2010}.

However, as recently pointed out by \citet{Gralla2019} and \citet{Narayan2019}, this spherical-infall scenario is highly fine-tuned, and in fact no longer viewed as relevant for M87*: the latest EHT constraints derived from the polarimetric image of M87* \citep{EHT7,EHT8} strongly favor ``magnetically arrested disk'' (MAD) models \citep{Narayan2003,Igumenshchev2003} with emission concentrated near the midplane.
Such models do not display the traditional ``shadow'' feature (i.e., a central brightness depression filling the critical curve), but rather a different ``inner shadow'' feature associated with the inner edge of the emission, which produces a smaller but even darker central brightness depression contained well within the critical curve \citep{Chael2021}.
For equatorial disks extending to the horizon, the inner edge of the emission in principle coincides with the lensed equatorial event horizon (though it may appear larger due to redshift effects).

To summarize, the boundary of the black hole shadow is a mathematical curve which is not in itself observable, and only happens to coincide with an image feature---the central brightness depression---in special configurations currently disfavored by EHT data on M87*.
On the other hand, the photon ring is a visible feature that always arises in images of a BH surrounded by an optically thin emission region.
Moreover, if this emission is gapped, then this photon ring decomposes into subrings that converge (exponentially in $n$) to the theoretical critical curve, which can be thought of as the $n\to\infty$ subring.
The most easily accessible $n=1$ and $n=2$ subrings are close to, but nonetheless still distinct from, the exact critical curve.
Therefore their shape is constrained to be similar, but not exactly equal, to that of the shadow.
This observation is key for the GLM method.

\subsection{Bound geodesics form the photon shell}

Before turning to the study of the observable photon ring shape, we first elucidate the shape of the theoretical image-plane curve comprising the null geodesics that asymptote to unstably bound orbits around a Kerr BH \citep{Bardeen1973}.
The region of spacetime containing these orbits---the photon shell---is extensively reviewed in \citet{Teo2003,Johnson2020}.
Here we describe it with Boyer-Lindquist coordinates $(t,r,\theta,\phi)$ on the spacetime of a Kerr BH with mass $M$ and angular momentum $J=a_*M$:
\begin{align}
	ds^2=\frac{\Delta}{\Sigma}\pa{\ed t-a_*\sin^2{\theta}\ed\phi}^2+\frac{\Sigma}{\Delta}\ed r^2+\Sigma\ed\theta^2\notag\\
	+\frac{\sin^2{\theta}}{\Sigma}\br{\pa{r^2+a_*^2}\ed\phi-a_*\ed t}^2,
\end{align}
with $\Delta(r)=r^2-2Mr+a_*^2$ and $\Sigma(r,\theta)=r^2+a_*^2\cos^2{\theta}$.
We will often use a dimensionless spin parameter $a\equiv a_*/M\in[-1,1]$.

A Kerr photon with four-momentum $p^\mu$ has a conserved energy-rescaled angular momentum and Carter constant
\begin{align}
    \label{eq:ConservedQuantities}
	\lambda=\frac{p_\phi}{-p_t},\qquad
	\eta=\frac{p_\theta^2}{p_t^2}-a_*^2\cos^2{\theta}+\lambda^2\cot^2{\theta}.
\end{align}

Bound photon orbits in Kerr are ``spherical'': they evolve at a constant Boyer-Lindquist radius in the range $\tilde{r}_-\leq r\leq\tilde{r}_+$, where
\begin{align}
	\tilde{r}_\pm=2M\br{1+\cos\pa{\frac{2}{3}\arccos\pa{\pm a}}}
\end{align}
is the radius of the retrograde (upper sign) or prograde (lower sign) circular-equatorial orbit.
As they rotate around the BH, these bound orbits undergo polar librations between angles $[\tilde{\theta}_-,\tilde{\theta}_+]$, where
\begin{align}
	\tilde{\theta}_\pm&=\arccos\pa{\mp\sqrt{\tilde{u}_+}}
	\gtrless\frac{\pi}{2},\\
	\label{eq:uCritical}
	\tilde{u}_\pm&=\frac{r}{a_*^2(r-M)^2}\left[-r^3+3M^2r-2a_*^2M\right.\notag\\
	&\qquad\qquad\qquad\left.\pm2\sqrt{M\Delta\pa{2r^3-3Mr^2+a_*^2M}}\right],
\end{align}
and their characteristic conserved quantities are given by
\begin{align}
	\label{eq:AngMomentum}
	\tilde{\lambda}=\frac{M\pa{r^2-a_*^2}-r\Delta}{a_*(r-M)},\qquad
	\tilde{\eta}=-a_*^2\tilde{u}_+\tilde{u}_-.
\end{align}
Bound orbits are either prograde or retrograde, depending on the sign of $\tilde{\lambda}$.
Orbits at generic radii $r\in[\tilde{r}_-,\tilde{r}_+]$ are not closed, as they do not return to the same location after a full rotation by $\Delta\phi=2\pi$, but rather explore the entire allowed region within their shell of fixed $r$; resonant orbits which are closed form an exceptional set of measure zero in the photon shell \citep{Wong2021}.
The photon shell occupies a 3D region of space given at any time $t$ by
$\tilde{r}_-\leq r\leq\tilde{r}_-$, $\theta_-\leq\theta\leq\theta_+$, and $0\leq\phi<2\pi$.
In the Schwarzschild case ($a=0$), it reduces to the sphere at $r=3M$.

\subsection{Asymptotically bound geodesics form the critical curve}
\label{subsec:CriticalCurve}

A photon that is initially outside the photon shell but has exactly the same conserved quantities $(\lambda,\eta)=(\tilde{\lambda},\tilde{\eta})$ as a photon bound at orbital radius $r$ asymptotes to that orbit in its far future or past.
The critical curve is defined as the theoretical curve in the sky of an observer corresponding to these asymptotically bound photons.
It delineates the region of photon capture by the BH (inside the curve) from that of photon escape (outside the curve).

For the case of an observer at large distance $D\gg M$ and inclination $i\neq0$ from the spin axis of a BH, \citet{Bardeen1973} showed that a photon with conserved quantities $(\lambda,\eta)$ appears in the sky at a position given by\footnote{Only photons with vanishing angular momentum $\lambda=0$ can pass over the poles.
Thus, for the special case $i=0$ of an on-axis observer, $\rho=D^{-1}\sqrt{\eta+a_*^2}$ and $\varphi_\rho=\phi_o$ is the photon's angle of arrival, while
the critical curve is the circle $\tilde{\rho}=D^{-1}\sqrt{a_*^2\pa{1-\tilde{u}_+\tilde{u}_-}}$ corresponding to the bound orbit at the unique radius $\tilde{r}_0$ such that $\tilde{\lambda}=0$, which is given in \citep{GrallaLupsasca2020a}, where the ``BH line of sight'' is also defined.
}
\begin{align}
	\label{eq:BardeenCoordinates}
	\rho&=\frac{1}{D}\sqrt{\eta+\lambda^2+a_*^2\cos^2{i}},\qquad
	\cos{\varphi_\rho}=-\frac{\lambda}{D\rho\sin{i}},
\end{align}
where, following \citet{Johnson2020}, we replaced Bardeen's original Cartesian coordinates $(\alpha,\beta)$ with (dimensionless) polar coordinates $(\rho,\varphi_\rho)$ that are centered about the ``line of sight to the black hole'' and such that $\varphi_\rho=90^\circ$ corresponds to the BH spin axis projected onto the plane perpendicular to this line of sight.
Hence, for such an observer, the critical curve is the parametric curve $C_\gamma=(\tilde{\rho},\tilde{\varphi}_\rho)$ obtained by tracing
\begin{subequations}
\label{eq:CriticalCurve}
\begin{align}
	\tilde{\rho}&=\frac{1}{D}\sqrt{\tilde{\lambda}^2+a_*^2\pa{\cos^2{i}-\tilde{u}_+\tilde{u}_-}},\\
	\label{eq:CriticalCurveAngle}
	\cos{\tilde{\varphi}_\rho}&=-\frac{\tilde{\lambda}}{D\tilde{\rho}\sin{i}},
\end{align}
\end{subequations}
over the radial extent $r\in[\tilde{r}_-,\tilde{r}_+]$ of the photon shell.
Counter-intuitively, the angle $\tilde{\varphi}_\rho$ is parameterized not by the angle $\phi$ around the BH, but rather by the radius $r$ within the photon shell.
This effect highlights the warped nature of the Kerr spacetime.

Since Eq.~\eqref{eq:CriticalCurveAngle} admits two solutions for every choice of $r$, each orbit in the photon shell corresponds to two angles $\tilde{\varphi}_\rho$ around the critical curve: this corresponds to the fact that an asymptotically bound photon can reach the observer on either an upwards or downwards libration.
Only the equatorial observer with $i=90^\circ$ can see the full set of orbits within the photon shell; other observers see only the subset of orbits that have sufficient inclination to reach them (that is, such that $\tilde{\rho}^2\geq0$).
Examples of critical curves for various BH spins and inclinations are plotted in many papers \citep[e.g., Fig.~1 in][]{Farah2020}.

Finally, we emphasize that the notion of a critical curve is not limited to observers far from the BH.
For instance, the critical curve for observers on timelike circular-equatorial orbits around a Kerr BH is analyzed in \citet{Gates2021}.

\subsection{Nearly bound photons form the photon ring}

A photon with conserved quantities $(\lambda,\eta)\approx(\tilde{\lambda},\tilde{\eta})$ very close to those of a photon bound at orbital radius $\tilde{r}$ can describe multiple oscillations in $\theta\in[\tilde{\theta}_-,\tilde{\theta}_+]$ near the photon shell before leaving it.
Since Kerr bound orbits are unstable, their nearby, nearly bound geodesics display an exponential rate of radial deviation with respect to their (fractional) number $N$ of polar half-orbits,
\begin{align}
	\ab{r-\tilde{r}}\propto e^{\gamma N},
\end{align}
where one full orbit is defined as a complete oscillation from a turning point $\theta_\pm$ back to itself, while fractional $N$ is precisely defined in Eq.~(36) of \citet{GrallaLupsasca2020a}.
The Lyapunov exponents governing the orbital instability depend on both the BH spin and photon shell radius \citep{Johnson2020}:
\begin{align}
	\gamma=\frac{4}{a_*}\sqrt{r^2-\frac{Mr\Delta}{(r-M)^2}}\int_0^1\frac{\ed t}{\sqrt{\pa{1-t^2}\pa{\tilde{u}_+t^2-\tilde{u}_-}}}.
\end{align}

By definition, when a nearly bound photon reaches a distant observer, it appears at a position close to the critical curve \eqref{eq:CriticalCurve}: $(\rho,\varphi_\rho)\approx(\tilde{\rho}(\tilde{r}),\tilde{\varphi}_\rho(\tilde{r}))$ where $\tilde{r}$ is the radius of the nearby bound orbit.
More precisely, its arrival position can be labeled by the coordinates $(\tilde{r},d)$, where $d$ denotes the perpendicular distance from the nearest point $(\tilde{\rho}(\tilde{r}),\tilde{\varphi}_\rho(\tilde{r}))$ on the critical curve.
\citet{Johnson2020} heuristically showed, and then \citet{GrallaLupsasca2020a} rigorously proved, that a light ray shot backwards into the geometry from position $(\tilde{r},d)$ in the sky executes
\begin{align}
	\label{eq:LogDivergence}
	N(\tilde{r},d)\approx-\frac{1}{\gamma(\tilde{r})}\log\ab{d}
\end{align}
half-orbits around the BH before either returning to asymptotic infinity, if it was outside the critical curve ($d>0$), or crossing the event horizon, if it was inside ($d<0$).
\citet{GrallaLupsasca2020a} also compute subleading $\mathcal{O}(d^0)$ corrections to $N(\tilde{r},d)$.

While Eq.~\eqref{eq:LogDivergence} is a purely geometric statement about the lensing behavior of a Kerr BH, it also constrains the structure of BH images.
To describe these constraints, we will from now on restrict our attention to emission regions that are optically thin, stationary and axisymmetric.
The assumption of stationarity and axisymmetry amounts to considering only time-averaged images (averaged over sufficiently long timescales), and the assumption of infinite optical depth allows one to neglect absorption effects; as discussed in Sec.~\ref{subsec:Shadow}, this approximation becomes exact in the limit of large observation frequency, and could already provide a good approximation at the frequencies of current or near-future observations of M87*.

Since the emission region around the BH is optically thin, a light ray that passes through it $2N$ times (each time at essentially the same inclination) collects roughly twice as many photons as a neighboring light ray that passes through it $N$ times.
That is,
\begin{align}
	\label{eq:IntensityPeak}
	I(\tilde{r},d)\propto N(\tilde{r},d)
	\sim-\frac{1}{\gamma(\tilde{r})}\log\ab{d}.
\end{align}
As such, the lensing formula \eqref{eq:LogDivergence} implies a mild (logarithmic) divergence in the observed intensity $I(\rho,\varphi_\rho)$ near the critical curve.
The photon ring is defined as the intensity bump caused by this divergence; in practice, it is cut off by absorption effects (finite optical depth) as $N$ grows large.
In summary: GR predicts that embedded within every image of a BH surrounded by an optically thin emission region, there lies a narrow photon ring.

Due to Eq.~\eqref{eq:LogDivergence}, any axisymmetric ring emitting isotropically produces an infinite sequence of lensed images within the photon ring \citep{GrallaLupsasca2020a}.
Successive images within this sequence are produced by photons describing increasingly many half-orbits around the black hole: if the first (direct) image arises from photons with fractional half-orbit number $0\le N_0<1$, then the $n^\text{th}$ lensed image will arise from photons with fractional half-orbit number $N\approx N_0+n$.
Since the entirety of the emission surrounding the black hole can be decomposed into such source rings, it follows that the photon ring consists of a series of lensed images of the main emission labeled by half-orbit number $n\in\mathbb{N}$.

Moreover, Eq.~\eqref{eq:LogDivergence} implies that these successive images are increasingly lensed towards the critical curve by an exponential demagnification factor: a source ring whose $n^\text{th}$ image appears at perpendicular distance $d_n(\tilde{r})$ from the critical curve will produce its $(n+1)^\text{th}$ image at an exponentially smaller distance
\begin{align}
	\label{eq:Demagnification}
	d_{n+1}(\tilde{r})=e^{-\gamma(\tilde{r})}d_n(\tilde{r}).
\end{align}
Depending on whether the emission region is gapped or not, this lensing pattern can result in one of two very distinctive image substructures within the photon ring.

If the BH is entirely immersed in its surrounding emission (as in the case of a spherically symmetric accretion flow onto the hole, for instance), then successive images of the emission region continuously blend together and the resulting intensity profile logarithmically rises to a smooth peak centered about the critical curve.
Such an intensity profile is displayed in Fig.~1 of \citet{Narayan2019}, with the corresponding image shown in Fig.~2 therein: a bright photon ring encircling a central brightness deficit (the interior of the critical curve).
Moreover, as shown in that paper, if the emitting matter is infalling, then beaming effects strongly attenuate the intensity inside the critical curve, creating the distinctive shadow effect reviewed in Sec.~\ref{subsec:Shadow} above.
The reason is that light rays appearing inside the critical curve cannot encounter a radial turning point, so any photons collected along such rays had to have been emitted opposite to the infalling matter's direction of motion, thereby incurring a strong redshift.

On the other hand, if the emission is gapped (as in the case of emission localized around the equator with gaps near the poles, for instance), then each successive image of the emission region displays a corresponding gap, and the resulting intensity profile no longer consists of a smooth logarithmic peak \eqref{eq:IntensityPeak}, but rather a sequence of distinct peaks accumulating near the critical curve.
Such an intensity profile is displayed in Fig.~1 of \citet{Gralla2019}, with the corresponding image shown in Fig.~5 therein.
In such images, the photon ring decomposes into a stack of distinct photon subrings, each of which is labeled by $n$ and corresponds to a single lensed image of the main emission, demagnified by $e^{-\gamma n}$.
As reviewed in Sec.~\ref{subsec:Shadow}, the latest EHT observations favor M87* models with near-equatorial emission, and thus we expect this subring structure to be present in actual images of M87*.

High-resolution, time-averaged, 230\,GHz images of M87* produced using state-of-the-art GRMHD simulations display at least $n=3$ subrings and numerically confirm this GR-predicted substructure \citep{Johnson2020}, which is also present in all the images that we obtained from our toy models of M87*.

\subsection{Photon subrings lie in lensing bands}
\label{subsec:LensingBands}

\begin{figure}[ht]
	\centering
	\includegraphics[width=\columnwidth]{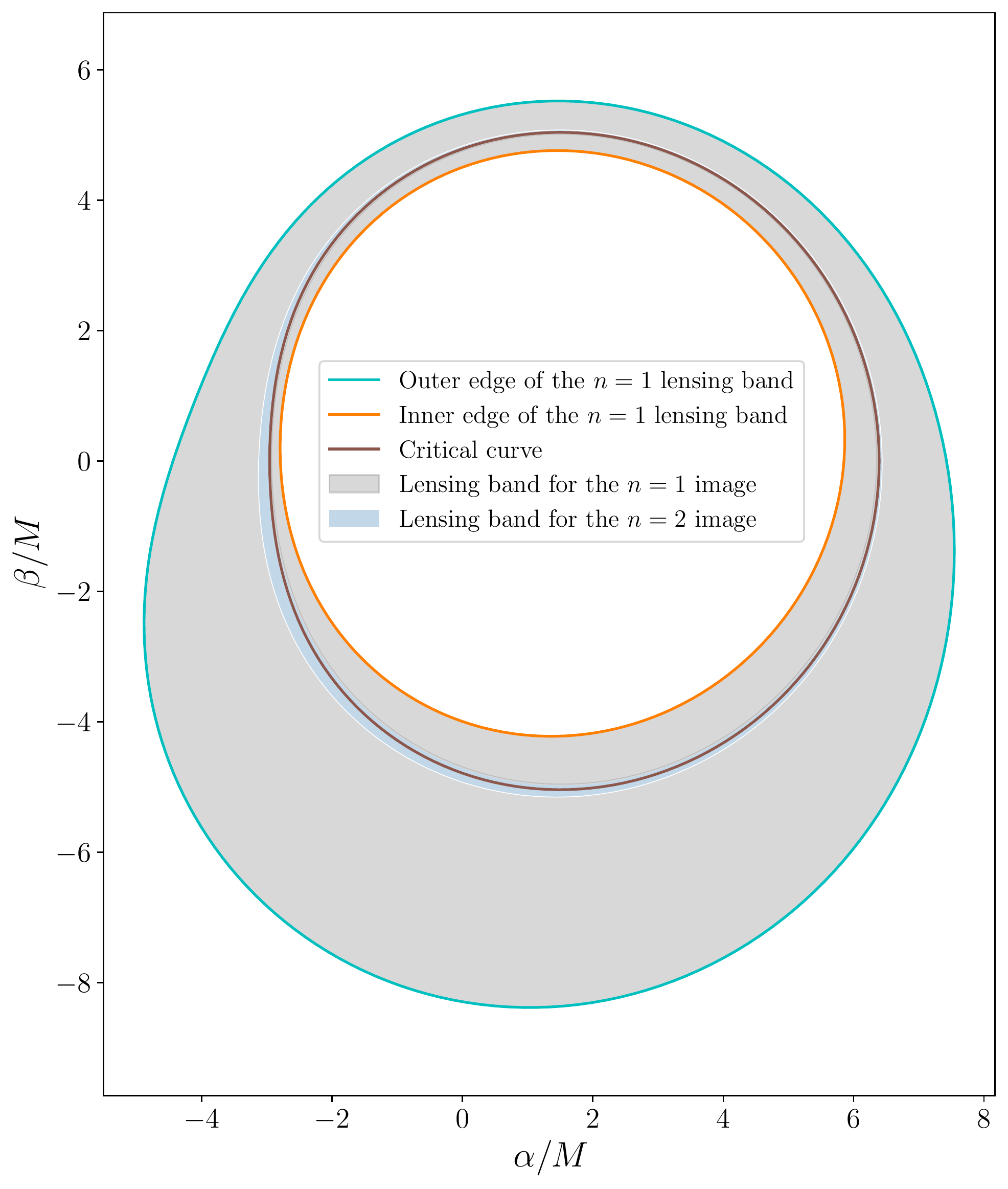}
	\caption{
	$n=1$ and $n=2$ lensing bands for a distant observer at inclination $i=45^\circ$ from a BH with spin $a=0.99$. The critical curve is given as a reference.
	}
	\label{fig:LensingBands}
\end{figure}

To study the subring structure of the photon ring, it is helpful to introduce the (purely geometric) notion of a lensing band \citep{GLM2020,Chael2021}: a Kerr observer's $n^\text{th}$ lensing band is defined as the region in the sky corresponding to geodesics that cross the equatorial plane at least $n+1$ times before escaping to asymptotic infinity (if $d>0$) or crossing the event horizon $r=r_+$ of the BH (if $d<0$), where the inner and outer horizon radii are
\begin{align}
	\label{eq:HorizonRadius}
	r_\pm=M\pm\sqrt{M^2-a^2}.
\end{align}
We present a general method for computing Kerr lensing bands in App.~\ref{app:LensingBands}.
Figure~\ref{fig:LensingBands} illustrates the $n=1$ and $n=2$ lensing bands for an observer at inclination $i=45^\circ$ from a BH of spin $a=0.99$.

Each $n>0$ lensing band consists of a region bounded by two concentric, closed curves, and therefore takes an annular shape.
The inner and outer edges of the $n^\text{th}$ lensing band correspond to the $n^\text{th}$ lensed images of the equatorial circles of radius $r=r_+$ and $r\to\infty$, respectively.
The inner edge is always contained within the critical curve, while the outer edge always encircles it on the outside.
According to Eq.~\eqref{eq:Demagnification}, these edges converge to the critical curve exponentially fast in $n$, with widths scaling as
\begin{align}
	\label{eq:Widths}
	w_{n+1}(\tilde{r})\approx e^{-\gamma(\tilde{r})}w_n(\tilde{r})
	\approx e^{-n\gamma(\tilde{r})}w_1(\tilde{r}),
\end{align}
where $w_1(\tilde{r})$ is the angle-dependent width of the first ($n=1$) lensing band (here parameterized by photon shell radius) and the first relation becomes exact in the limit of large $n\to\infty$.
In short, the lensing bands form a stacked sequence of annular shapes, each straddling the critical curve, with the $(n+1)^\text{th}$ lensing band strictly contained within the $n^\text{th}$ one.

From now on, we further restrict our attention to a thin disk of isotropic emitters filling the equatorial plane.
In this setting, the lensing bands are particularly useful because they coincide with the photon subrings: the $n^\text{th}$ lensed image of the disk, which produces the $n^\text{th}$ photon subring, exactly fills the $n^\text{th}$ lensing band.
It follows that the photon subrings also display a stacked structure with widths scaling as in Eq.~\eqref{eq:Widths}.
Hence, measuring the angle-dependent ratio $w_{n+1}(\tilde{r})/w_n(\tilde{r})\approx e^{-\gamma(\tilde{r})}$ of successive subrings could (in this setting) allow for the Lyapunov exponents $\gamma(\tilde{r})$ to be experimentally determined \citep{Johnson2020}.

\subsection{Interferometric signature of a narrow curve in the sky}

Having described the subring structure of BH images, we next turn to a description of the corresponding subring structure in the visibility (Fourier) domain.
We begin by reviewing some facts about the interferometric signature of a narrow feature in the sky.

We consider an infinitely thin, closed light curve in the sky.
Assuming this plane curve is convex, we may parameterize it by normal angle $\varphi\in[0,2\pi)$ as $\vec{r}(\varphi)=(\alpha(\varphi),\beta(\varphi))$, where we temporarily revert to Cartesian coordinates $(\alpha,\beta)$.
(Non-convex curves decompose into multiple such segments $\vec{r}_i(\varphi)$, but we will not need this generalization here.)
The inward-pointing unit normal is $\hat{\vec{n}}(\varphi)=-(\cos{\varphi},\sin{\varphi})$.
Following \citet{GrallaLupsasca2020c}, we define the ``projected position'' $f(\varphi)$ by
\begin{align}
	\label{eq:ProjectedPosition}
	f(\varphi)\equiv-\vec{r}(\varphi)\cdot\hat{\vec{n}}(\varphi)
	=\alpha(\varphi)\cos{\varphi}+\beta(\varphi)\sin{\varphi}.
\end{align}
This function encodes the whole curve $\vec{r}(\varphi)=(\alpha(\varphi),\beta(\varphi))$, which may be reconstructed as
\begin{subequations}
\begin{align}
	\alpha(\varphi)&=f(\varphi)\cos{\varphi}-f'(\varphi)\sin{\varphi},\\
	\beta(\varphi)&=f(\varphi)\sin{\varphi}+f'(\varphi)\cos{\varphi}.
\end{align}
\end{subequations}
The projected position $f(\varphi)$ thus encodes all other properties of the curve, such as its radius of curvature $\mathcal{R}(\varphi)=f(\varphi)+f''(\varphi)$.
The real motivation for introducing the quantity $f(\varphi)$ is that it naturally connects to the interferometric signature of the curve.
To see this, we first note that the projected position can be uniquely decomposed into parity-even and parity-odd parts
\begin{align}
	f(\varphi)=\frac{d_\varphi}{2}+C_\varphi,
\end{align}
where
\begin{subequations}
\begin{align}
	d_\varphi&=f(\varphi)+f(\varphi+\pi),\\
	C_\varphi&=\frac{1}{2}\br{f(\varphi)-f(\varphi+\pi)}.
\end{align}
\end{subequations}
The even part $d_\varphi=d_{\varphi+\pi}$ is the ``projected diameter'' of the curve at angle $\varphi$, while the odd part $C_\varphi=-C_{\varphi+\pi}$ is its centroid motion.
\citet{GrallaLupsasca2020c} illustrate several examples of this shape decomposition.

We now arrive at the key point: if the curve is not infinitely thin, but still very narrow relative to its diameter ($w\ll d_\varphi$), then its Fourier transform $V(u,\varphi)$ (where $(u,\varphi)$ are polar coordinates in the Fourier plane) exhibits a simple behavior in the regime
\begin{align}
	\label{eq:UniversalRegime}
	\frac{1}{d}\ll u\ll\frac{1}{w},
\end{align}
where it adopts the ``universal" form, valid to leading order in an expansion in $0<w/d\ll1$ \citep{Gralla2020},
\begin{align}
	\label{eq:UniversalVisibility}
	V(u,\varphi)\approx\frac{e^{-2\pi iC_\varphi u}}{\sqrt{u}}\br{\alpha_L(\varphi)e^{-\frac{i\pi}{4}}e^{i\pi d_\varphi u}+\alpha_R(\varphi)e^{\frac{i\pi}{4}}e^{-i\pi d_\varphi u}},
\end{align}
with $\alpha_{L,R}(\varphi)=\alpha_{R,L}(\varphi+\pi)>0$ encoding the intensity profile of the curve and ensuring that $V(u,\varphi+\pi)=V^*(u,\varphi)$, as required by definition [Eq.~\eqref{eq:ComplexVisibility} below].
In the context of interferometry, the complex function $V(u,\varphi)$ is known as the radio visibility, with the visibility plane $(u,\varphi)$ usually referred to as the baseline plane.
Intuitively, short baselines $u\ll1/d$ are not sufficient to resolve the shape of a bright curve, which then just appears as a blob, while on very long baselines $u\gg1/w$, the smooth profile of the curve has been resolved and its signal therefore dies off exponentially; on the other hand, in the regime \eqref{eq:UniversalRegime}, the curve appears infinitely thin and as a result, its signal follows a very weak $u^{-1/2}$ power-law fall-off \eqref{eq:UniversalVisibility}.

In the context of realistic high-frequency VLBI observations, it is in practice significantly easier to measure the amplitude $|V|$ of the complex visibility---the ``visamp''---than its absolute phase, as the latter is more susceptible to contamination by a variety of atmospheric and instrumental effects.
While the full complex visibility \eqref{eq:UniversalVisibility} encodes both $d_\varphi$ and $C_\varphi$, the visibility amplitude depends on $d_\varphi$ only, and obeys $|V(u,\varphi+\pi)|=|V(u,\varphi)|$:
\begin{align}
	\label{eq:UniversalAmplitude}
	\ab{V(u,\varphi)}&\approx\sqrt{\frac{\alpha_L^2(\varphi)+\alpha_R^2(\varphi)+2\alpha_L(\varphi)\alpha_R(\varphi)\sin\pa{2\pi d_\varphi u}}{u}}.
\end{align}
To summarize, the clearest interferometric signature of a bright curve in the sky is its visibility amplitude \eqref{eq:UniversalAmplitude}, which is strong in the universal regime \eqref{eq:UniversalRegime} and only encodes partial information about the shape, namely its projected diameter $d_\varphi$ (but not the full shape information, which also requires the centroid motion $C_\varphi$ that is only available in the harder to measure visibility phase).

\begin{figure}[ht]
	\centering
	\includegraphics[width=\columnwidth]{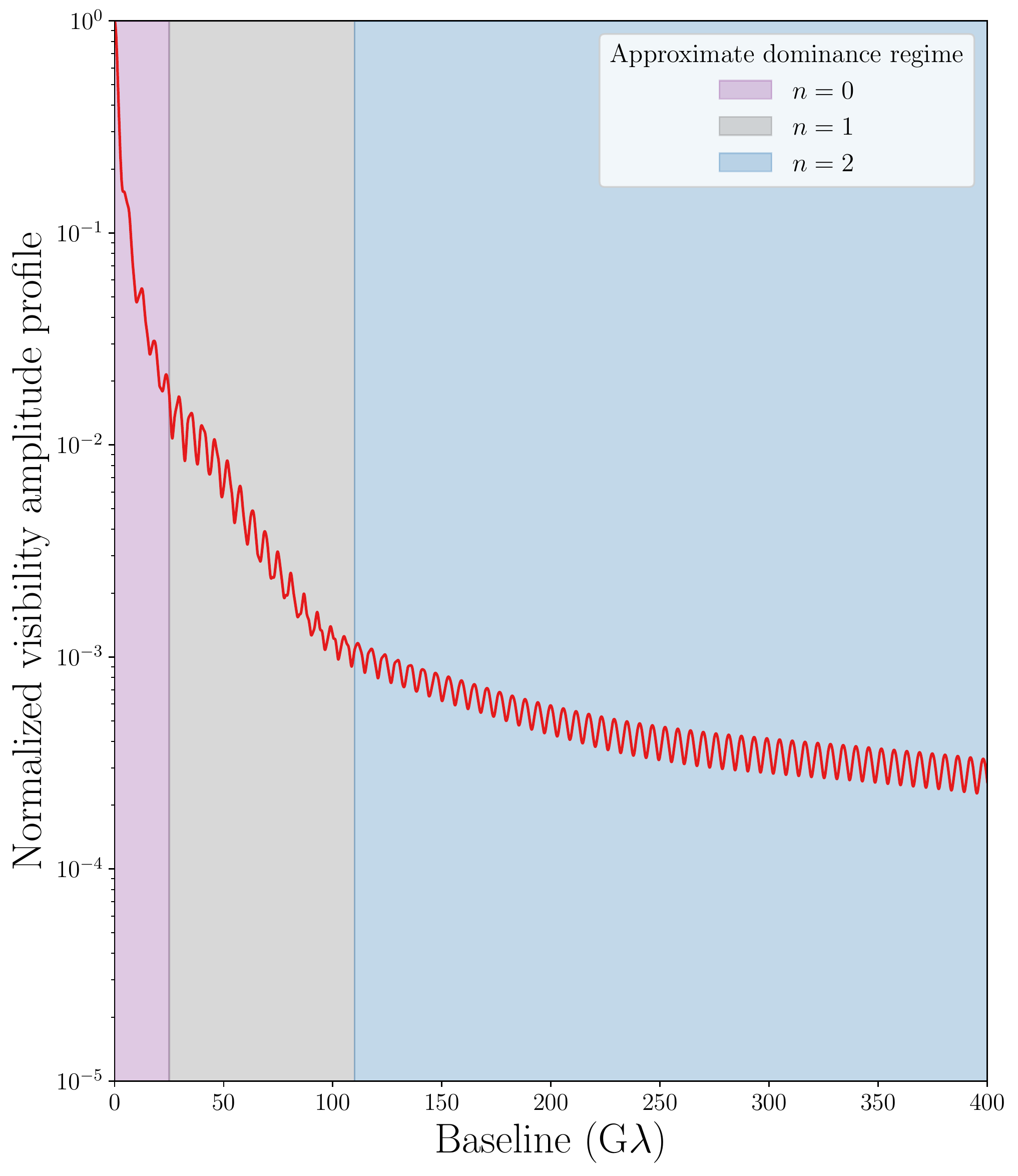}
	\caption{
	Example of a visibility amplitude profile at fixed angle $\varphi=0^\circ$ in the baseline plane, obtained from a BH image simulated with \texttt{Gyoto} and exhibiting the clean ringing expected from the narrow photon ring.
	Three regimes \eqref{eq:SubringRegime} are visible: in the range $[0,20]$\,G$\lambda$, the direct ($n=0$) image dominates; for $u\in[20,120]$\,G$\lambda$, the first ($n=1$) subring prevails; beyond 120\,G$\lambda$, the $n=2$ ring signature takes over.
	}
	\label{fig:TypicalVisibility}
\end{figure}

\subsection{Interferometric signature of the photon ring}
\label{subsec:Subrings}

Combining the results of the previous sections, we are now in a position to describe the interferometric signature produced by the photon ring.
We expect the photon subrings to be the only narrow features present in time-averaged BH images, since other fine features (such as emission ropes and other flares) should be transient and wash out after sufficient time-averaging.
Thus, we expect that the $n^\text{th}$ subring will dominate the visibility amplitude in the regime  \citep{Johnson2020}
\begin{align}
	\label{eq:SubringRegime}
	\frac{1}{w_{n-1}}\ll u\ll\frac{1}{w_n},
\end{align}
where, according to Eq.~\eqref{eq:Widths}, $w_n(\tilde{r})=e^{-\gamma(\tilde{r})}w_{n-1}(\tilde{r})$.
Beyond, its signal starts to die off exponentially and the $(n+1)^\text{th}$ subring begins to dominate.
This results in a distinctive cascade structure illustrated in Fig.~\ref{fig:TypicalVisibility}.
In each range \eqref{eq:SubringRegime}, the $n^\text{th}$ subring produces a damped oscillation \eqref{eq:UniversalAmplitude} governed by its projected diameter $d_\varphi^{(n)}$.
All that remains for us to do is to describe the diameters $d_\varphi^{(n)}$, beginning with the projected diameter $\tilde{d}_\varphi\equiv d_\varphi^{(\infty)}$ of the critical curve, which they exponentially converge to in the limit $n\to\infty$.

\subsection{Theoretical shape of the critical curve}
\label{subsec:CriticalCurveShape}

A ``phoval'' (for ``photon ring oval'') is a geometric shape with
\begin{subequations}
\label{eq:Phoval}
\begin{align}
	\label{eq:Circlipse}
	\frac{d_\varphi}{2}&=R_0+\sqrt{R_1^2\sin^2{\varphi}+R_2^2\cos^2{\varphi}},\\
	\label{eq:Centroid}
	C_\varphi&=\pa{X-\chi}\cos{\varphi}+\arcsin\pa{\chi\cos{\varphi}}.
\end{align}
\end{subequations}
More precisely, this is a family of shapes parameterized by five parameters $R_0$, $R_1$, $R_2$, $X$, and $\chi$, with the interesting property that the Kerr critical curve is extremely well-approximated by some member of the family for any choice of observer inclination and BH spin \citep{GrallaLupsasca2020c}.

We briefly summarize the geometric interpretation of these parameters.
First, the projected position $f_{\rm circ}(\varphi)=R_0$ describes a circle centered at the origin with radius $R_0$.
Next, the projected position $f_{\rm ellipse}^2(\varphi)=R_1^2\sin^2{\varphi}+R_2^2\cos^2{\varphi}$ describes an ellipse centered at the origin with radii $R_1$ and $R_2$.
Adding these two functions results in the projected position $f_{\rm circlipse}(\varphi)=d_\varphi/2$ given in Eq.~\eqref{eq:Circlipse}, which defines a shape known as a ``circlipse.''
It follows from the last paragraph that the projected diameter of the critical curve is always exquisitely close to that of a circlipse,
\begin{align}
    \label{eq:CriticalCurveShape}
	\tilde{d}_\varphi\equiv d_\varphi^{(\infty)}\approx
	2R_0+2\sqrt{R_1^2\sin^2{\varphi}+R_2^2\cos^2{\varphi}},
\end{align}
with the best-fit parameters $R_0$, $R_1$, and $R_2$ depending in a rather intricate way on the mass-to-distance ratio $M/D$, BH spin $a$, and observer inclination $i$.
Finally, adding the centroid motion \eqref{eq:Centroid} to the projected function produces the full phoval shape, which is translated along the $\alpha$ axis (orthogonal to the spin axis projected onto the plane perpendicular to the line of sight) by a horizontal offset $X\in\mathbb{R}$, and carries a left-right asymmetry induced by a parameter $\chi\in[-1,1]$ that matches the warping of the critical curve at high inclinations and high spins.
With these additional parameters, the best-fit phoval reproduces the Kerr critical curve to a part in $10^5$ over the vast majority of parameter space, and to a part in $10^3$ in the extremal limit $a\to1$ \citep{GrallaLupsasca2020c}.
Away from extremality, the (parity-odd) centroid motion gives a minor correction to the shape of the critical curve, which is almost entirely encoded in its (parity-even) circlipse diameter.

A critical curve with best-fit circlipse parameters $R_0$, $R_1$, and $R_2$ has diameters $d_\perp\equiv d_0$ and $d_\parallel\equiv d_{\pi/2}$ along the directions perpendicular ($\alpha$ axis) and parallel ($\beta$ axis) to the projected spin
\begin{align}
	\label{eq:Diameters}
	d_\perp=2\pa{R_0+R_2},\qquad
	d_\parallel=2\pa{R_0+R_1},
\end{align}
which define a fractional asymmetry \citep{Johnson2020}
\begin{align}
	f_A\equiv 1-\frac{d_\perp}{d_\parallel}
	=\frac{R_1-R_2}{R_0+R_1}
	\in[0,1].
\end{align}
The range $0\leq f_A\leq 1$ follows from the observation that $d_\perp\leq d_\parallel$ over all parameter space, which also implies that $R_2\leq R_1$.
Moreover, we can see from Eq.~\eqref{eq:Circlipse} that $d_\perp$ and $d_\parallel$ are the minimal and maximal diameters of the circlipse, respectively.

Finally, we point out that the two quantities $d_\parallel$ and $f_A$ encode the BH spin $a$ and observer inclination $i$, as illustrated in Fig.~7 of \citet{Johnson2020}.
The map $(a,i)\to(d_\parallel,f_A)$ is not exactly bijective, however, but rather two-to-one, as $(a,i)$ and $(-a,i+\pi)$ both map to the same $(d_\parallel,f_A)$.
Putting aside this discrete twofold degeneracy (which can be broken by other means), this suggests that one could in principle infer a BH's spin and inclination from the projected diameter $\tilde{d}_\varphi$ of its critical curve [Eq.~\eqref{eq:CriticalCurveShape}], or even just its minimum and maximum $d_\perp$ and $d_\parallel$ (without any need for the parameters $X$ and $\chi$ encoding its centroid).

However, this suggestion is misleading because the critical curve is purely theoretical and not in itself observable.
Nonetheless, the photon subrings---which are observable---converge to it exponentially fast and moreover, while the first subring is still noticeably different to the naked eye, the second subring tracks it quite closely.
It is therefore tempting to extract the BH spin and inclination from the projected diameter $d_\varphi^{(2)}$.
Unfortunately, we have found this approach to be only moderately successful, as the residual astrophysics-dependence of the $n=2$ ring shape introduces significant uncertainties (see Sec.~\ref{subsec:ModelSurveyResults} below).

\subsection{Testing GR with the photon ring shape}
\label{subsec:Test}

As we have explained, the projected diameter of the Kerr critical curve depends only on the spacetime geometry: $\tilde{d}_\varphi$ encodes the BH spin and inclination, and is fully determined by them.
However, this theoretical curve, and hence its diameter $\tilde{d}_\varphi$, are not directly observable.
Instead, what we may (in principle) observe is a sequence of photon subrings labeled by half-orbit number $n$ and approaching the critical curve as $n\to\infty$, with projected diameters $d_\varphi^{(n)}\to d_\varphi^{(\infty)}=\tilde{d}_\varphi$.
Thus, in the large-$n$ limit, the rings enter a ``universal'' regime in which their dependence on astrophysical conditions drops out and they converge to the GR-predicted, astrophysics-independent shape $\tilde{d}_\varphi$.
This suggests that a shape measurement of a ``large-$n$'' subring could provide a test of GR and a means to infer the BH parameters.

The first ($n=1$) subring is certainly not yet in the universal regime, as the photons comprising it are not sufficiently bent and its shape still carries a substantial imprint of the astrophysical details of the surrounding source.
On the other hand, the shape of the higher $n\ge3$ subrings is much less sensitive on astrophysics, but these rings are also much harder to detect experimentally, for at least two reasons: first, their signature only comes to dominate on exponentially long baselines [Eq.~\eqref{eq:SubringRegime}], which (given present observation frequencies) could only be accessed using a space element at unrealistic separation from the Earth \citep[e.g., Fig.~5 of][]{Johnson2020}; second, this problem is compounded by the need for exceptional sensitivity, as the flux in each subring is also exponentially suppressed in $n$ [Eq.~\eqref{eq:Widths}].
As such, the second subring seems to be the most promising target for a shape test of GR, as it may sit in the ``sweet spot'' where it remains accessible with current or near-future technology \citep{GLM2020}, while $n=2$ is ``sufficiently large'' that $d_\varphi^{(2)}\approx\tilde{d}_\varphi$.

\citet{GLM2020} initiated a quantitative investigation of this idea and found that in a wide class of toy models of M87* (subject to the limitations listed in the introduction), the $n=2$ subring indeed produced the universal visibility amplitude \eqref{eq:UniversalAmplitude}.
Moreover, they showed that it is possible to recover $d_\varphi^{(2)}$ from this signature, even in the presence of simulated instrument noise.
They found that, while the $n=2$ ring is still noticeably different from its $n\to\infty$ limit (the critical curve), it nonetheless adopts the same shape: more precisely, even though the $n=2$ ring and the critical curve have different projected diameters $d_\varphi^{(2)}\neq\tilde{d}_\varphi$, they both take the shape of a circlipse.
That is, $d_\varphi^{(2)}$ must also follow the functional form \eqref{eq:CriticalCurveShape}.

We emphasize that, while the best-fit radii $R_0$, $R_1$, and $R_2$ for $\tilde{d}_\varphi$ are determined only by the BH parameters $M/D$ and $a$, as well as the observer inclination $i$, the best-fit radii $R_0$, $R_1$, and $R_2$ for $d_\varphi^{(2)}$ additionally depend on astrophysical conditions as well.
This is illustrated in Fig.~7 of \citet{GLM2020}, which shows how these best-fit parameters vary with the astrophysical model, keeping the BH parameters fixed.
Thus, the $n=2$ ring carries some astrophysics-dependence (which makes it harder to extract the BH parameters from its diameter), but it is weak enough that $d_\varphi^{(2)}$ does not break out of the functional form
\begin{align}
	\label{eq:FunctionalForm}
	\frac{d_\varphi^{(2)}}{2}\approx R_0+\sqrt{R_1^2\sin^2\pa{\varphi-\varphi_0}+R_2^2\cos^2\pa{\varphi-\varphi_0}},
\end{align}
which includes an additional parameter $\varphi_0$ to account for the fact that the orientation of the image is generically not aligned with the Bardeen coordinate system.
This observation led \citet{GLM2020} to propose a novel GR test based on the $n=2$ ring shape: a space-based VLBI mission targeting M87* could extract $d_\varphi^{(2)}$ and check to what extent it follows the form \eqref{eq:FunctionalForm}.

If the deviation of $d_\varphi^{(2)}$ from its best-fit circlipse is large, then the Kerr hypothesis fails the test; if it is small, then one can report a test of the Kerr hypothesis and GR with a precision given by the root-mean-square deviation of $d_\varphi^{(2)}$ from its best fit.
With their simulated experimental forecast, \citet{GLM2020}  attained a stringent, sub-percent (0.04\%) level of precision.

Finally, we wish to clarify in what sense we regard the GLM method as a test of the Kerr hypothesis.
Such a test can either be:
\begin{itemize}
	\item[\textbullet] a \textit{consistency test} aiming to confirm that some observable is consistent with the prediction of the Kerr spacetime, in which case there is no notion of model comparison; or,
	\item[\textbullet] a \textit{model-comparison test} aiming to compare predictions of the Kerr hypothesis with alternative predictions derived from other spacetime geometries, so as to determine which one best explains the data (in the Bayesian sense).
\end{itemize}

In this article, we discuss the GLM method as a consistency test of the Kerr hypothesis and thus do not consider any other spacetime but that of Kerr.
Performing an unambiguous model-comparison test of the Kerr hypothesis is a difficult challenge, as one must be careful to properly take into account all complex astrophysical degeneracies \citep[e.g.,][]{Bauer2021}.

Additional work is needed to establish the viability of this GLM method for a wider range of configurations.
In this paper, we explore whether the method remains viable when some of the limitations of the original GLM analysis are removed; this will be the object of our parameter survey in Secs.~\ref{sec:EmissionSurvey} and \ref{sec:SpinSurvey} below.

\section{Implementation} 
\label{sec:Implementation}

To check the robustness and precision of the Kerr hypothesis test reviewed in Sec.~\ref{sec:Theory}, we simulated high-resolution images of Kerr BHs using various types of emission and confirmed that they all display a photon ring with $n=1$ and $n=2$ subrings.
Then, we inferred the $n=2$ ring diameter $d_\varphi^{(2)}$ from its characteristic damped oscillation \eqref{eq:UniversalAmplitude} in the visibility amplitude on long baselines, and verified that it follows the predicted functional form \eqref{eq:FunctionalForm}, thereby establishing the viability of the GLM method as a test of the Kerr hypothesis.
We also studied the discrepancy between the $n=2$ subring and the critical curve, and checked that a phoval with the minimal and maximal diameters measured from the $n=2$ ring could fit within the $n=2$ lensing band.
Our simulations were performed with the relativistic ray tracing code \texttt{Gyoto} (General relativitY Orbit Tracer of the \textit{Observatoire de Paris}), whose details are presented in \citet{Vincent2011}.

\subsection{Image simulation}

\texttt{Gyoto} simulates an image, that is, a map $I(\rho,\varphi_\rho)$ of the specific intensity in the observer sky.
To reproduce the complex visibility $V(u,\varphi)$ that is directly sampled via VLBI observations, we had to additionally compute its 2D Fourier transform,
\begin{align}
	\label{eq:ComplexVisibility}
	V(u,\varphi)=\int_0^\infty\int_0^{2\pi}I(\rho,\varphi_\rho+\varphi)e^{-2\pi iu\rho\cos{\varphi_\rho}}\rho\ed\rho\ed\varphi_\rho,
\end{align}
which by definition satisfies $V(u,\varphi+\pi)=V^*(u,\varphi)$.
Instead of computing this 2D FT, we followed \citet{GLM2020} and  made use of the projection-slice theorem to directly compute $V(u,\varphi)$ along slices of fixed angle $\varphi$ in the Fourier plane: for each $\varphi$, we computed the Radon projection (i.e., the integrals of $I(\rho,\varphi)$ along lines perpendicular to the slice of constant $\varphi$ across the image), and then applied a 1D fast Fourier transform (FFT) to obtain the visibility $V(u,\varphi)$ evaluated at that angle $\varphi$.

We made the following assumptions in our computations: 
\begin{itemize}
	\item[\textbullet] BH mass: $M=6.2\times10^9M_\odot$, in the range of values for M87* favored by stellar dynamical measurements \citep[e.g.,][]{Gebhardt2011,EHT6};
	\item[\textbullet] distance between the BH and observer: $D=16.9$\,Mpc, in the range of values for M87* \citep[e.g.,][]{EHT6};
	\item[\textbullet] observation wavelength: $\lambda=1.3$\,mm, matching the 230\,GHz frequency of current observations  \citep{EHT2};
	\item[\textbullet] we neglected absorption effects, a suitable approximation for optically thin accretion flows, as discussed in Sec.~\ref{sec:Theory} above;
	\item[\textbullet] we used Bardeen's coordinates \eqref{eq:BardeenCoordinates} in the sky (such that the projected spin axis points in the direction $\varphi_\rho=90^\circ$);
	\item[\textbullet] we terminated the ray tracing of geodesics after their third equatorial crossing, so as to avoid parasitic pixels produced by an under-resolved, partially imaged $n=3$ ring.
\end{itemize}
From these values of $M$ and $D$, one obtains a conversion ratio from microarcseconds to units of $M$ consistent with EHT priors \citep{EHT6}:
\begin{align}
	(M/D)_{\rm M87^*}=3.6212\,\mu{\rm as}.
\end{align}
We took our field of view to be of 180\,$\mu{\rm as}\approx50M$ and ray traced most images with a resolution of $10000\times10000$ pixels, though to resolve the $n=2$ ring, we found it necessary in some cases (when the ring is highly peaked) to double the linear resolution.

\subsection{Image treatment and analysis}

\subsubsection{Apodization}

We find that the visibility amplitude profile $|V(u,\varphi)|$ obtained from the FFT of a Radon slice indeed presents (on sufficiently long baselines) the expected periodicity linked to the $n=2$ ring.
Without further treatment, however, this ringing signature does not quite take the ``clean'' form \eqref{eq:UniversalAmplitude} because it is ``polluted'' by some other oscillation with higher period and lesser amplitude.

This effect arises from an artificial discontinuity in the Radon projection, which is set to zero outside our field of view but does not exactly vanish at its edges, where it instead decays to a tiny (but still nonzero) intensity that is $\lesssim10^{-4}$ times the maximum intensity.
Though small, this step function nonetheless produces its own ringing on long baselines, with periodicity set by the size of the field of view, and this interferes with our signal.

To eliminate this spurious artifact, one solution would be to increase the field of view until the intensity at the edges is so small that this effect vanishes.
However, in order to still resolve the $n=2$ ring, these larger images would have to be ray traced at a correspondingly higher resolution.
This would be too costly, especially since the bulk of the computation would be dedicated to exterior pixels of very little interest to us.
An ingenious way to overcome this difficulty would be to use adaptative ray tracing, either by employing algorithms designed for this purpose \citep[e.g.,][]{Gelles2021}, or else by decomposing the total image into a sum of multiple layers, each with its own resolution.
The latter approach was adopted by \citet{GLM2020}, who divided the image into layers labeled by half-orbit number $n$: since the $n^\text{th}$ layer only has nonzero pixels in the $n^\text{th}$ lensing band, which is exponentially small, they were able to exponentially increase the resolution in each layer while keeping fixed the number of pixels computed in each layer.

Instead of computing images with an unnecessarily large field of view or using adaptive ray tracing, we dealt with this effect by simply multiplying the Radon projection by a window function.
This method is known as apodization.

More specifically, we used the window function (also known as an apodization or tapering function)
\begin{align}
	W(x)=p(r(1-x))\,p(r(1+x)),
\end{align}
where $x\in[-1,1]$ ranges over the Radon projection, $\epsilon=1/r$ is the range of the cutoff,	and $p$ is a ``smooth plateau'' function,
\begin{align}
	p(x)=\frac{f(x)}{f(x)+f(1-x)},\qquad
	f(x)=\left\{
		\begin{array}{ll}
			e^{-1/x^2} & \mbox{if }x>0,\\
			0 & \mbox{if }x\leq0.
		\end{array}
	\right.
\end{align}
This function is infinitely smooth ($C^\infty$), exactly equal to unity for $x\in[-1+\epsilon,1-\epsilon]$, and exactly vanishes at the edges $x=\pm 1$.
As a result, multiplying the Radon transform with it smoothens the FFT while retaining the periodicity that we are interested in.

\subsubsection{Visibility amplitude profile fit}

Once the visibility amplitude of a given image has been obtained, the next step in the GLM test of the Kerr hypothesis is to extract the diameter $d_\varphi^{(2)}$ of the $n=2$ ring at every angle $\varphi_\rho$ around the image.
This diameter can be inferred from the ringing signature displayed by $|V(u,\varphi)|$ at the corresponding angle $\varphi=\varphi_\rho$ in the Fourier plane, which is predicted to take the universal form \eqref{eq:UniversalAmplitude} in the baseline regime \eqref{eq:SubringRegime} appropriate to the $n=2$ ring.
Na\"ively, one would simply determine the best-fit parameters $\alpha_L$, $\alpha_R$, and $d_\varphi$, and take the latter to provide the desired measurement of $d_\varphi^{(2)}$.

\begin{figure*}[hbtp]
	\centering
	\includegraphics[width=\textwidth]{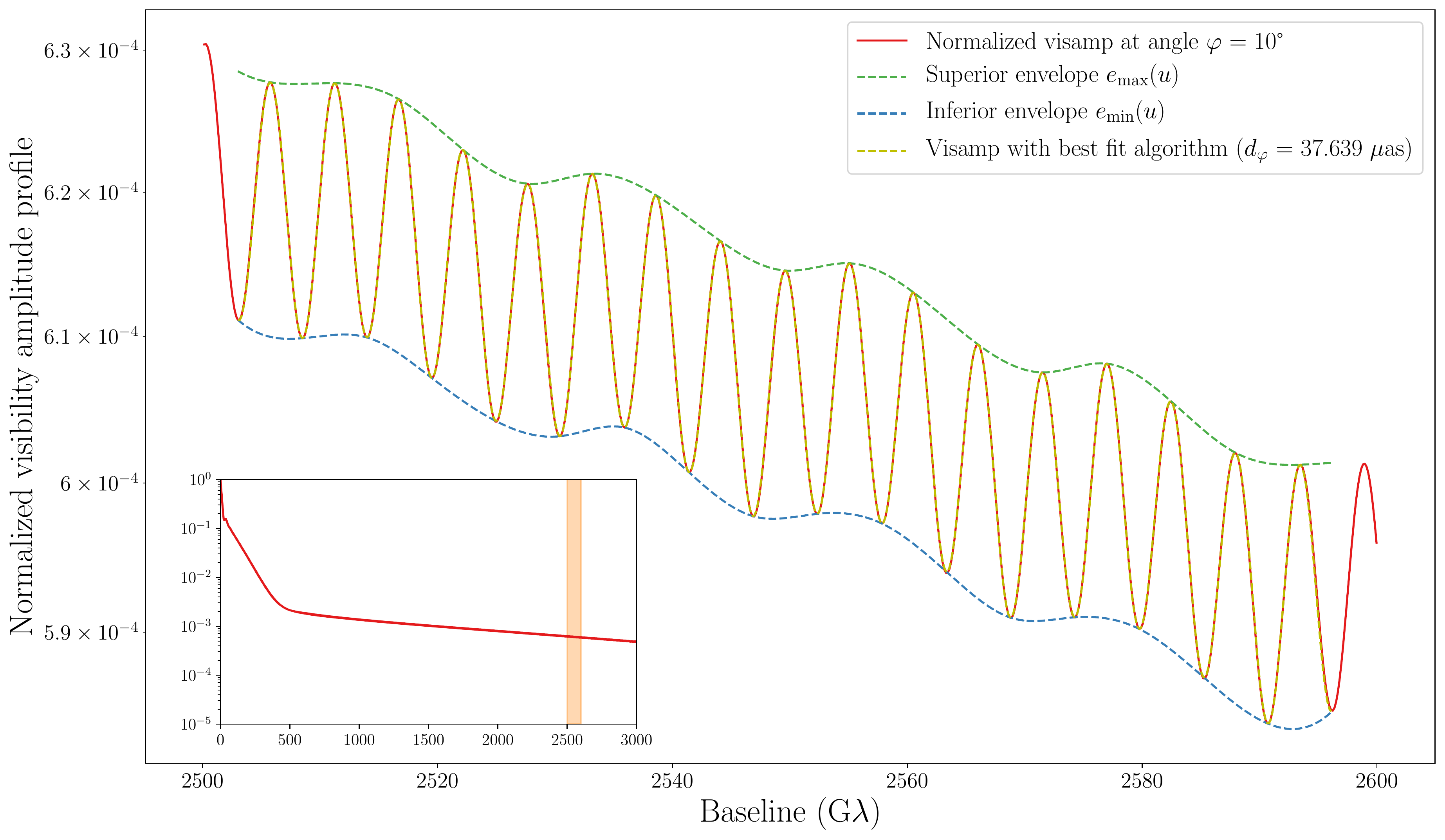}
	\caption{
	Example of a visibility amplitude profile with its envelope and best-fit model curve.
	Here, the naïve fitting method would fail because the envelope has complex, non-monotonic variations; yet, the periodicity is clearly visible and amenable to extraction via our refined fitting method.
	}
	\label{fig:ExampleFit}
\end{figure*}

In practice, however, a simple fitting method---such as the \texttt{FindFit} routine implemented in \texttt{Mathematica}---works only when the signal is extremely ``clean,'' in the sense that $|V(u,\varphi)|$ displays no artifacts and follows the precise functional form \eqref{eq:UniversalAmplitude}.
This happens in ideal cases where the image perfectly resolves the photon ring and the baselines chosen for the fit correspond exactly to the domain \eqref{eq:SubringRegime} over which the contributions from the $n=2$ subring are dominant.
Unfortunately, these cases are quite rare and the signal is typically less clean.
As a result, the na\"ive fitting methods fail, even if---as in Fig.~\ref{fig:ExampleFit}---the periodicity is still manifestly present.
A more robust approach is clearly called for.

Markov chain Monte-Carlo (MCMC) methods offer such an approach and prove effective even when the signal is buried in noise.
For instance, the data simulated by \citet{GLM2020} for their experimental forecast contained so much noise so as to render the periodicity invisible to the eye (see panels (a) \& (c) in their Fig.~8), and yet an MCMC analysis enabled the true ring diameter to be extracted with great precision (see panels (b) \& (d) in their Fig.~8).
While extremely robust, MCMC methods are also computationally intensive and thus a ``middle-ground'' approach is more desirable, especially for models without noise.

We devised such a procedure, which proved highly effective.
We now describe it using Fig.~\ref{fig:ExampleFit} as an illustrative example:
\begin{enumerate}
	\item For each $\varphi$, choose a baseline window in which to perform the fit (in Fig.~\ref{fig:ExampleFit}, we chose the range $u\in[2500,2600]$\,G$\lambda$).
	\item Identify the local extrema (both maxima and minima) of the visibility amplitude profile within this window.
	\item Interpolate between maxima to obtain the superior envelope  $e_{\rm max}(u)$ of the amplitude (green dashed curve in Fig.~\ref{fig:ExampleFit}).
	\item Interpolate between minima to obtain the inferior envelope $e_{\rm min}(u)$ of the amplitude (blue dashed curve in Fig.~\ref{fig:ExampleFit}).
	\item Using a simple fitting routine, obtain the ring diameter $d_\varphi$ as the parameter $d$ for which the refined model
	\begin{align}
	\label{eq:RefinedModel}
		V_{\rm fit}(u;d)=\sqrt{\alpha_L^2(u)+\alpha_R^2(u)+2\alpha_L(u)\alpha_R(u)\sin\pa{2\pi du}}
	\end{align}
	best fits the data, where the functions $\alpha_{L/R}(u)$ are defined as
	\begin{align}
	\label{eq:alphaLR}
		\alpha_{L/R}(u)=\frac{e_{\rm max}(u)\pm e_{min}(u)}{2}.
	\end{align}
\end{enumerate}
The precise interpolation method used to obtain the envelopes $e_{\rm max/min}(u)$ is not important; we resorted to cubic interpolation with the \texttt{interp1d} routine provided with \texttt{SciPy}.
Likewise, the particular fitting method used in the last step is not important either, as the model is by construction very close to the data; we made use of the \texttt{curve\_fit} routine, also provided with \texttt{SciPy}.

Some words of explanation are in order.
First, the use of the refined model \eqref{eq:RefinedModel} allows us to consider signals that display the same periodic behavior as that predicted by the formula \eqref{eq:UniversalAmplitude}, but with variable extrema $e_{\rm min}$ and $e_{\rm max}$.
This relaxation significantly improves the method's robustness and enables it to determine the periodicity of even moderately resolved rings, which can display non-monotonic modulations in their envelopes.
We thus expect the method to remain effective even at lower resolution, thereby enabling a noticeable speed-up of parameter surveys.
A trade-off of the method is that, in order to attain sufficient precision on the determination of the periodicity, we are required to examine the signal over a relatively larger interval of baselines; in practice, we used a window of 100\,G$\lambda$, corresponding to $\sim 15$ periods.

This method also derives significant power from the flexible fall-off rate built into the visamp model \eqref{eq:RefinedModel}, which generalizes the $u^{-1/2}$ power-law fall-off exhibited by the analytic formula \eqref{eq:UniversalAmplitude} to arbitrary damping rates.
Such a step was already taken by \citet{Johnson2020}, who noticed that their signal presented damped oscillations with envelope $e_{\rm max}(u)=u^{-1/2}e^{-(uw)^\zeta}$, with two additional parameters $w$ and $\zeta$ needed to obtain a good fit (see their Fig.~4).
The present approach is more general still.
The use of an envelope tailored to the oscillation damping provides more robustness and allows one to ignore other ringing patterns that could arise from pixel effects, or even beating produced by interactions with other signals of larger periodicity: as shown in Fig.~\ref{fig:ExampleFit}, this method works even when such beats produce a local increase of the visamp.
Its drawback is that astrophysical noise (ignored here) may result in ill-defined envelope functions \eqref{eq:alphaLR}.

We also wish to emphasize that Eq.~\eqref{eq:UniversalVisibility} is only the leading term in an expansion in powers of $0<w/d\ll1$.
Interestingly, preliminary analysis suggests that subleading corrections set by the angle-dependent, nonzero thickness of the ring $w(\varphi)$ could also predict a beating pattern; analytic study of these corrections will be the subject of future work.

Finally, we note that while the model \eqref{eq:RefinedModel} is symmetric under $L\leftrightarrow R$ interchange, this symmetry is broken by the definitions \eqref{eq:alphaLR}, according to which $\alpha_L$ measures the oscillations' mean and $\alpha_R$ their amplitude, with $\alpha_R<\alpha_L$ by convention.

\begin{figure*}[hbtp]
	 \centering
	 \includegraphics[width=\textwidth]{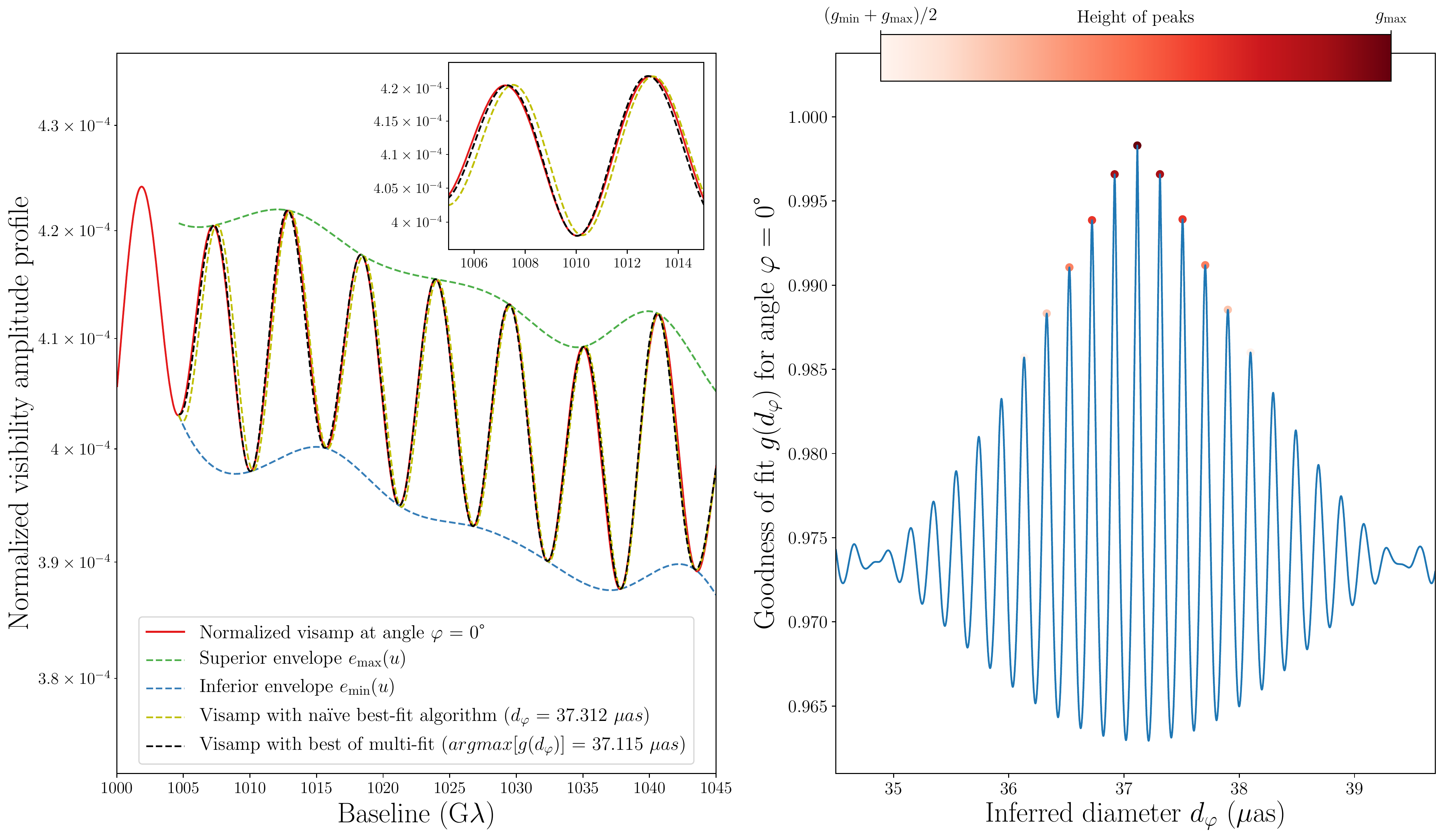}
	 \caption{Example of a model requiring a multi-peak fitting technique.
	 (Left) Fit comparison: na\"ively fitting the model $V_{\rm fit}(u;d)$ [Eq.~\eqref{eq:RefinedModel}] to the visamp profile $|V(u)|$ (solid red curve) yields an excellent but suboptimal fit (dashed yellow curve), since a better one exists (dashed black curve).
	 (Right) Goodness of fit $g(d_\varphi)$ as a function of fitting diameter [Eq.~\eqref{eq:GoodnessOfFit}].
	 The absolute best-fit corresponds to the global maximum of $g(d_\varphi)$, while other local maxima correspond to good but subopitmal fits.
	 These peaks are approximately periodic with a separation $\Delta d_\varphi\approx0.2\,\mu$as that exactly matches $1/u_w$, since our baseline window is at $u_w\sim1000\,$G$\lambda$.
	 }
	 \label{fig:MultiPeakFit}
\end{figure*}

\subsubsection{Multi-peaked distribution for the diameter}

The above fitting procedure returns a ``best-fit'' diameter $d_\varphi$ that minimizes the normalized root-mean-square deviation (RMSD)
\begin{align}
	\label{eq:RMSD}
	\rmsd_u(d)=\frac{\sqrt{\av{\br{V_{\rm fit}(u;d)-\ab{V(u)}}^2}_u}}{\av{V_{\rm fit}(u;d)}_u},
\end{align}
where $\av{\cdot}_u$ denotes an average over the chosen baseline window.

Often, this $d_\varphi$ is only a local (rather than global) minimum of $\rmsd_u(d)$, in which case it is not the absolute best-fit diameter; even so, it still gives a good fit so long as $0<\rmsd_u(d_\varphi)\ll1$.
This is illustrated with the example in Fig.~\ref{fig:MultiPeakFit}, for which the above procedure yields a diameter $d_\varphi=37.312\,\mu$as that is a local (but not global) minimum of Eq.~\eqref{eq:RMSD}.
Nevertheless, the resulting model $V_{\rm fit}(u;d_\varphi)$ (dashed yellow curve) closely tracks the visamp $|V(u)|$ (solid red curve), as indicated by their small deviation (shown in the left panel inset) and as measured by the correspondingly small numerical value of $\rmsd_u(d_\varphi)$.

Still, the global minimum of $\rmsd_u(d)$ is $d_\varphi=37.115\,\mu$as, and with this diameter the model $V_{\rm fit}(u;d_\varphi)$ (dashed black curve) provides an even better fit to $|V(u)|$: their deviation (left inset) is even narrower, as measured by the smaller value of $\rmsd_u(d_\varphi)$.

By definition, the global minimum of $\rmsd_u(d)$ is always the absolute best-fit diameter.
Equivalently, it is also determined as the global maximum of the ``goodness-of-fit'' measure
\begin{align}
	\label{eq:GoodnessOfFit}
	g(d_\varphi)\equiv e^{-\rmsd_u(d_\varphi)}.
\end{align}
We show $g(d_\varphi)$ for the above example in the right panel of Fig.~\ref{fig:MultiPeakFit}, where we recognize the tallest peak to be the global maximum $d_\varphi=37.115\,\mu$as, with the next local maximum corresponding to the na\"ive best-fit parameter $d_\varphi=37.312\,\mu$as.
Surprisingly, we also observe an entire periodic sequence of local maxima (with several providing a good fit) separated by a gap of $\Delta d_\varphi\approx0.2\,\mu$as.

This multi-peak structure was already observed by \citet{GLM2020} in their experimental forecast.
Its physical origin can be intuitively understood as follows.

Over a sufficiently narrow baseline window, one may view $u\approx u_w\gg1/d$ as fixed.
Within such a window, the universal form \eqref{eq:UniversalAmplitude} of the visibility amplitude is periodic in the diameter, as $|V(u)|$ is approximately invariant under shifts $d_\varphi\to d_\varphi+k/u_w$ for integer $k\in\mathbb{Z}$.
As a result, the probability distribution for $d_\varphi$ is multi-peaked, with peaks separated roughly by $\Delta d_\varphi\approx 1/u_w$.
The example in Fig.~\ref{fig:MultiPeakFit} confirms this argument: the observed gap of $\Delta d_\varphi\approx0.2\,\mu$as between peaks corresponds to a baseline length
\begin{align}
	\label{eq:Units}
	u_w\approx\frac{1}{2\times10^{-7}\,\text{arcsec}}
	\approx\frac{1.03\times10^{12}}{\text{rad}}
	\approx1030\,\text{G}\lambda,
\end{align}
which, as expected, lies well within the chosen baseline window of $u\in[1000,1045]\,$G$\lambda$.

Another intuitive way to understand this degeneracy comes from the observation that the number of damped oscillations (or ``hops'') from the origin $u=0$ to the baseline window $u\approx u_w$ is
\begin{align}
	N_w\approx u_wd_\varphi.
\end{align}
Hence, shifts $d_\varphi\to d_\varphi+k/u_w$ are equivalent to $N_w\to N_w+k$, and this degeneracy in $d_\varphi$  implies that the number $N_w$ of hops from the origin cannot be measured with perfect precision using only the visibility amplitude on very long baselines $u\approx u_w$:
adding or substracting a few units to this period number $N_w$ would barely shift the extrema within our fitting window, and as a result the fit $V_{\rm fit}(u,d_\varphi+k/u_w)$ remains acceptable for several values of $k\in\mathbb{Z}$.

In practice, we accounted for this degeneracy by adjuting our fitting procedure to keep track of several of the most prominent peaks of the multi-peak distribution, as follows:
\begin{enumerate}
	\item For each $\varphi$, obtain a first estimate of $d_\varphi$ by computing the mean distance between local maxima of the visamp $|V(u)|$.
	\item Pick an interval of some width $\Delta d_\varphi\gtrsim10/u_w$ centered around this estimate and plot the goodness of fit $g(d_\varphi)$ defined by Eqs.~\eqref{eq:RMSD}--\eqref{eq:GoodnessOfFit}, as in the right panel of Fig.~\ref{fig:MultiPeakFit}.
	\item Find the peaks of $g(d_\varphi)$ in this interval and keep the values of $d_\varphi$ corresponding to local maxima above a chosen peak height, together with the associated goodness of fit $g(d_\varphi)$.
\end{enumerate}
The resulting list should contain several diameters $d_\varphi$ that locally maximize $g(d_\varphi)$, including the global maximum.
It is important to note, however, that the latter is not always the ``true'' diameter $d_\varphi$ of the ring as measured from its image, which can sometimes correspond to one of the smaller peaks in $g(d_\varphi)$.
There are also cases where the multi-peaked distribution $g(d_\varphi)$ is relatively flat, so we do not obtain a good measurement at that angle $\varphi$.
Despite such failures, one can usually still infer the true diameter $d_\varphi$ by carrying out this analysis at multiple angles, as we next explain.

\subsubsection{Circlipse fit and test of the Kerr hypothesis}
\label{subsec:MultiFit}

\begin{figure*}[hbtp]
	\centering
	\includegraphics[width=\textwidth]{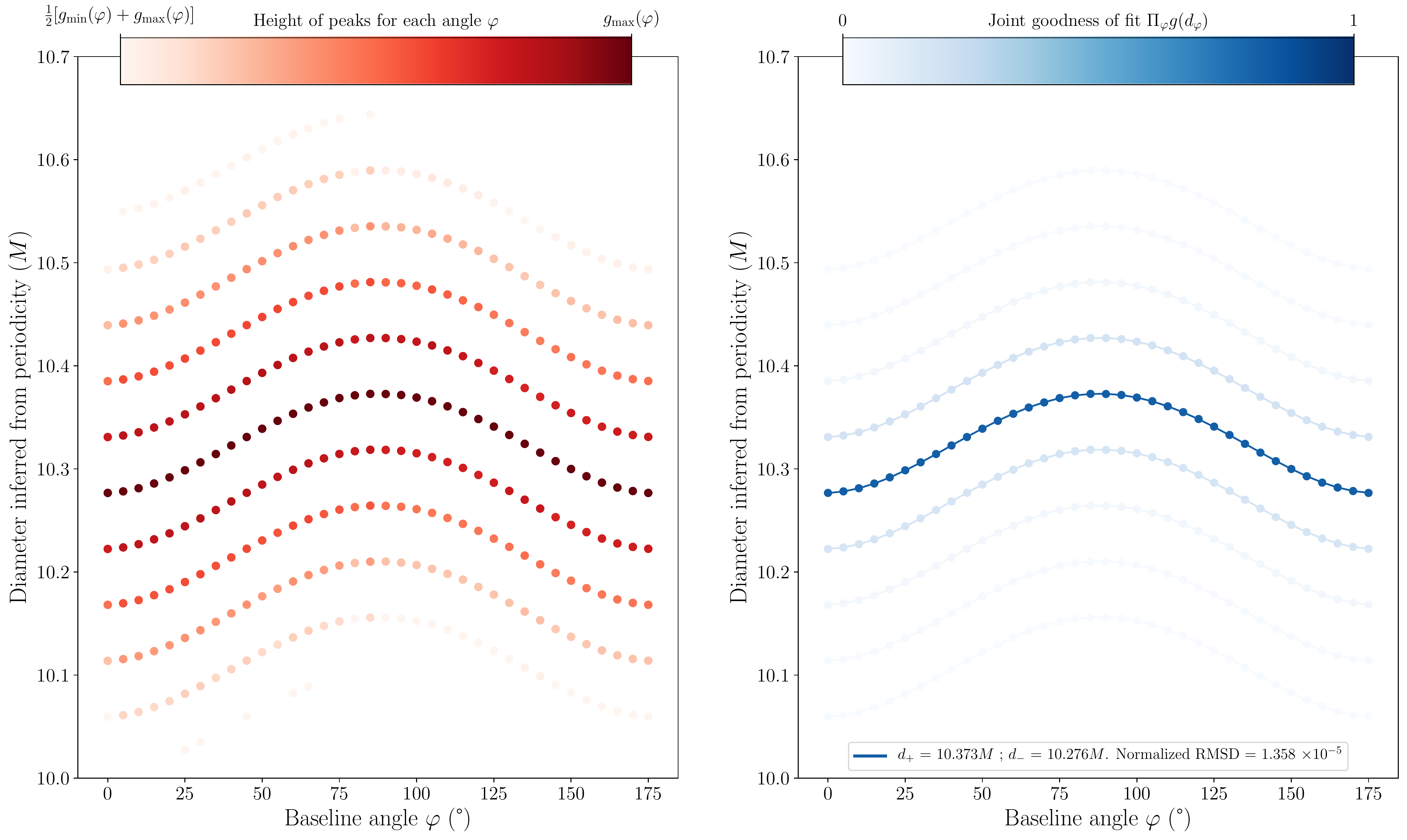}
	\caption{
	Example of a successful circlipse fit and Kerr hypothesis test.
	(Left) Every $5^\circ$, we determine the possible diameters $d_\varphi$ of the $n=2$ ring by maximizing the goodness-of-fit measured $g(d_\varphi)$ [Eq.~\eqref{eq:GoodnessOfFit}].
	For each angle $\varphi$, we obtain a periodic set of diameters separated by $\Delta d_\varphi\approx1/u_w$, with likelihood proportional to $g(d_\varphi)$ and indicated by the coloring of the data point.
	(Right) Circlipse fit for each of the possible rings.
	The darkest circlipse is most likely to be the true shape of the image ring, since the corresponding ``joint goodness of fit'' far exceeds that of the other solutions.
	}
	\label{fig:MultiFit}
\end{figure*}

Having determined the absolute best-fit ring diameter $d_\varphi$ at every angle $\varphi$ around the image, the final step of the GLM test is to fit  it to the GR-predicted functional form \eqref{eq:FunctionalForm} of the ring,
\begin{align}
	\label{eq:CirclipseShape}
	d_{\rm fit}(\varphi;\vec{R})=2R_0+2\sqrt{R_1^2\sin^2\pa{\varphi-\varphi_0}+R_2^2\cos^2\pa{\varphi-\varphi_0}},
\end{align}
with parameters $\vec{R}=\cu{R_0,R_1,R_2,\varphi_0}$.
In practice, we carried out this fitting by examining 36 angles in the range $[0^\circ,180^\circ)$ spaced at regular intervals of $5^\circ$. That is, we used $\varphi\in\cu{0^\circ,5^\circ,\ldots,175^\circ}$.
The best-fit parameters $\vec{R}_{\rm fit}$ are then obtained by minimizing
\begin{align}
	\label{eq:CirclipseFit}
	\rmsd_\varphi(\vec{R})=\frac{\sqrt{\av{\big[d_{\rm fit}(\varphi;\vec{R})-d_\varphi\big]^2}_\varphi}}{\av{d_{\rm fit}(\varphi;\vec{R})}_\varphi},
\end{align}
where $\av{\cdot}_\varphi$ denotes an average over the chosen baseline angles.

If $0<\rmsd_\varphi(\vec{R}_{\rm fit})\ll1$, then the circlipse fit is good and the Kerr hypothesis passes the GLM test at a level of precision given by the normalized root-mean-square deviation $\rmsd_\varphi(\vec{R}_{\rm fit})$.\footnote{
\citet{GLM2020} use the width of the peak in $g(d_\varphi)$ as a measure of the uncertainty $\sigma_\varphi$ in the inferred diameter $d_\varphi$, enabling them to use a standard chi-squared metric [their Eq.~(15)] to assess goodness of fit.
}
Otherwise, the fit is poor and the Kerr hypothesis fails the test.

As explained in the last section, this na\"ive approach is not robust because of the multi-peaked distribution for $d_\varphi$: keeping only the absolute best-fit diameter $d_\varphi$ may yield a suboptimal fit, since it may force us to use a wrong diameter at some angles $\varphi$.

However, we found a simple way to remedy this by using the data collected at all angles simultaneously, as follows:
\begin{enumerate}
	\item For each angle $\varphi$, compute $g(d_\varphi)$, identify several peak values of $d_\varphi$, and display them all together as in Fig.~\ref{fig:MultiFit}.
	\item Separate these peaks into multiple circlipse-shaped subsets $\mathcal{C}_i$, each with ``joint goodness-of-fit'' measure
	\begin{align}
		g(\mathcal{C}_i)\equiv\prod_\varphi g(d_{\varphi}).
	\end{align}
	\item Finally, fit each $\mathcal{C}_i$ to a circlipse shape \eqref{eq:CirclipseShape} with normalized $\rmsd(\mathcal{C}_i)\equiv\rmsd_\varphi(\vec{R}_{\rm fit})$ given by Eq.~\eqref{eq:CirclipseFit}.
\end{enumerate}
This ``multi-fit method'' results in several circlipses $\mathcal{C}_i$ that are not equally likely, as measured by their different values of $g(\mathcal{C}_i)$.
In favorable examples such as the one in Fig.~\ref{fig:MultiFit}, the likeliest $\mathcal{C}_i$ (i.e., the one with maximal $g(\mathcal{C}_i)$, shown in dark blue in the right panel) does turn out to be the ``true'' circlipse shape of the actual $n=2$ ring in the image.\footnote{However, we note that even in such examples, the true circlipse does not minimize $\rmsd(\mathcal{C}_i)$, which tends to monotonically decrease with the scale of $d_\varphi$ due to its normalization.
That is, larger ring diameters provide better fits to the circlipse shape according to the measure \eqref{eq:CirclipseFit}.
}

That said, there is no reason for this to be the case in general, and in some cases the ``true'' circlipse in the underlying image may well differ from the likeliest circlipse $\mathcal{C}^{\rm max}$ as determined by this analysis.
Of course, there is no other way to infer the true circlipse in a real experiment, so in principle we must take it to be the likeliest one $\mathcal{C}^{\rm max}$, even if it does not provide the best fit to the circlipse shape, that is, even if $\rmsd(\mathcal{C}^{\rm max})\neq\min_i\rmsd(\mathcal{C}_i)$.

In practice, the least favorable examples present a handful of  circlipses $\mathcal{C}_i^{\rm max}$ sharing similar near-maximal likelihoods $g(\mathcal{C}_i)$.
In such cases, we may only infer the true ring diameter up to a degeneracy of a few periods $\Delta d_\varphi\approx1/u_w$, but we may still report a test of the Kerr hypothesis at a level of precision given by the maximal $\rmsd(\mathcal{C}_i^{\rm max})$ within the set of best-fit circlipses $\mathcal{C}_i^{\rm max}$.

\section{Importance of the choice of baseline window}
\label{sec:BaselineWindow}

Before reporting the results of our parameter surveys, we first describe some new and important features of the GLM method that we noticed in the course of our investigation.
In order to determine the diameter $d_\varphi^{(2)}$ of the $n=2$ ring, it is crucial to sample the visamp $|V(u,\varphi)|$ in the appropriate regime \eqref{eq:SubringRegime}, that is, on baselines long enough to resolve the width of the $n=1$ ring (but not that of the $n=2$ ring) at image angle $\varphi_\rho=\varphi$:
\begin{align}
	\label{eq:BaselineCondition}
	\frac{1}{w_1(\varphi_\rho)}\ll u\ll\frac{1}{w_2(\varphi_\rho)}.
\end{align}
It is only in this regime that the signature of the $n=2$ ring can dominate the visamp and produce damped periodic oscillations \eqref{eq:UniversalAmplitude} that encode the ring diameter $d_\varphi^{(2)}$.
In the example of Fig.~\ref{fig:TypicalVisibility}, this requires us to choose a baseline window $u\gtrsim100\,$G$\lambda$.

The angle-dependence of condition \eqref{eq:BaselineCondition} is important.
At low inclinations, this dependence is relatively weak, and the baseline threshold $b_2(\varphi)\approx1/w_1(\varphi)$ past which the $n=2$ ring dominates $|V(u,\varphi)|$ is approximately constant in $\varphi$.
As a result, there exist baseline windows for which a ring diameter measurement works for all angles $\varphi$.
On the other hand, at high inclinations, $b_2(\varphi)$ may vary so much that no choice of baseline can satisfy the condition \eqref{eq:BaselineCondition} for all angles simultaneously; worse, the visamp may even cease to encode the $n=2$ ring diameter altogether.

These difficulties can be quantitatively fleshed out as follows.
The GLM method can only succeed for a fixed choice of baseline window $u\approx u_w$ that satisfies Eq.~\eqref{eq:BaselineCondition} for all angles, so that
\begin{align}
	\label{eq:BaselineThresholds}
	\max_\varphi b_2(\varphi)\ll u_w\ll\min_\varphi b_3(\varphi).
\end{align}
Here, $b_n(\varphi)\approx1/w_{n-1}(\varphi)$ denotes the threshold past which the visamp contribution from the $n^\text{th}$ ring overtakes its predecessor's.
These thresholds are well-defined when there is a clear transition between the baseline regimes in which each subring's signature dominates; however, these are not always cleanly delineated.
To understand why, we recall that $|V(u,\varphi+\pi)|=|V(u,\varphi)|$, which implies that the baseline thresholds must obey (when they exist)
\begin{align}
	\label{eq:BaselinePeriodicity}
	b_n(\varphi+\pi)=b_n(\varphi).
\end{align}
On the other hand, the subring widths $w_n(\varphi_\rho)$ are only subject to the periodicity constraint $w_n(\varphi_\rho+2\pi)=w_n(\varphi_\rho)$ and may display significant variation with image angle $\varphi_\rho\in[0,2\pi)$.
In particular, $w_1(\varphi_\rho)$ and $w_1(\varphi_\rho+\pi)$ may be very different at high inclinations, in which case $b_2(\varphi_\rho)$ fails to be sharply defined.
Worse, while it must always be the case that $w_2(\varphi_\rho)<w_1(\varphi_\rho)$, it may happen that $w_2(\varphi_\rho)\sim w_1(\varphi_\rho+\pi)$ for some angles $\varphi_\rho$, in which case the periodic ringing of the visamp $|V(u,\varphi)|$ at $\varphi=\varphi_\rho$ is affected by both the $n=1$ and $n=2$ rings but encodes neither's diameter.
We now describe these inclination-dependent effects in detail.

\subsection{Low to moderate inclinations}

\begin{figure*}[hbtp]
	\centering
	\includegraphics[width=\textwidth]{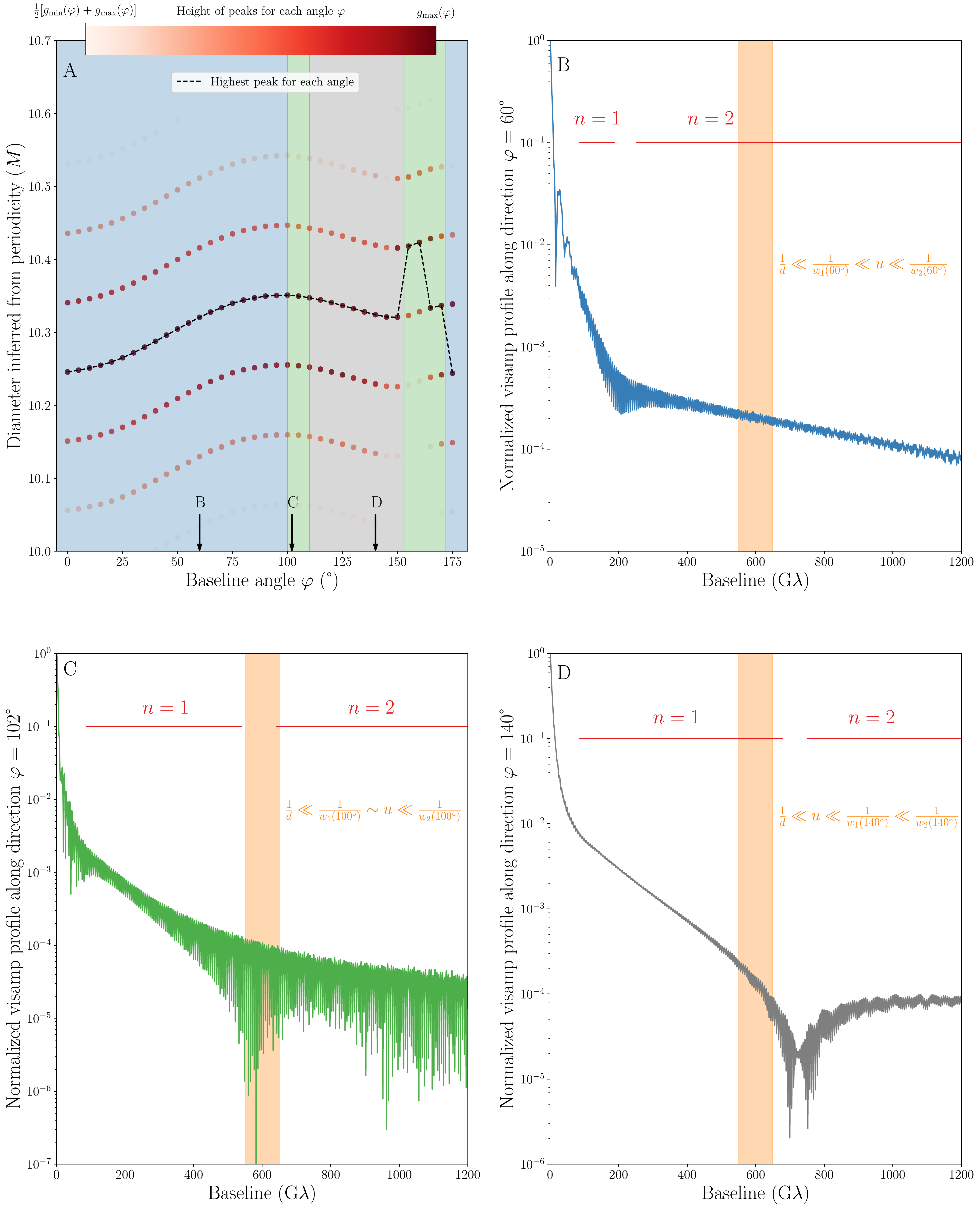}
	\caption{
	Measurements of the photon ring shape carried out over insufficiently long baseline windows display characteristic transition effects.
	In this example, the BH spin is $a=0.5$, the observer inclination is $i=45^\circ$, and the emission profile \eqref{eq:EmissionProfile} has parameters $\mu=3r_+/2$, $\gamma=-1$, and $\sigma=0.5M$.
	(A) Projected diameters $d_\varphi$ inferred from the periodicity of the visamp in the range $[550,650]\,$G$\lambda$ via the multi-fit method (Sec.~\ref{sec:Implementation}).
	With this baseline choice, $d_\varphi$ does not follow the circlipse shape \eqref{eq:FunctionalForm} because it only tracks $d_\varphi^{(2)}$ within the blue zone, which is separated by green transition zones from the gray zone where it corresponds to $d_\varphi^{(1)}$.
	The visamp profile $|V(u,\varphi)|$ over this baseline window (shown as an orange band) is dominated by the $n=2$ ring at $\varphi=60^\circ$ in the blue zone (B), by the $n=1$ ring at $\varphi=140^\circ$ in the gray zone (D), and by neither at $\varphi=100^\circ$ (C).
	} 
	\label{fig:BaselineChoice}
\end{figure*}

At low-to-moderate inclinations, each subring produces a distinct signature in its own baseline regime, and so the threshold $b_2(\varphi)$ exists.
This is the case illustrated in Figs.~\ref{fig:TypicalVisibility} and \ref{fig:BaselineChoice}.
According to Eq.~\eqref{eq:Widths}, the next thresholds are related to it by
\begin{align}
	\label{eq:ThresholdScaling}
	b_{n+1}(\varphi)\approx e^{\gamma(\varphi)}b_n(\varphi),
\end{align}
an approximate Kerr-lens equation that becomes exact as $n\to\infty$ and is already an excellent approximation for $n\gtrsim2$.

On the other hand, one cannot determine $b_2(\varphi)$ from first principles, as the first subring's width $w_1(\varphi_\rho)$ must be computed for each model separately.
We may thus rewrite Eq.~\eqref{eq:BaselineThresholds} as
\begin{align}
	\label{eq:BaselineTest}
	\max_\varphi b_2(\varphi)\ll u_w\ll\min_\varphi e^{\gamma(\varphi)}b_2(\varphi),
\end{align}
where $b_2(\varphi)$ is model-determined (astrophysics-dependent) and the demagnification factor $e^{\gamma(\varphi)}$ is universal (GR-predicted).

Letting $\gamma_0=\min_\varphi\gamma(\varphi_\rho)$ denote the minimal demagnification factor, we see that a sufficient (though not necessary) condition for the existence of a baseline window $u_w$ satisfying Eq.~\eqref{eq:BaselineTest} is
\begin{align}
	\max_\varphi b_2(\varphi)\ll e^{\gamma_0}\min_\varphi b_2(\varphi).
\end{align}
At low inclinations, where $b_2(\varphi)\approx b_0$ is relatively flat and $e^{\gamma_0}$ takes values of approximately $10\sim20$, this sufficient condition is manifestly satisfied, guaranteeing the existence of a suitable choice of baseline window over which to carry the measurement.
At higher inclinations, however, $e^{\gamma_0}\sim1$ (it is in fact exactly unity for equatorial inclination and maximal spin $a=1$).
Moreover, $b_2(\varphi)$ varies significantly, so this condition is clearly violated.
This does not necessarily imply that Eq.~\eqref{eq:BaselineThresholds} cannot be satisfied, but it does mean that some choices of baseline window $u_w$ that are suitable at a given angle $\varphi$ may not be suitable at other angles.

This effect is illustrated in Fig.~\ref{fig:BaselineChoice}, which displays a model from our parameter survey with BH spin $a=0.5$ and moderate inclination $i=45^\circ$.
Panels B--D illustrate the angle-dependence of the threshold $b_2(\varphi)$, which varies from $b_2(60^\circ)\approx200\,$G$\lambda$ to $b_2(140^\circ)\approx700\,$G$\lambda$.
If one were to extract a diameter $d_\varphi$ from the periodicity of the visamp in the baseline window $u_w\approx600\,$G$\lambda$, then one would be measuring the $n=2$ ring diameter $d_\varphi^{(2)}$ at $\varphi=60^\circ$ (panel B) and the $n=1$ ring diameter $d_\varphi^{(1)}$ at $\varphi=140^\circ$ (panel D).
At an intermediate angle of $\varphi=102^\circ$, one would land in a transition region between the $n=1$ and $n=2$ regimes in which the visamp periodicity encodes neither ring's diameter, but rather some ``superposition'' of the two.

Of course, at moderate inclinations, this issue can always be avoided by selecting a larger baseline window.
For the example of Fig.~\ref{fig:BaselineChoice}, any choice $u_w\gtrsim1000\,$G$\lambda$ is sufficiently large: over such baseline windows, the visamp periodicity always encodes the $n=2$ ring diameter (until the $n=3$ signal takes over on even larger baselines $u\gtrsim2000\,$G$\lambda$).
Meanwhile, measurements of $d_\varphi$ on insufficiently long baselines lead to tell-tale transitions between $d_\varphi^{(1)}$ and $d_\varphi^{(2)}$ that inevitably result in the characteristic, non-periodic pattern of data points shown in panel A of Fig.~\ref{fig:BaselineChoice}.

The reason is the following.
For the angles in the blue zone, $d_\varphi=d_\varphi^{(2)}$ is the ``true'' diameter of the $n=2$ ring's image, while for the angles in the gray zone, $d_\varphi=d_\varphi^{(1)}$ is the ``true'' diameter of the $n=1$ ring's image.
These regions are separated by green zones in which neither ring dominates the signal, so that $d_\varphi$ is equal to neither $d_\varphi^{(2)}$ nor $d_\varphi^{(1)}$: in such regions, $d_\varphi$ does not track the diameter of any actual image feature, and in particular it need no longer be a continuous curve.
As a result, the absolute best-fit diameter $d_\varphi$---defined to maximize Eq.~\eqref{eq:GoodnessOfFit}---is allowed to ``jump'' across circlipse bands (and indeed, one can see two such jumps in the second green zone $150^\circ<\varphi<175^\circ$ of panel A).
In fact, such jumps are necessary to ensure the periodicity of the physical ring diameter $d_\varphi^{(2)}$ in the blue zone, which must necessarily remain invariant under shifts $\varphi\to\varphi+\pi$.

On the other hand, the continuous curves $\mathcal{C}_i$ that the human eye naturally joins points into need not be periodic as $\varphi\to\varphi+\pi$;
it is only the totality $\cu{C_i}$ of such curves that needs to respect that periodicity.
It is therefore perfectly acceptable for these curves to continuously wrap into their successor as one looks up across panel A of Fig.~\ref{fig:BaselineChoice}.
While these curves $C_i$ are still separated by a gap $\Delta d_\varphi\approx1/u_w$, they are manifestly not circlipses, and their lack of periodicity is an indicator that the baseline window is inadequate for a full measurement of $d_\varphi^{(2)}$.

To summarize, at low inclinations $i\lesssim20^\circ$, the baseline regimes corresponding to each subring are sharply delineated; in particular, the baseline threshold $b_2(\varphi)$ exists and is relatively flat, so that if a baseline window $u_w$ is suitable for a measurement of the $n=2$ ring diameter $d_\varphi^{(2)}$ at some angle $\varphi$, then it is almost surely suitable at every other angle around the baseline plane.
At higher but still moderate inclinations $i\lesssim50^\circ$, the baseline threshold $b_2(\varphi)$ still exists but now exhibits substantial variation.
Because of this, some baseline windows $u_w$ may be suitable for a measurement of the $n=2$ ring diameter $d_\varphi^{(2)}$ at certain angles $\varphi$, but not others, where they instead pick up the $n=1$ ring diameter $d_\varphi^{(1)}$ or some unphysical ``superposition'' of the two.
Nonetheless, it should be possible to tell if this is happening from the shape of $d_\varphi$, which would display the general pattern shown in panel A of Fig.~\ref{fig:BaselineChoice}.
Moreover, at these inclinations, this difficulty can always be avoided by selecting a sufficiently large baseline window.

\subsection{High inclinations}

At higher inclinations $i\gtrsim60^\circ$, subrings do not always produce clearly delineated signatures in individual baseline regimes, and as a result the baseline threshold $b_2(\varphi)$ does not exist at all angles $\varphi$.
Physically, this can occur because the baseline threshold must be $\pi$-periodic [Eq.~\eqref{eq:BaselinePeriodicity}], while the subring images need only be $2\pi$-periodic.
In particular, while $w_2(\varphi_\rho)<w_1(\varphi_\rho)$ for every angle $\varphi_\rho$ around the image, at high inclinations it is usually the case that $w_2(\varphi_\rho)\sim w_1(\varphi_\rho+\pi)$ for angles $\varphi_\rho$ opposite the spin axis.
In that case, the visamp $V(u,\varphi)$ at baseline angle $\varphi=\varphi_\rho$ (or equivalently, $\varphi=\varphi_\rho+\pi$) is sensitive to the diameters of both the $n=1$ and $n=2$ subrings.

\begin{figure*}[hbtp]
	\centering
	\includegraphics[width=\textwidth]{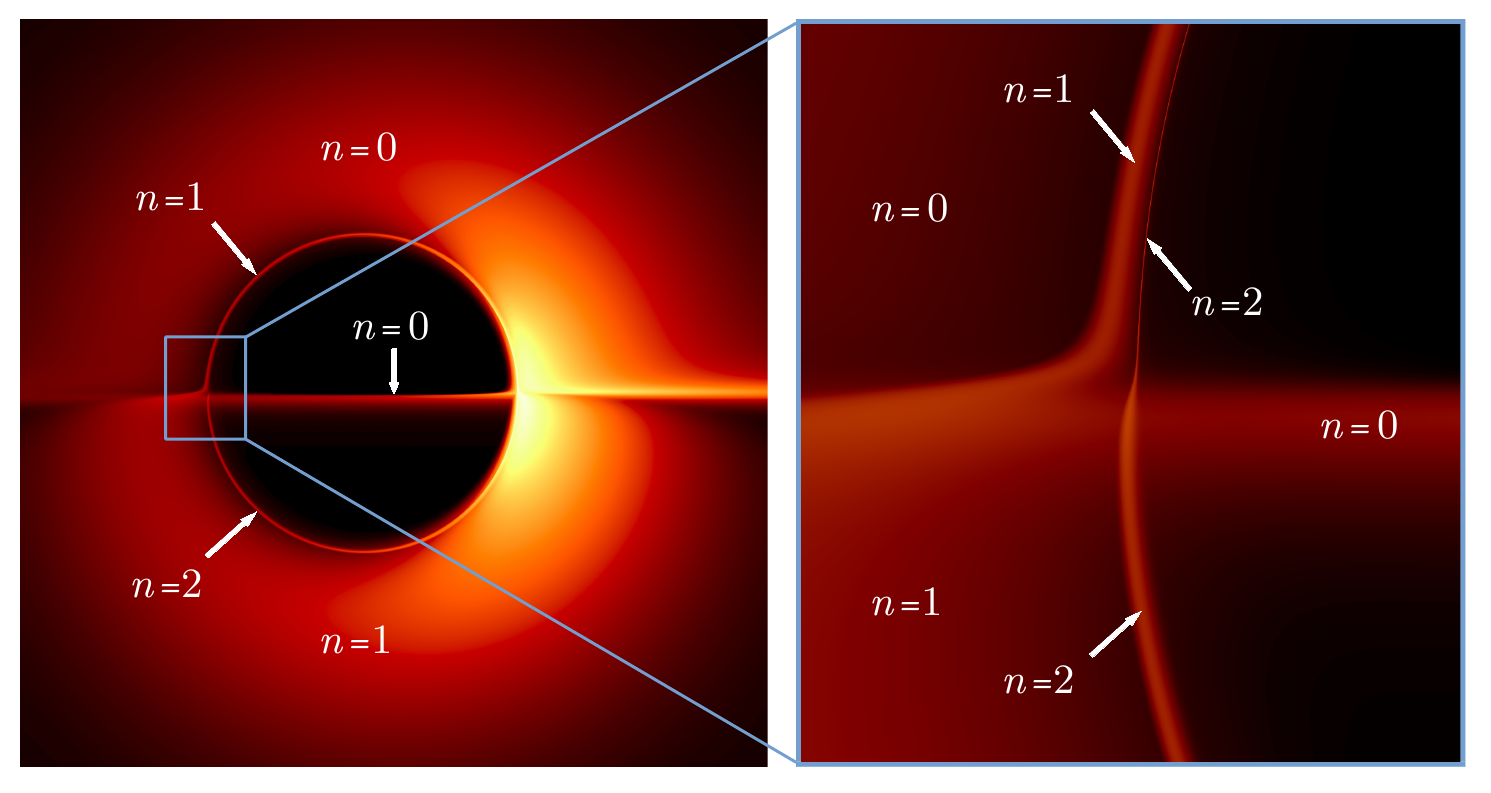}
	\caption{
	(Left) \texttt{Gyoto} log-scale image of a BH with spin $a=0.5$ and inclination $i=89^\circ$, surrounded by a thin disk with emission profile \eqref{eq:EmissionProfile} with parameters $\mu=r_-$, $\gamma=-2$, and $\sigma=1.5M$.
	The bright horizontal line is the direct $n=0$ image of the disk viewed almost edge-on.
	(Right) Zoom-in around a region with $\beta\approx0$.
	The width of the $n=1$ subring for $\beta>0$ is of the same order of magnitude as the width of the $n=2$ ring for $\beta <0$.
	}
	\label{fig:Scrambling}
\end{figure*}

\begin{figure}[ht]
	\centering
	\includegraphics[width=\columnwidth]{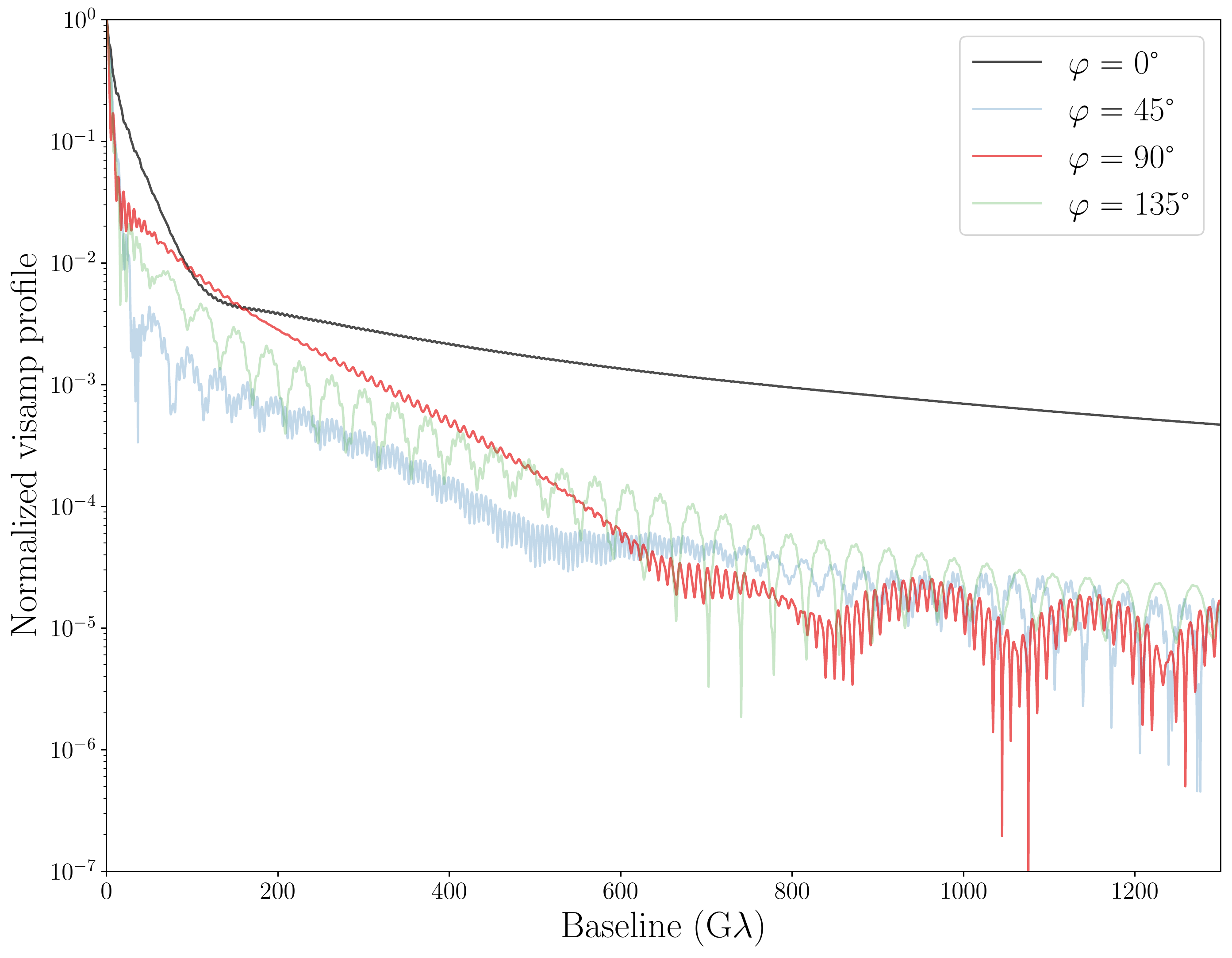}
	\caption{
	Visibility amplitude profiles $|V(u,\varphi)|$ of the image in Fig.~\ref{fig:Scrambling}.
	At $\varphi=90^\circ$, the ringing signature is sensitive to the projected diameter $d_\varphi$ of a hybrid ring composed of the upper half of the $n=1$ ring and lower half of the $n=2$ ring, as well as to the distance between this hybrid ring and the horizontal line of direct $n=0$ emission.
	These visamps are not exact as they neglect emission from outside the field of view of Fig.~\ref{fig:Scrambling}.
	}
	\label{fig:ScrambledVisamp}
\end{figure}

This phenomenon is clearly visible in the example shown in Fig.~\ref{fig:Scrambling}, which corresponds to a BH of spin $a=0.5$ observed from a near-equatorial inclination $i=89^\circ$.
Since $w_2(\varphi_\rho)\approx w_1(\varphi_\rho+\pi)$ for $\varphi_\rho=90^\circ$, it follows that the visamp profile $|V(u,90^\circ)|$, which is plotted in red in Fig.~\ref{fig:ScrambledVisamp}, cannot display a sharply delineation between regimes dominated by the $n=1$ and $n=2$ subrings.
Nonetheless, this profile still displays oscillations.
According to Eq.~\eqref{eq:UniversalVisibility}, their periodicity ought to be set by the diameter of a ring of fixed width in the image, which in this case must be the composite ring consisting of the upper half of the $n=1$ ring joined together with the lower half of the $n=2$ ring.
In other words, we would expect the visamp for this model to still adopt the universal form \eqref{eq:UniversalAmplitude} on long baselines, but with a projected diameter $d_\varphi$ corresponding to a hybrid $n=1\&2$ ring of uniform width.
In practice, this expectation is not quite realized in this model because of the direct ($n=0$) image of the near-side of the equatorial disk, which cuts across the photon ring.
Since this horizontal line is itself thin and bright like the hybrid $n=1\&2$ ring, the vertical intensity cross-section consists of three spikes;
as a result, the visamp profile at angles $\varphi\approx90^\circ$ is sensitive to multiple projected diameters, including that between the upper half of the $n=1$ ring and the $n=0$ emission line, and that between the $n=0$ emission line and the lower half of the $n=2$ ring.
These diameters are roughly half that of the hybrid ring, and indeed the visamp consists of a beating pattern with roughly twice the period expected for the signature of the photon ring.

\section{Parameter survey for the emission profile}
\label{sec:EmissionSurvey}

\subsection{Description of the models}
\label{subsec:EmissionModel}

In this paper, we restrict our attention to the ``thin-disk'' emission models presented in Sec.~III.A of \citet{GLM2020} and Sec.~3.2 of \cite{Chael2021}, which consist of emission profiles that are stationary, axisymmetric, and confined to the equatorial plane.
The emitters forming the disk are assumed to describe stable, circular-equatorial orbits all the way down to the radius of the innermost stable circular orbit (ISCO), past which they follow the prescription of \cite{Cunningham1975} for infall to the horizon.
This specifies the redshift factor $g$ of light rays emitted from the disk, whose observed (bolometric) intensity is then computed as
\begin{align}
	\label{eq:EquatorialApproximation}
	I(\alpha,\beta)=\sum_{n=0}^{N(\alpha,\beta)}f_nJ\pa{r_s^{(n)}}g^4\pa{r_s^{(n)},\alpha},
\end{align}
where $J(r)$ is an arbitrary radial emission profile, $r_s^{(n)}=r_s^{(n)}(\alpha,\beta)$ denotes the radius at which a light ray crosses the equatorial plane for the $n^\text{th}$ time after being shot back into the geometry from position $(\alpha,\beta)$ in the observer sky, and $N+1$ is the total number of times the light ray intersects the equatorial plane.\footnote{Analytic formulas for $r_s^{(n)}$ and $N$ are derived by \citet{GrallaLupsasca2020a,GrallaLupsasca2020b} and are succintly summarized in App.~\ref{app:LensingBands}.
}
Here, we also introduced a ``fudge factor''
\begin{align}
	f_n=\begin{cases}
		1&\quad n=0,\\
		1.5&\quad n\in\cu{1,2},\\
		0&\quad n\ge3,
	\end{cases}
\end{align}
whose effect is to remove $n\ge3$ subrings from the image (which would be underresolved), while enhancing the intensity of the $n=1$ and $n=2$ subrings relative to the direct $n=0$ image.
This enhancement can account for the effects of geometric thickness and results in images whose intensity cross-sections match those observed in time-averaged GRMHD-simulated images \citep{Johnson2020,Chael2021}.
In this sense, the equatorial model \eqref{eq:EquatorialApproximation} provides a good approximation for realistic accretion flows, and its simplicity and relatively low computational cost make it ideal for the survey of emission profiles undertaken here.

\begin{figure*}[hbtp]
	\centering
	\includegraphics[width=\textwidth]{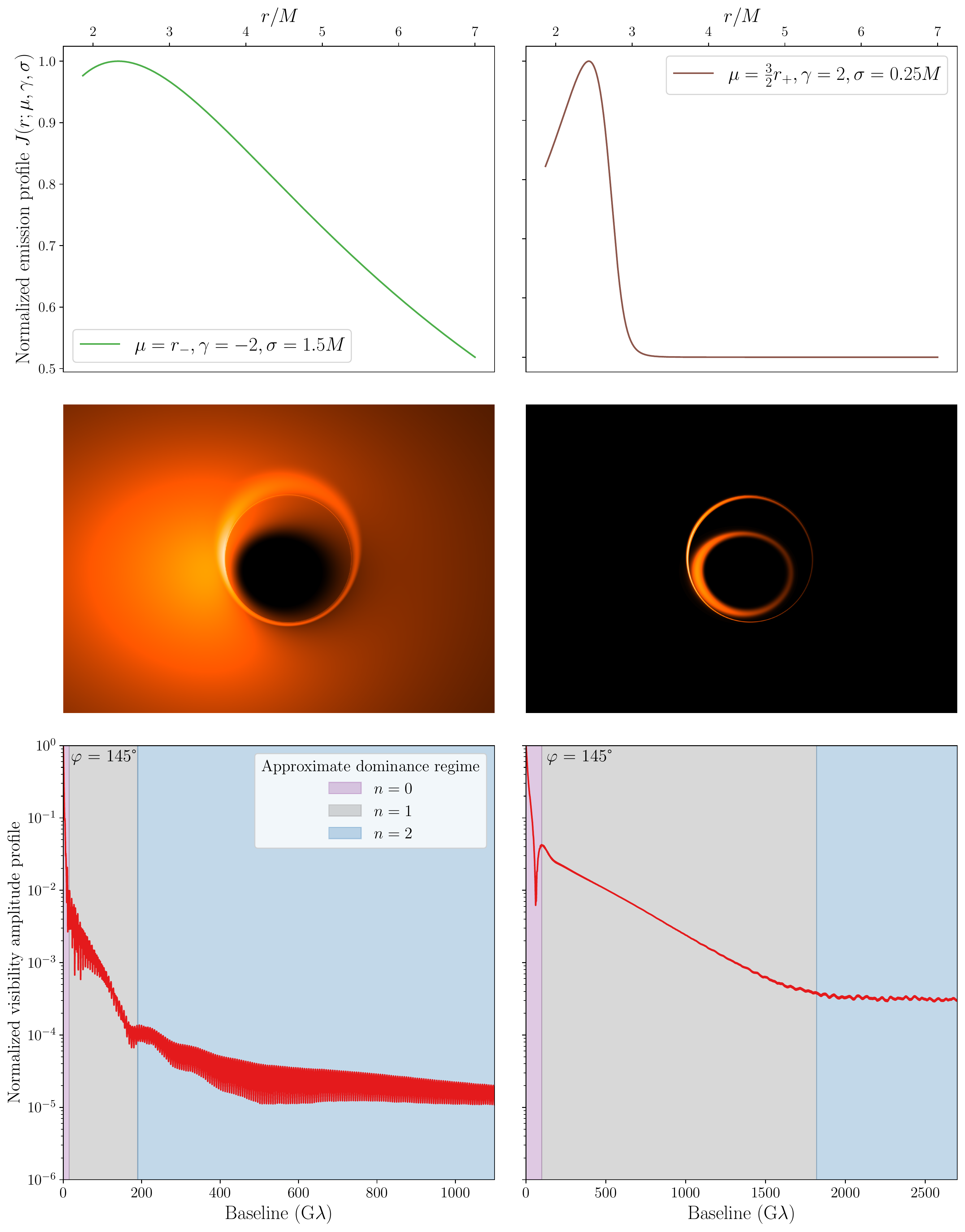}
	\caption{
	Impact of the emission profile on the observed ring thickness and baseline domains.
	We show two extreme cases:
	(Left) A broad emission profile, leading to large rings and an early transition between $n=1$ to $n=2$ regimes in the visibility amplitude profile with $b_2(145^\circ)\approx200\,$G$\lambda$.
	(Right) A highly peaked emission profile, leading to thin rings (the $n=2$ subring is not even visible in the image) and a late transition between the $n=1$ and $n=2$ regimes with $b_2(145^\circ)\gtrsim1800\,$G$\lambda$.
	}
	\label{fig:EmissionProfiles}
\end{figure*}

While the equatorial source profile $J(r)$ can in principle take any form, in this survey we considered only profiles of the form
\begin{align}
	\label{eq:EmissionProfile}
	J(r;\mu,\gamma,\sigma)=\frac{e^{-\frac{1}{2}\br{\gamma+\arcsinh\pa{\frac{r-\mu}{\sigma}}}^2}}{\sqrt{\pa{r-\mu}^2+\sigma^2}}.
\end{align}
This is a three-parameter subset of Johnson's $S_U$-distribution (the fourth parameter $\delta$ is set to unity), already adopted by \citet{GLM2020} for its versatility: this functional form provides a simple way to generate smooth profiles with emission typically concentrated near the event horizon of the BH, and displaying substantial variation with changes in $\mu$, $\gamma$, and $\sigma$.
In short, these three parameters respectively determine a location (the peak of the distribution), a shape (its asymmetry), and a scale (its width).
We now describe in detail how their variation affects ring width.

First, $\mu$ influences the location of the emission profile's peak.
We considered values in the range $[r_-,2r_+]$, where $r_\pm$ denote the horizon radii given in Eq.~\eqref{eq:HorizonRadius}.
At the lower end of this range (left column of Fig.~\ref{fig:EmissionProfiles}), the peak is very close to or behind the event horizon, but it moves farther out near the upper end of the range (right column of Fig.~\ref{fig:EmissionProfiles}).
The observed thickness of the photon subrings is not significantly affected by this variation.
	
Second, $\gamma$ controls the profile's asymmetry.
We considered values in the range $[-2,2]$.
When $\gamma<0$ (left column of Fig.~\ref{fig:EmissionProfiles}), the steep end of the distribution is to the left and (mostly) cut off by the horizon, leaving only a gentle-sloping tail in the emission profile.
When $\gamma>0$ (right column of Fig.~\ref{fig:EmissionProfiles}), the reverse is true: the profile's steep end lies to the right and remains visible outside the horizon, while its gentle-sloping side is cut off.
Hence, the rings tend to become thinner as $\gamma$ increases.
	
Third, $\sigma$ encodes the profile's width.
We considered values in the range $[0.25M,1.5M]$.
At the upper end of this range (left column of Fig.~\ref{fig:EmissionProfiles}), the profile is broad enough to produce thick rings, which makes the GLM test easy to carry out on a baseline window $u\in\br{450,550}\,$G$\lambda$ for all our considered values of $\mu$ and $\gamma$.
At the lower end (right column of Fig.~\ref{fig:EmissionProfiles}), the profile is a steep spike (provided that the distribution's peak lies outside the horizon) that produces narrower rings: as a result, the $n=1$ ring dominates on longer baselines, and a measurement of the $n=2$ ring diameter may require (in the most extreme cases) a baseline window $u_w\gtrsim2000\,$G$\lambda$.

Images with thinner rings have a larger baseline threshold $b_2(\varphi)$ for carrying out a measurement of the $n=2$ ring diameter $d_\varphi^{(2)}$.
This trend is illustrated in Fig.~\ref{fig:EmissionProfiles} with two extreme examples that display maximally thick rings (left column) and maximally thin rings (right column) within our considered parameter range.

\subsection{Choice of parameters for the survey}
\label{subsec:ParameterChoice}

In order to assess the robustness of the GLM method described in Secs.~\ref{sec:Theory} and \ref{sec:Implementation}, we conducted a survey over the emission-profile parameter space $\cu{\mu,\gamma,\sigma}$.
We used ``intermediate'' values for the BH spin and inclination of $a=0.5$ and $i=45^\circ$, and examined a total of 100 emission models corresponding to all the possible combinations of parameters with values presented in Table~\ref{tbl:EmissionSurvey}.

\begin{table}[h!]
	\centering
	\caption{Parameter values considered in the emission profile survey.}
	\label{tbl:EmissionSurvey}
	\begin{tabular}{|c|c|}
	\hline
	Parameter & Considered values \\
	\hline
	$\mu$ & $\cu{r_-,r_+/2,r_+,3r_+/2,2r_+}$ \\
	$\gamma$ & $\cu{-2,-1,0,1,2}$ \\
	$\sigma/M$ & $\cu{0.25, 0.5, 1., 1.5}$ \\
	\hline
	\end{tabular}
	\tablefoot{$r_\pm$ denote the radii \eqref{eq:HorizonRadius} of the outer and inner horizons.
	}
\end{table}

As a cross-check, we repeated the survey using the currently favored values of BH spin and inclination for M87*.
The power in the relativistic jet produced by M87* suggests that this BH is highly spinning, since most low-spin GRMHD models do not give rise to a sufficiently powerful jet \citep{EHT1}.
We therefore took $a=0.94$, the highest value considered in EHT simulations \citep{EHT5}.
We also assumed that the spin axis of the BH and of its accretion disk are aligned with its forward jet, which has a measured inclination of $i\approx17^\circ$ \citep{Walker2018}.
Since $r_-\approx r_+/2$ for this value of BH spin, we examined the same combinations of parameters except for those with $\mu=r_+/2$; this amounted to a total of 80 emission profiles in our survey.

\subsection{Results}
\label{subsec:ModelSurveyResults}

\begin{figure*}[hbtp]
	\centering
	\includegraphics[width=.95\textwidth]{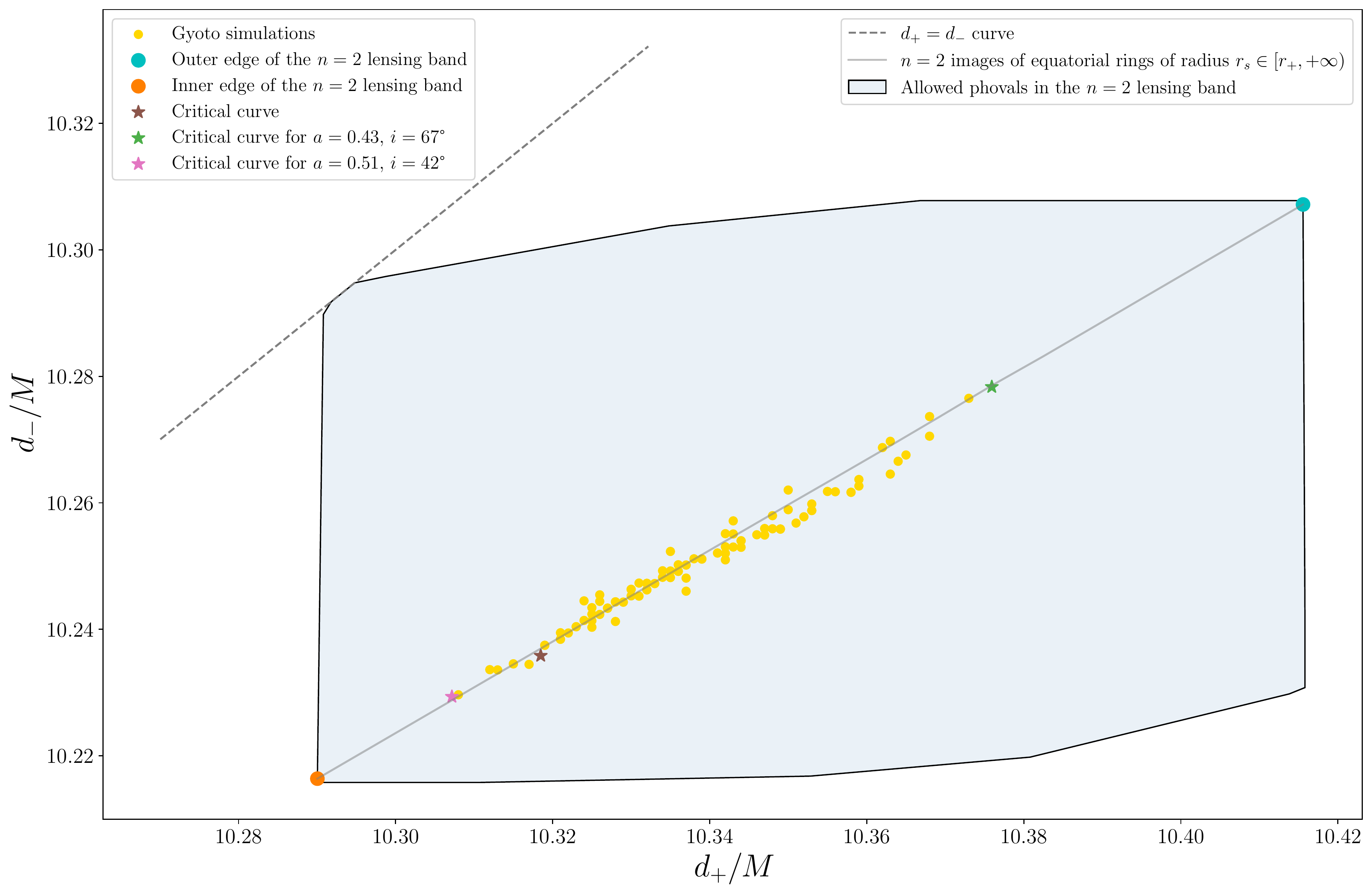}
	\caption{
	Results of the survey of emission profiles in the ``intermediate'' case with spin $a=0.5$ and inclination $i=45^\circ$.
	}
	\label{fig:IntermediateSpinSurvey}
\end{figure*}

\begin{figure*}[hbtp]
	\centering
	\includegraphics[width=.95\textwidth]{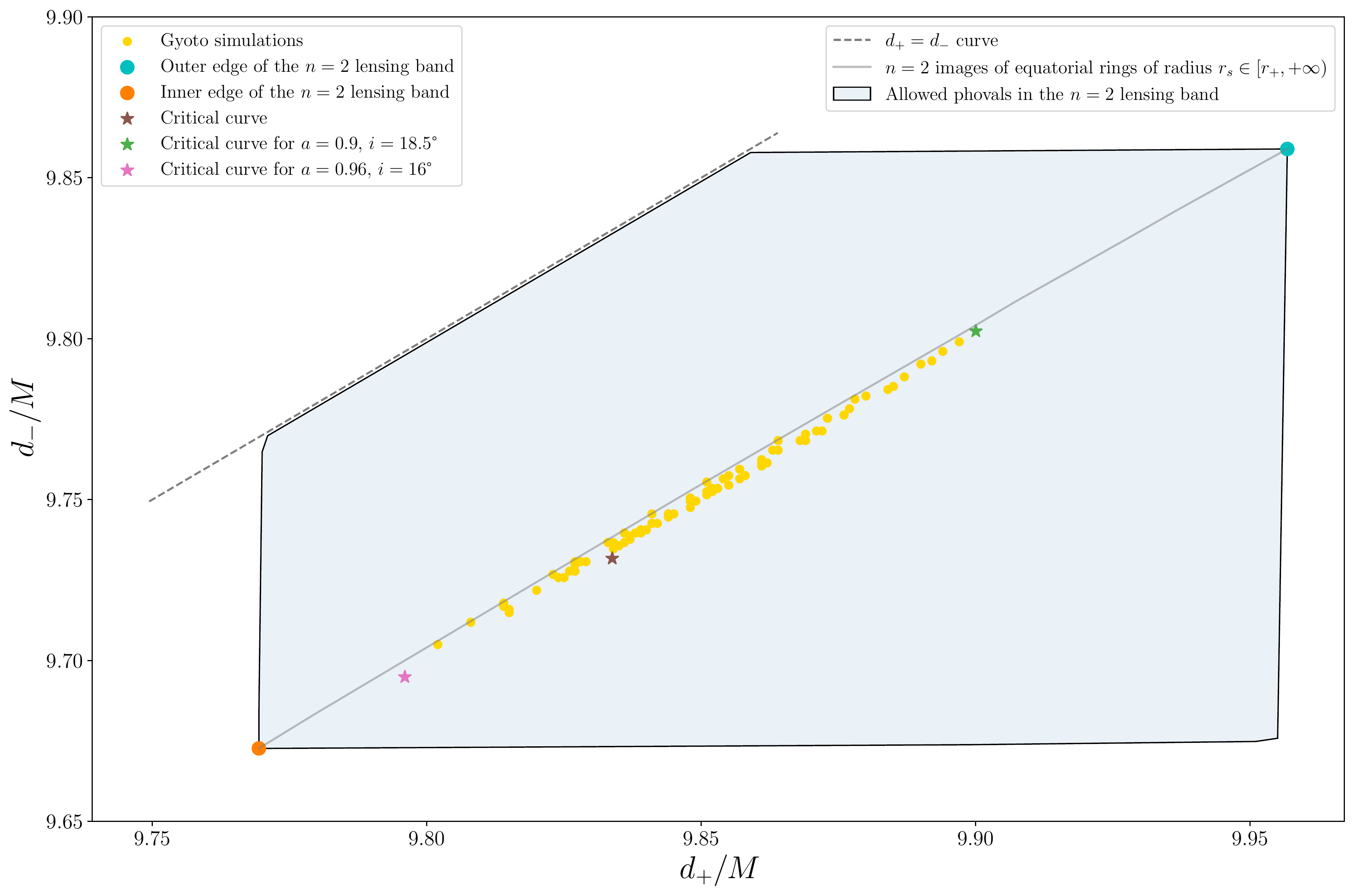}
	\caption{
	Results of the survey of emission profiles in the best-guess case for M87* with spin $a=0.94$ and inclination $i=17^\circ$.
	}
	\label{fig:SurveyM87}
\end{figure*}

We investigated these models to address the following questions:
\begin{enumerate}
	\item[Q1.] Is the GLM test possible in principle?
	In particular, does $d_\varphi^{(2)}$ follow the GR-predicted functional form \eqref{eq:CirclipseShape} of a circlipse?
	\item[Q2.] If so, how do the best-fitting circlipse parameters $\vec{R}_{\rm fit}$ vary with the astrophysical profile?
	In particular, what can we learn about the BH parameters from the photon ring shape? 
\end{enumerate}
Our survey led us to the following answers:
\begin{enumerate}
	\item[A1.] \textbf{Yes.}
	In all the models considered here, we identified a regime in which the $n=2$ ring dominates the visamp and extracted the projected ring diameter $d_\varphi^{(2)}$ from the signal's periodicity.
	For reasons discussed in Sec.~\ref{sec:BaselineWindow}, this required using different baseline lengths, sometimes as low as $u\gtrsim150\,$G$\lambda$ (Fig.~\ref{fig:TypicalVisibility}) or as high as $u\gtrsim2000\,$G$\lambda$ (bottom right panel of Fig.~\ref{fig:EmissionProfiles}).
	We always found an excellent fit to the circlipse shape \eqref{eq:CirclipseShape} as measured by the normalized RMSD \eqref{eq:CirclipseFit}: in all cases,
	\begin{align}
		\rmsd_\varphi(\vec{R}_{\rm fit})<10^{-4}.
	\end{align}
	\item[A2.] By analogy with Eq.~\eqref{eq:Diameters}, it is helpful to repackage the best-fit parameters $\vec{R}_{\rm fit}=\cu{R_0,R_1,R_2,\varphi_0}$ describing the projected diameter of the $n=2$ ring into maximal/minimal diameters\footnote{
	We use the notation $(d_+,d_-)$ instead of $(d_\parallel,d_\perp)$ to emphasize that these diameters describe the $n=2$ ring rather than the critical curve.
	Another difference is that $d_+\neq d_{\pi/2}$ and $d_-\neq d_0$ because of the rotation parameter $\varphi_0$ in Eq.~\eqref{eq:CirclipseShape}, though this angle is typically extremely small.
}
	\begin{align}
		\label{eq:Projection}
		d_+=d_{\varphi_0+\pi/2}^{(2)}=2\pa{R_0+R_1},\quad
		d_-=d_{\varphi_0}^{(2)}=2\pa{R_0+R_2}
	\end{align}
	of the associated circlipse shape ($d_-\le d_+$ since $R_1\ge R_2$ by definition).
	We plot the set of $(d_+,d_-)$ obtained from all our emission models in Figs.~\ref{fig:IntermediateSpinSurvey} and \ref{fig:SurveyM87}, and provide mean values and standard deviations for $(d_+,d_-)$ in Tables \ref{tbl:SpreadIntermediateSpin} and \ref{tbl:SpreadM87}.
	These results give a good overview of the variabilty in the circlipse shape obtained from different astrophysical source profiles.
	We return to the question of spin inference in Table~\ref{tbl:ErrorBars} below.
\end{enumerate}

\begin{table}[h!]
	\centering
	\caption{
	Mean values and standard deviations for the inferred $n=2$ ring diameters $d_\pm$ in the ``intermediate'' case $a=0.5$ and $i=45^\circ$. 
	}
	\label{tbl:SpreadIntermediateSpin}
	\begin{tabular}{|c|c c|}
	\hline
	& Mean value & Standard deviation \\
	\hline
	$d_+/M$ & 10.338 & 0.014 \\
	$d_-/M$ & 10.250 & 0.009 \\
	\hline
	\end{tabular}
	\tablefoot{The critical curve has $d_\parallel=10.319M$ and $d_\perp=10.236M$.}
\end{table}

\begin{table}[h!]
	\centering
	\caption{
	Mean value and standard deviations for the inferred $n=2$ ring diameters $d_\pm$ for the ``best guess for M87*'' case $a=0.94$ and $i=17^\circ$.
	}
	\label{tbl:SpreadM87}
	\begin{tabular}{|c|c c|}
	\hline
	& Mean value & Standard deviation \\
	\hline
	$d_+/M$ & 9.849 & 0.022 \\
	$d_-/M$ & 9.751 & 0.022 \\
	\hline
	\end{tabular}
	\tablefoot{The critical curve has $d_\parallel=9.834M$ and $d_\perp=9.733M$.}
	\vspace{-4mm}
\end{table}

As discussed in Sec.~\ref{subsec:Test}, one expects the $n=2$ photon ring to closely track the critical curve.
This is verified in practice, though we find that the choice of astrophysical source profile still has a significant impact on the shape of the $n=2$ ring.
In most of our models, the $n=2$ ring is slightly larger in diameter than the critical curve, so that $d_+>d_\parallel$ and $d_->d_\perp$.
In particular, we find that broader emission profiles produce thicker $n=2$ rings and therefore a larger difference in these diameters.
Conversely, narrow profiles that produce thinner rings lead to diameters that are closer to (and sometimes, even smaller than) those of the critical curve.
Either way, conflating the shape of the $n=2$ ring with that of the critical curve can lead to erroneous conclusions.

Nevertheless, it is possible to approximate the critical curve diameters $(d_\parallel,d_\perp)$ by $(d_+,d_-)$, and to then use these parameters to estimate the BH spin and inclination (assuming a mass prior) as described in Sec.~\ref{subsec:CriticalCurveShape}.
However, the precision of this spin and inclination determination is significantly limited by the spread of results for the $n=2$ ring diameters at a given $(a,i)$ value.
To illustrate this, Figs.~\ref{fig:IntermediateSpinSurvey} and \ref{fig:SurveyM87} also display diameters of critical curves that are compatible with the measured $n=2$ ring shape, but which correspond to $(a,i)$ values that are different from the ``true'' values of the BH spin and inclination used in the model (green and pink stars).
The constraints on these BH parameters derived from the inferred $n=2$ ring diameters have substantial error bars, for which we provide lower bounds in Table~\ref{tbl:ErrorBars}.

\begin{table}[h!]
	\centering
	\caption{
	Lower bounds on the uncertainties that we can expect from a measurement of BH spin and inclination using the $n=2$ ring diameters.
	}
	\label{tbl:ErrorBars}
	\begin{tabular}{|c|c c|}
	\hline
	& Spin &  Inclination \\
	\hline
	Intermediate case ($a=0.5$, $i=45^\circ$) & $\pm0.07$ & $\pm22^\circ$ \\
	M87* ``best guess'' ($a=0.94$, $i=17^\circ$) & $\pm0.04$ & $\pm1.5^\circ$ \\
	\hline
	\end{tabular}
	\vspace{-4mm}
\end{table}

\subsection{Range of allowed phovals within the \texorpdfstring{$n=2$}{n=2} lensing band}
\label{subsec:LensingBandsPhovals}

These results highlight that the critical curve is not necessarily the most relevant concept for the analysis of the $n=2$ photon ring shape.
For our thin-disk models, a more fruitful approach is to consider the $n=2$ lensing band described in Sec.~\ref{subsec:LensingBands}, since it contains all the possible geometric shapes of the $n=2$ photon ring for a given BH spin and inclination.
The critical curve lies within this band as well, but so do the critical curves of BHs with neighboring spins and inclinations, as Figs.~\ref{fig:IntermediateSpinSurvey} and \ref{fig:SurveyM87} indicate.

Since Sec.~\ref{subsec:Test}, we have used ``photon ring shape'' to refer to the simplest observational target for a measurement of the $n=2$ ring: its projected diameter $d_\varphi^{(2)}$, which matches that of a circlipse [Eq.~\eqref{eq:Circlipse}].
In this section, we study the full geometric shape of the $n=2$ ring, which typically includes centroid motion effects [Eq.~\eqref{eq:Centroid}] and is therefore not a circlipse in general.
Instead, we expect this geometric shape to be close to a phoval [Eq.~\eqref{eq:Phoval}]; that is, the $n=2$ ring's projected position should closely match
\begin{align}
	\label{eq:PhovalProjectedPosition}
	f_{\rm ph}(\varphi;\vec{R})=R_0&+\sqrt{R_1^2\sin^2\pa{\varphi-\varphi_0}+R_2^2\cos^2\pa{\varphi-\varphi_0}}\\
	&+\pa{X-\chi}\cos\pa{\varphi-\varphi_0}+\arcsin\br{\chi\cos\pa{\varphi-\varphi_0}},\notag
\end{align}
for some choice of parameters $\vec{R}=\cu{R_0,R_1,R_2,X,\chi,\varphi_0}$.
(Here, as in Eq.~\eqref{eq:FunctionalForm}, we have included a rotation angle $\varphi_0$ to account for the arbitrary orientation of the BH spin axis in the image.)
This expectation is based on the observation (reviewed in Sec.~\ref{subsec:CriticalCurveShape}) that the Kerr critical curve is always a phoval to high accuracy, together with the fact that the photon rings converge very rapidly (exponentially fast in $n$) to the critical curve [Eq.~\eqref{eq:Demagnification}].

Since we expect the geometric shape of the $n=2$ photon ring to be that of a phoval, it is interesting to consider the range of all phoval shapes that can possibly fit inside the $n=2$ lensing band.
Since phovals are described by six parameters, it is convenient to project this 6D region of allowed phoval shapes down onto the 2D plane of diameters $(d_+,d_-)$ via the projection \eqref{eq:Projection}.
The resulting region in the $(d_+,d_-)$ plane gives a good sense of the allowed range of possible $n=2$ ring shapes.

For the two choices of BH spin and inclination in our survey of emission profiles, this projection results in the blue regions displayed in Figs.~\ref{fig:IntermediateSpinSurvey} and \ref{fig:SurveyM87}.
Their numerical computation is nontrivial and we relegate the details to App.~\ref{app:PhovalRegions}.
We note that since $d_-\leq d_+$, these regions must lie below the diagonal.\footnote{We also note that these regions are in general different from the set of $(d_+,d_-)$ corresponding to the circlipse shapes that fit within the $n=2$ lensing band.
These two notions only coincide at low inclination, where the centroid motion is negligible and the allowed phovals essentially reduce to circlipses---this explains the square shape of the blue region in the $i=17^\circ$ case of Fig.~\ref{fig:SurveyM87}.
At higher inclinations, the lensing band is no longer circlipse-shaped, so the set of circlipses it can contain becomes an increasingly smaller subregion of the blue region of allowed phovals.
}

As expected, all the emission profiles in our survey produce $n=2$ photon rings that lie within the $n=2$ lensing band, with diameters $(d_+,d_-)$ lying within the blue region of allowed phoval shapes that fit inside the band.
The band's outer and inner edges can themselves be fitted to the phoval shape \eqref{eq:PhovalProjectedPosition} with excellent agreement, and their diameters $(d_+,d_-)$---shown in Figs.~\ref{fig:IntermediateSpinSurvey} and \ref{fig:SurveyM87} as blue and orange dots, respectively---lie on opposite corners of this blue region.
Remarkably, all of our astrophysical source profiles produce phovals whose diameters $(d_+,d_-)$ lie within a surprisingly narrow subregion of the full allowed region.

More precisely, these diameters (the gold dots in Figs.~\ref{fig:IntermediateSpinSurvey} and \ref{fig:SurveyM87}) form a relatively straight line connecting the corners of the blue region that correspond to the outer and inner edges of the lensing band.
In addition, they are not uniformly scattered along this line; rather, they are tightly grouped near its center.
Both of these facts may be intuitively understood as follows.
We recall from Sec.~\ref{subsec:LensingBands} that the inner and outer edges of the $n^\text{th}$ lensing band are the $n^\text{th}$ lensed images of the equatorial circles of radius $r=r_+$ (the event horizon) and $r\to\infty$, respectively.
Images of intermediate radii $r_+<r<\infty$ form curves that fill the lensing band and interpolate between its inner and outer edges.
From Fig.~6 of \citet{GrallaLupsasca2020a}, these curves roughly appear to be dilated versions of each other (with $\gamma(\varphi)$ controlling the angular-dependent dilation) and indeed, we find that they map to a series of points in the $(d_+,d_-)$ plane tracing a diagonal line that connects the blue and orange dots (shown in gray in Figs.~\ref{fig:IntermediateSpinSurvey} and \ref{fig:SurveyM87}): the same line that the gold dots are concentrated around (see App.~\ref{app:PhovalRegions} for details of its computation).

Since an equatorial disk is a superposition of equatorial rings, the  $n=2$ photon ring is also a superposition of the $n=2$ images of said rings.
That is, its position in the $(d_+,d_-)$ plane should be a weighted average of points along the diagonal line, and must therefore also lie on this line.
Moreover, this position may be associated to an ``effective'' radius of equatorial emission lying somewhere in the line's middle, away from its extremities.
We intend to further explore this connection in future work.

\section{Parameter survey for BH spin and inclination}
\label{sec:SpinSurvey}

\subsection{Choice of parameters for the survey}
\label{subsec:SpinChoice}

To better assess the robustness of the GLM method described in Secs.~\ref{sec:Theory} and \ref{sec:Implementation}, we also conducted a parameter survey over BH spins and inclinations $(a,i)$.
Simulating 100 different emission models for each $(a,i)$, as we did in Sec.~\ref{subsec:ParameterChoice}, would have been prohibitively expensive here; instead, we focused on a handful of ``representative'' emission profiles of the form \eqref{eq:EmissionProfile} that spanned the full range of outcomes observed in that survey:
\begin{itemize}
	\item[\textbullet] a ``typical 1'' profile for which the diameters $(d_+,d_-)$ of the $n=2$ ring tended to be near their average value (that is, near their mean value averaged over all emission models);
	\item[\textbullet] an ``overestimated'' profile for which the diameters $(d_+,d_-)$ of the $n=2$ ring were among the largest;
	\item[\textbullet] an ``underestimated'' profile for which the diameters $(d_+,d_-)$ of the $n=2$ ring were among the smallest;
	\item[\textbullet] a ``typical 2'' profile for which they were slightly subaverage;
	\item[\textbullet] a ``narrow'' profile with a sharp peak (small $\sigma$) for which the circlipse fit was more difficult (required longer baselines) and led to the smallest values for $(d_+,d_-)$ when successful.
\end{itemize}
Table~\ref{tbl:SpinSurvey} lists the parameters $\cu{\mu,\gamma,\sigma}$ of these emission profiles.

\begin{table}[h!]
	\centering
	\caption{
	Parameters of the five representative emission profiles selected for the survey over BH spins and inclinations.
	}
	\label{tbl:SpinSurvey}
	\begin{tabular}{|c|c c c|} 
	\hline
	Profile & $\mu$ & $\gamma$ & $\sigma/M$ \\
	\hline
	Typical 1 & $r_+$ & $0$ & $1$ \\
	Overestimated & $r_-$ & $-2$ & $1.5$ \\
	Underestimated & $r_+$ & $2$ & $0.5$ \\
	Typical 2 & $r_-$ & $2$ & $1$ \\
	Narrow & $\frac{3}{2}r_+$ & $2$ & $0.25$ \\
	\hline
	\end{tabular}
\end{table}

In this survey, we considered all the combinations $(a,i)$ with  $a\in\{0,0.25,0.5,0.75,0.99\}$ and $i \in \{1^\circ,22^\circ,45^\circ,67^\circ,89^\circ\}$ (avoiding the fine-tuned cases $0^\circ$ and $90^\circ$).
In the Schwarzschild case $a=0$, the inclination is defined relative to the equatorial disk, whose spin is always aligned with the BH spin.

\subsection{Results}
\label{subsec:SpinSurveyResults}

\begin{figure*}[hbtp]
	\centering
	\includegraphics[width=0.98\textwidth]{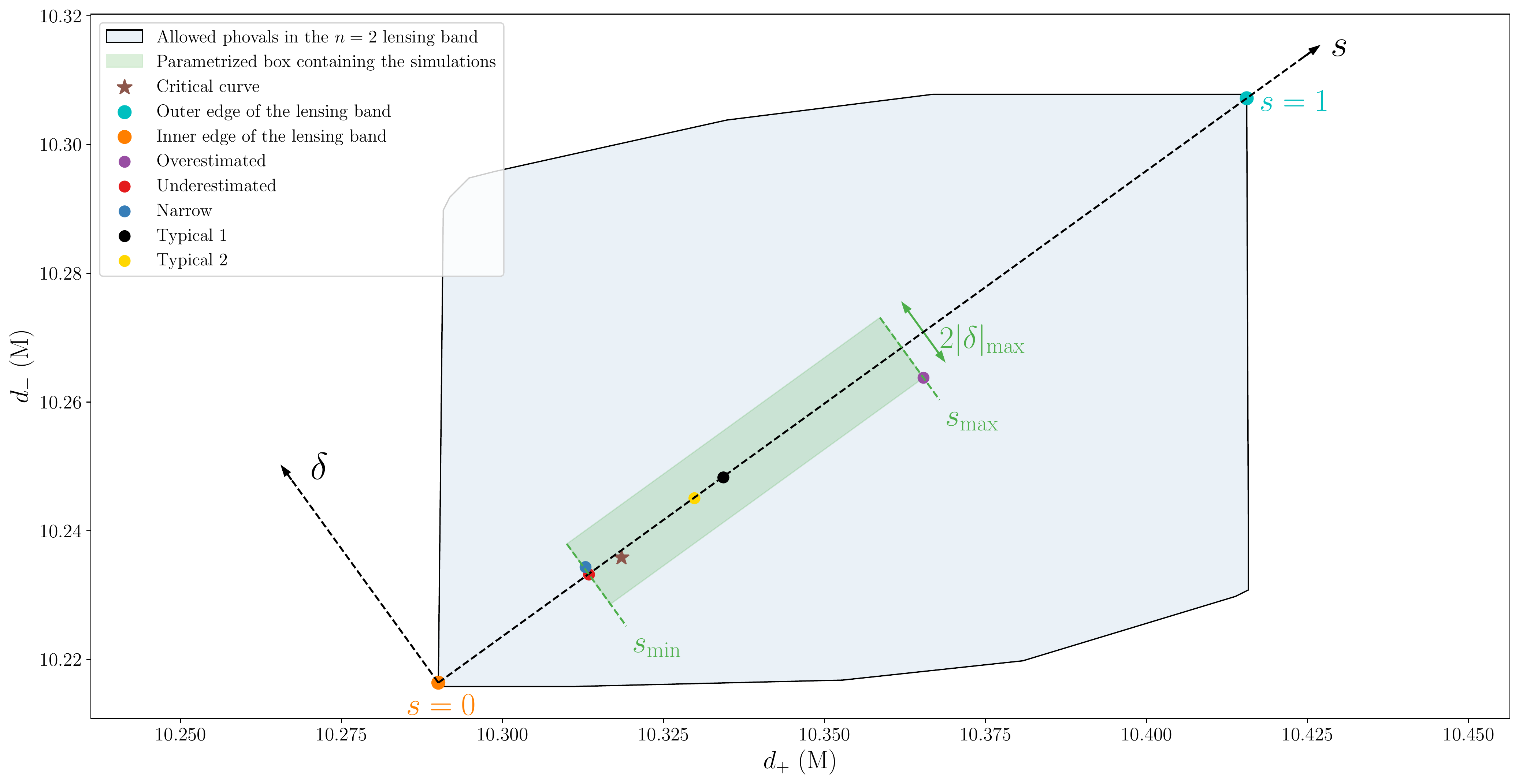}
	\caption{
	Parametrization of the region in the $(d_+,d_-)$ plane containing all the astrophysical photon ring shapes (here, for $a=0.5$ and $i=45^\circ$).
	}
	\label{fig:BoundingBox}
\end{figure*}

We investigated these models to address the two questions posed in Sec.~\ref{subsec:ModelSurveyResults}.
Our survey led us to the following answers:
\begin{enumerate}
	\item[A1.] \textbf{Yes, up to $i\lesssim45^\circ$.}
	In all the models considered here, we identified a regime in which the $n=2$ ring dominates the visamp and extracted the projected ring diameter $d_\varphi^{(2)}$ from the signal's periodicity.
	For reasons discussed in Sec.~\ref{sec:BaselineWindow}, the diameters inferred from images at inclinations of $i=67^\circ$ and $i=89^\circ$ sometimes presented features that did not allow for a full circlipse fit as described in Secs.~\ref{sec:Theory} and \ref{sec:Implementation}.
	However, for low-to-moderate inclinations $i\lesssim45^\circ$, we always found an excellent fit to the circlipse shape \eqref{eq:CirclipseShape} as measured by the normalized RMSD \eqref{eq:CirclipseFit}: in all such cases,
	\begin{align}
		\rmsd_\varphi(\vec{R}_{\rm fit})<5\times10^{-4}.
	\end{align}
	For most configurations, these best fits were obtained in a window $u\in\br{1000,1100}\,$G$\lambda$.
	However, the ``narrow'' profile produced $n=2$ rings that were so thin that they only became dominant on much longer baselines, requiring the fit to be carried out in a baseline window $u_w\approx2000\,$G$\lambda$.
	This also became necessary at $i=45^\circ$ for the ``underestimated'' profile.
	\item[A2.] As in Sec.~\ref{subsec:ModelSurveyResults}, we extracted a pair of diameters $(d_+,d_-)$ for each emission model.
	This enabled us to produce, for each BH spin and inclination, a plot like the one shown in Fig.~\ref{fig:BoundingBox} for the case of spin $a=0.5$ and inclination $i=45^\circ$.
	Since our models were handpicked to form a ``representative'' sample spanning the range of possible diameters $(d_+,d_-)$ that one can obtain for each spin and inclination, it is informative to draw a bounding box containing them (like the green box in Fig.~\ref{fig:BoundingBox}) and then measure its size.
	The dimensions of these bounding boxes are listed for each $(a,i)$ in Table~\ref{tbl:BoundingBox} below.
	As we explain in Sec.~\ref{subsec:InverseProblem}, this analysis bears relevance for the ``inverse problem'' of how to infer BH spin and inclination from the $n=2$ photon ring shape.
\end{enumerate}

\begin{table*}[hbtp]
	\centering
		\caption{
	Triplets $(s_{\rm min},s_{\rm max},|\delta|_{\rm max}/M)$ describing the region of the $(d_+,d_-)$ plane containing all the models computed with \texttt{Gyoto}.
	}
	\label{tbl:BoundingBox}
	\begin{tabular}{|l|c c c|} 
	\hline
	Spin & $i=1^\circ$ & $i=22^\circ$ & $i=45^\circ$ \\
	\hline
	$a=0$ & $(0.23, 0.59, 3\times10^{-3})$ & $(0.19, 0.57, 9\times10^{-4})$ & $(0.18, 0.56, 5\times10^{-3})$ \\
	$a=0.25$ & $(0.21, 0.58, 2\times10^{-3})$ & $(0.19, 0.58, 5\times10^{-4})$ & $(0.19, 0.57, 4\times10^{-3})$ \\
	$a=0.5$ & $(0.23, 0.60, 5\times10^{-3})$ & $(0.20, 0.59, 9\times10^{-4})$ & $(0.19, 0.58, 6\times10^{-3})$ \\
	$a=0.75$ & $(0.19, 0.61, 1\times10^{-4})$ & $(0.20, 0.60, 2\times10^{-3})$ & $(0.19, 0.43, 3\times10^{-3})$ \\
	$a=0.99$ & $(0.23, 0.73, 4\times10^{-4})$ & $(0.23, 0.75, 8\times10^{-3})$ & $(0.06, 0.94, 2\times10^{-3})$ \\
	\hline
	\end{tabular}
    \tablefoot{$s_{min}$ and $s_{max}$ are affine parameters along the line connecting the edges of the lensing band, while $\delta$ is a distance perpendicular to this line.}
    \vspace{-2mm}
\end{table*}

We first note that this survey over BH spins and inclinations confirms one of our crucial observations from Sec.~\ref{subsec:LensingBandsPhovals}: for each choice of $(a,i)$, all our emission models produce $n=2$ rings with diameters $(d_+,d_-)$ that are very tightly clustered near the line connecting the inner and outer edges of the $n=2$ lensing band.
In other words, the five observed ring shapes corresponding to the models in Table~\ref{tbl:SpinSurvey} occupy a very small subregion (green box in Fig.~\ref{fig:BoundingBox}) of the full region of allowed phoval shapes (blue box).
As explained in Sec.~\ref{subsec:LensingBandsPhovals}, diameters along this line correspond to $n=2$ images of equatorial circles of different radii $r_s\in[r_+,\infty)$, so we may equivalently say that the $n=2$ rings produced by our emission models have the same diameters $(d_+,d_-)$ as images of equatorial rings of emission concentrated near a fixed radius $r_s$.

Rather than pursue this association with ``effective'' emission radii, we employed a more agnostic parameterization of this line in the $(d_+,d_-)$ plane and used an affine parameter $s$ such that $s=0$ corresponds to the inner edge of the lensing band and $s=1$ to its outer edge.
We also introduced $\delta$, a perpendicular distance from the line, thereby defining a map $(d_+,d_-)\leftrightarrow(s,\delta)$.
Then, the diameters produced by our models are all contained in a bounding box extending from some $s_{\rm min}$ to some $s_{\rm max}$ in the direction parallel to the line, with a perpendicular width of $|\delta|_{\rm max}$ on each side on the line.
For each $(a,i)$, the resulting triplet $(s_{\rm min},s_{\rm max},|\delta|_{\rm max})$ characterizes the spread of the ring diameters obtained from our models.
These triplets are listed in Table~\ref{tbl:BoundingBox}.

We note that, aside from the extreme case with $a=0.99$, $i=45^\circ$, we typically have $s_{\rm min}\gtrsim0.18$ and $s_{\rm max}\lesssim0.75$.
Hence, our models produce rings that are concentrated away from the edges of the lensing band, as already observed in Sec.~\ref{subsec:LensingBandsPhovals}.
This is consistent with an ``effective'' emission radius $2M\lesssim r_s\lesssim7M$, in line with the choice of emission peak locations in Table~\ref{tbl:SpinSurvey}.

\subsection{Inferring BH spin and inclination from photon ring shape}
\label{subsec:InverseProblem}

In this survey, we have solved the ``forward problem'' of how to compute the diameters $(d_+,d_-)$ of the $n=2$ photon ring given a choice of emission profile, and a BH spin and inclination $(a,i)$.
Purely geometric considerations constrain the observed photon ring to lie within the $n=2$ lensing band, which corresponds to the blue regions in Figs.~\ref{fig:IntermediateSpinSurvey}, \ref{fig:SurveyM87}, and \ref{fig:BoundingBox}.
Perhaps suprisingly, we have found that the ring shapes produced by our emission models occupy a much smaller subregion (the green bounding box in Fig.~\ref{fig:BoundingBox}) of this full region of geometrically allowed shapes.

This observation is highly relevant to the ``inverse problem'' of how to infer BH spin and inclination from a measurement of the $n=2$ ring diameters $(d_+,d_-)$.
Given such a data point, one could try to determine which blue regions ($n=2$ lensing bands) it is contained within.
The answer would provide the set of $(a,i)$ for which lensing would allow the observed photon ring shape.

Evidently, given the considerable size of these blue regions, these purely geometric constraints on $(a,i)$ (which are derived from lensing alone) are relatively weak.
However, if one were to make additional assumptions about the astrophysics of the source (such as the class of emission models being considered), then one would obtain narrower subregions for each $(a,i)$ (the smaller green boxes), and a given data point would therefore be compatible with a smaller range of BH spins and inclinations.

Since our five models were picked to span the full range of $s$ observed in our survey over emission profiles in Sec.~\ref{sec:EmissionSurvey}, we expect $s_{\rm min}$ and $s_{\rm max}$ to be close to the values they would take in more general situations, at least so long as the emission peaks near an equatorial source ring at small radius outside the horizon.
On the other hand, our values of $|\delta|_{\rm max}$ are probably less reliable, and would likely increase if we were to consider geometrically thick models in which the emission is not well-approximated by an equatorial source ring.

Thus, we do not claim here to have definitively computed the bounding boxes for ``realistic'' choices of astrophysical profiles.
Rather, we hope to have provided useful organizing principles for tackling the problem of spin inference: the choice between blue and green boxes clarifies the interplay between geometric and astrophysical assumptions, with the additional constraining power provided by the latter made manifest in the much smaller size of the green regions relative to the blue ones.

An additional benefit of this framework is that it enables us to quantify the uncertainty in a given spin estimate by computing the range of $(a,i)$ that produce green or blue boxes containing a given measurement $(d_+,d_-)$ of the $n=2$ photon ring shape.
We leave a more detailed examination of this problem to the future.

\section{Discussion}
\label{sec:Discussion}

\subsection{Limitations of this work}
\label{subsec:Limitations}

We have revisited the proposal made by \citet{GLM2020} for a test of the Kerr hypothesis based on the photon ring shape.
We have established the robustness of their proposed test in a large class of models, provided that the BH inclination is not too high.
This confirms that M87* holds great promise as an observational target for this test.
However, as pointed out in the introduction, this conclusion remains subject to some caveats.

First, we have only considered simple equatorial emission models.
It is important to test whether the GLM method remains viable in the presence of more realistic accretion flows (such as fluctuating, geometrically thick structures), for which effects such as absorption can become very significant (for instance, the interferometric signature of the photon ring is suppressed as the emission region grows thicker and more opaque).
Encouraging preliminary results in this direction will be reported elsewhere \citep{Vincent2022}.
	
Second, we have not presented a complete, quantitative study of the range of baseline lengths for which the method works.
In particular, we have not identified an ``ideal'' baseline window in which the GLM test succeeds for the largest set of configurations (astrophysical profiles, BH spins, and inclinations), and we have largely ignored practical concerns in our theoretical study of the baseline domain.
With the current wavelength of observations (1.3\,mm), a space antenna would have to be almost 1.5 million km away from Earth---close to $L_2$!---to sample the visibility at $u_w\sim1000\,$G$\lambda$ \citep{Johnson2020}.
	
Third, we have not included any noise in our simulations.
Generating mock data with both instrumental and astrophysical noise is crucial, not only for demonstrating the feasibility of this proposed test, but also for developing a more robust, Bayesian approach to the multi-fit method described in Sec.~\ref{sec:Implementation}.
In their original analysis, \cite{GLM2020} included instrument noise but not did not model astrophysical fluctuations, which can have a considerable impact on the photon ring signatures present in instantaneous snapshots.
A new code tailored to include such fluctations---\texttt{AART}, for ``Analytic Adaptive Ray Tracing''---has recently been developed \citep{CardenasAvendano2022} to tackle this specific problem and will inform the requirements for (incoherent) time-averaging.

While amplitude gain errors can be problematic for VLBI data analyses \citep[e.g.,][]{EHT4,EHT6}, it is worth noting that they may not pose a significant problem for a measurement of the photon ring diameter, which would only require measuring the periodicity of oscillations in the visibility amplitude (rather than their absolute scale).

\subsection{Perspectives: Inferring black hole spin and inclination}

If a future space mission targeting M87* successfully measured the shape of its $n=2$ photon ring and confirmed that it matched the Kerr prediction \eqref{eq:CirclipseShape}, then the natural next step would be to assume the Kerr nature of the source and try to infer the BH spin and inclination $(a,i)$ from the measured photon ring shape.

Our survey over emission profiles in Sec.~\ref{sec:EmissionSurvey} shows that this shape displays a residual dependence on the astrophysical source profile (even for the $n=2$ ring).
Though small, this dependence nonetheless introduces significant uncertainty in estimates of the spin based on the $n=2$ ring shape alone, as evidenced by the lower bounds in Table~\ref{tbl:ErrorBars}.
(This uncertainty is undoubtedly much greater for spin estimates based on the $n=1$ ring shape.)

Spin estimates can be tightened with additional assumptions about the nature of the emission.
As discussed in Secs.~\ref{subsec:LensingBandsPhovals} and \ref{subsec:SpinSurveyResults}, the photon ring shapes produced by our emission models inhabit only a small subset of the full range of geometric shapes allowed by BH lensing.
Thus, assuming that an observed photon ring shape is produced by a model within this class considerably narrows down the set of BH spins and inclinations compatible with the measurement, excluding many more values of $(a,i)$ than could be ruled out based on the Kerr geometry alone.

It would be extremely interesting to quantify the reduction in uncertainty enabled by different astrophysical assumptions.
This requires one to tackle the inverse problem of how to determine the values of $(a,i)$ that are compatible with a given photon ring shape; we hope that our framing of this problem in Sec.~\ref{subsec:InverseProblem} will prove useful for future discussions of spin inference.
We note that other approaches based on the identification of the photon ring with the critical curve either fail to quantify the significant systematic error in their spin estimates (see Figs.~\ref{fig:IntermediateSpinSurvey} and \ref{fig:SurveyM87}), or else they effectively incorporate some version of our analysis.

Finally, spin estimates can also be substantially improved by combining the constraints derived from multiple image features.
Early work suggests that measuring both the photon ring shape and that of the central brightness depression could significantly constrain BH spin and inclination \citep{Chael2021}.
Of course, simultaneously resolving more than one ring (such as the $n=1$ and $n=2$ subrings, for instance) should also enable sharper spin estimates, though this question has so far only been explored in the simplest cases with circular subrings \citep{Broderick2021}.

\subsection{Perspectives: Other photon ring tests}

The prospect of measuring multiple subrings naturally opens the door to even more stringent tests of the Kerr hypothesis (or more precise techniques for determining the BH parameters).

For instance, comparing the widths of successive subrings at different angles around the image could in principle provide a measurement of the angle-dependent demagnification factors $e^{-\gamma(\varphi)}$, and therefore of the Lyapunov exponents $\gamma(\tilde{r})$ that offer a direct probe of the Kerr geometry \citep{Johnson2020,GrallaLupsasca2020a}.
If the demagnification factors were not $e^{-\gamma(\varphi)}$, then this might suggest that the spacetime is not Kerr \citep{Wielgus2021}.
Conversely, if $\gamma(\varphi)$ did match the form predicted from Kerr, then these exponents would provide a function's worth of constraints on the BH spin and inclination $(a,i)$, which could allow these parameters to be inferred with great precision.

Likewise, polarimetric imaging of successive subrings could also be used to test the Kerr hypothesis, or conversely, to yield an estimate of the BH parameters $(a,i)$ \citep{Himwich2020}, while also providing a probe of the plasma and magnetic field around the BH \citep{EHT8}.
Lastly, the photon ring substructure predicts a characteristic pattern of autocorrelations whose detection could also provide a window into the Lyapunov exponents $\gamma$ (and other Kerr critical exponents $\tau$ and $\delta$), even if the rings themselves remain unresolved \citep{Hadar2021}.
A related promising idea involves timing signatures such as quasi-periodic oscillations \citep{Johannsen2010}.

Finally, a wide variety of other techniques for testing the Kerr hypothesis and strong-field GR via VLBI observations have been proposed \citep[e.g.,][]{Psaltis2019}, many of which do not rely upon the photon ring.
Similarly, some constraints on BH inclination can be derived from other image features, such as its brightness asymmetry \citep{Medeiros2021}.

\subsection{Open questions and future directions}

We have thoroughly investigated the shape of the $n=2$ photon ring and its corresponding interferometric signature in a certain class of equatorial models.
We can now confirm the prediction that the ring should follow a circlipse shape at low-to-moderate inclinations, but have yet to obtain a precise characterization of its shape at high inclinations, where its image can display kinks or discontinuities (Fig.~\ref{fig:Scrambling}) that produce characteristic jumps in its signature in the visibility domain (Fig.~\ref{fig:BaselineChoice}).
Understanding these patterns is crucial for devising a test of the Kerr hypothesis based on the photon ring shape that remains valid at high inclinations.

Reversing the logic, one can assume the Kerr nature of the source and try to infer its parameters from the photon ring shape.
We hope to further explore the question of spin inference by solving the inverse problem described in Sec.~\ref{subsec:InverseProblem}.
This approach could most likely also be used to infer the BH mass with much greater precision than the EHT has so far achieved without being able to probe higher spatial frequencies.

As discussed in Sec.~\ref{subsec:Limitations}, it is important to determine whether our conclusions generalize to models with geometrically thick emission, with noise created by inevitable source fluctuations, with different astrophysical conditions (e.g., a radially infalling flow rather than a circularized one), and also with tilted disks that are misaligned with the BH spin.
For instance, our observation that the measured diameters $(d_+,d_-)$ of the $n=2$ ring lie on a straight line connecting the inner and outer edges of the $n=2$ lensing band could well depend sensitively on any one of these assumptions, with direct consequences for estimates of BH spin.

Finally, the most pressing task is to revisit all these questions in the context of the first ($n=1$) subring, which is likely to be observed first.
What is the predicted shape of the $n=1$ ring?
How well can the BH parameters be recovered from its shape?

From a theoretical standpoint, these questions are harder to tackle because the $n=1$ ring is likely too thick to display a universal regime \eqref{eq:UniversalRegime}; that is, its width-to-diameter ratio $w/d$ is likely not small enough for the separation of scales \eqref{eq:UniversalRegime} to arise.
If so, then there is no baseline regime in which its interferometric signature follows the universal form \eqref{eq:UniversalAmplitude}, and it may well be necessary to compute the subleading correction in $w/d$ to this visibility amplitude.
We take a first stab at these issues in App.~\ref{app:FirstRing}.

\section{Summary}
\label{sec:Conclusion}

Images of a BH are typically dominated by a bright, narrow ring encircling a central brightness depression.
This ``photon ring'' decomposes into a sequence of self-similar subrings indexed by half-orbit number $n$, with each subring consisting of a full, lensed image of the main emission surrounding the BH (Sec.~\ref{sec:Theory}).
Here, we analyzed a method proposed by \citet{GLM2020} to test the Kerr hypothesis via an interferometric measurement of the $n=2$ ring shape.
To check the viability of this shape test, we applied it to a wide range of simulated images of a Kerr BH surrounded by a thin, equatorial disk (Sec.~\ref{sec:Implementation}), varying both the emission profile (Sec.~\ref{sec:EmissionSurvey}) and the BH spin and inclination (Sec.~\ref{sec:SpinSurvey}).
These two parameter surveys confirm the robustness of the test at low-to-moderate inclinations $i\lesssim45^\circ$, indicating that M87* (which likely lies at an inclination $i\approx17^\circ$ from Earth) is a promising target for such a test via space-VLBI observation.

At higher inclinations $i\gtrsim45^\circ$, our study uncovers the emergence of qualitatively new phenomena that complicate ring shape measurements  (Sec.~\ref{sec:BaselineWindow}).
These effects arise because the ring width develops significant angular variation, which impacts the range of baselines dominated by the ring's interferometric signature.
In particular, we find that observations made at fixed baseline length in the visibility domain cannot measure the shape of a single ring at every angle: such observations are sensitive to a fixed ring width and must therefore either jump between the $n=1$ and $n=2$ rings (at moderately high inclinations; Fig.~\ref{fig:BaselineChoice}), or else pick up the signature of a ``hybrid $n=1\&2$ ring'' of roughly uniform width (at very high inclinations; Fig.~\ref{fig:Scrambling}).
A more refined, sophisticated shape test has yet to be devised for these situations.

Our investigation also reveals that for a given choice of BH spin and inclination, the observable photon ring shape depends noticeably on the astrophysical properties of the emitting source.
In particular, the most directly accessible ($n=1$ and $n=2$) rings can differ significantly from the theoretical critical curve, and in many models they are in fact closer to the critical curves of neighboring BH spins and inclinations (Figs.~\ref{fig:IntermediateSpinSurvey} \& \ref{fig:SurveyM87}).
As such (and as discussed throughout Secs.~\ref{sec:EmissionSurvey} and \ref{sec:SpinSurvey}), the lensing bands introduced in Sec.~\ref{subsec:LensingBands} are more relevant than the critical curve to the problem of BH spin inference from the photon ring shape.
We have outlined a promising approach for tackling this problem in Sec.~\ref{subsec:InverseProblem}, with some encouraging preliminary results (Fig.~\ref{fig:BoundingBox}).

\section*{Acknowledgements}
{All of the images in this paper were ray traced on the \texttt{tycho} computer cluster of the Paris Observatory.
We thank Samuel Gralla, Alejandro C\`ardenas-Avenda\~no, Andrei Lobanov, and Hengrui Zhu for helpful comments and discussions.
A.L. also gratefully acknowledges support from Will and Kacie Snellings, and M.W. thanks Alexandra Elbakyan for her contributions to the open science initiative.}

\bibliography{RingVariation}

\begin{thebibliography}{60}
\expandafter\ifx\csname natexlab\endcsname\relax\def\natexlab#1{#1}\fi

\bibitem[{{Bardeen}(1973)}]{Bardeen1973}
{Bardeen}, J.~M. 1973, in Black Holes (Les Astres Occlus), ed. C.~{Dewitt} \&
  B.~S. {Dewitt} (Gordon and Breach Science Publishers), 215--239

\bibitem[{{Bauer} {et~al.}(2022){Bauer}, {C{\'a}rdenas-Avenda{\~n}o}, {Gammie},
  \& {Yunes}}]{Bauer2021}
{Bauer}, A.~M., {C{\'a}rdenas-Avenda{\~n}o}, A., {Gammie}, C.~F., \& {Yunes},
  N. 2022, \apj, 925, 119

\bibitem[{{Beckwith} \& {Done}(2005)}]{Beckwith2005}
{Beckwith}, K. \& {Done}, C. 2005, \mnras, 359, 1217

\bibitem[{{Broderick} {et~al.}(2022){Broderick}, {Tiede}, {Pesce}, \&
  {Gold}}]{Broderick2021}
{Broderick}, A.~E., {Tiede}, P., {Pesce}, D.~W., \& {Gold}, R. 2022, \apj, 927,
  6

\bibitem[{{C{\'a}rdenas-Avenda{\~n}o}
  {et~al.}(2022){C{\'a}rdenas-Avenda{\~n}o}, {Lupsasca}, \&
  {Zhu}}]{CardenasAvendano2022}
{C{\'a}rdenas-Avenda{\~n}o}, A., {Lupsasca}, A., \& {Zhu}, H. 2022, arXiv
  e-prints, arXiv:2211.07469

\bibitem[{{Chael} {et~al.}(2021){Chael}, {Johnson}, \& {Lupsasca}}]{Chael2021}
{Chael}, A., {Johnson}, M.~D., \& {Lupsasca}, A. 2021, \apj, 918, 6

\bibitem[{{Cunha} \& {Herdeiro}(2018)}]{Cunha2018}
{Cunha}, P. V.~P. \& {Herdeiro}, C. A.~R. 2018, General Relativity and
  Gravitation, 50, 42

\bibitem[{{Cunningham}(1975)}]{Cunningham1975}
{Cunningham}, C.~T. 1975, \apj, 202, 788

\bibitem[{{Doeleman} {et~al.}(2019){Doeleman}, {Blackburn}, {Dexter}, {Gomez},
  {Johnson}, {Palumbo}, {Weintroub}, {Farah}, {Fish}, {Loinard}, {Lonsdale},
  {Narayanan}, {Patel}, {Pesce}, {Raymond}, {Tilanus}, {Wielgus}, {Akiyama},
  {Bower}, {Broderick}, {Deane}, {Fromm}, {Gammie}, {Gold}, {Janssen},
  {Kawashima}, {Krichbaum}, {Marrone}, {Matthews}, {Mizuno}, {Rezzolla},
  {Roelofs}, {Ros}, {Savolainen}, {Yuan}, {Zhao}, {Blackburn}, {Doeleman},
  {Dexter}, {Gomez}, {Johnson}, {Palumbo}, {Weintroub}, {Farah}, {Fish},
  {Loinard}, {Lonsdale}, {Narayanan}, {Patel}, {Pesce}, {Raymond}, {Tilanus},
  {Wielgus}, {Akiyama}, {Bower}, {Broderick}, {Deane}, {Fromm}, {Gammie},
  {Gold}, {Janssen}, {Kawashima}, {Krichbaum}, {Marrone}, {Matthews}, {Mizuno},
  {Rezzolla}, {Roelofs}, {Ros}, {Savolainen}, {Yuan}, \& {Zhao}}]{Doeleman2019}
{Doeleman}, S., {Blackburn}, L., {Dexter}, J., {et~al.} 2019, in Bulletin of
  the American Astronomical Society, Vol.~51, 256

\bibitem[{{Event Horizon Telescope Collaboration}(2019{\natexlab{a}})}]{EHT1}
{Event Horizon Telescope Collaboration}. 2019{\natexlab{a}}, \apjl, 875, L1

\bibitem[{{Event Horizon Telescope Collaboration}(2019{\natexlab{b}})}]{EHT2}
{Event Horizon Telescope Collaboration}. 2019{\natexlab{b}}, \apjl, 875, L2

\bibitem[{{Event Horizon Telescope Collaboration}(2019{\natexlab{c}})}]{EHT3}
{Event Horizon Telescope Collaboration}. 2019{\natexlab{c}}, \apjl, 875, L3

\bibitem[{{Event Horizon Telescope Collaboration}(2019{\natexlab{d}})}]{EHT4}
{Event Horizon Telescope Collaboration}. 2019{\natexlab{d}}, \apjl, 875, L4

\bibitem[{{Event Horizon Telescope Collaboration}(2019{\natexlab{e}})}]{EHT5}
{Event Horizon Telescope Collaboration}. 2019{\natexlab{e}}, \apjl, 875, L5

\bibitem[{{Event Horizon Telescope Collaboration}(2019{\natexlab{f}})}]{EHT6}
{Event Horizon Telescope Collaboration}. 2019{\natexlab{f}}, \apjl, 875, L6

\bibitem[{{Event Horizon Telescope Collaboration}(2021{\natexlab{a}})}]{EHT7}
{Event Horizon Telescope Collaboration}. 2021{\natexlab{a}}, \apjl, 910, L12

\bibitem[{{Event Horizon Telescope Collaboration}(2021{\natexlab{b}})}]{EHT8}
{Event Horizon Telescope Collaboration}. 2021{\natexlab{b}}, \apjl, 910, L13

\bibitem[{{Falcke} {et~al.}(2000){Falcke}, {Melia}, \& {Agol}}]{Falcke2000}
{Falcke}, H., {Melia}, F., \& {Agol}, E. 2000, \apjl, 528, L13

\bibitem[{{Farah} {et~al.}(2020){Farah}, {Pesce}, {Johnson}, \&
  {Blackburn}}]{Farah2020}
{Farah}, J.~R., {Pesce}, D.~W., {Johnson}, M.~D., \& {Blackburn}, L. 2020,
  \apj, 900, 77

\bibitem[{{Gates} {et~al.}(2021){Gates}, {Hadar}, \& {Lupsasca}}]{Gates2021}
{Gates}, D. E.~A., {Hadar}, S., \& {Lupsasca}, A. 2021, \prd, 103, 044050

\bibitem[{{Gebhardt} {et~al.}(2011){Gebhardt}, {Adams}, {Richstone}, {Lauer},
  {Faber}, {G{\"u}ltekin}, {Murphy}, \& {Tremaine}}]{Gebhardt2011}
{Gebhardt}, K., {Adams}, J., {Richstone}, D., {et~al.} 2011, \apj, 729, 119

\bibitem[{{Gelles} {et~al.}(2021){Gelles}, {Prather}, {Palumbo}, {Johnson},
  {Wong}, \& {Georgiev}}]{Gelles2021}
{Gelles}, Z., {Prather}, B.~S., {Palumbo}, D.~C.~M., {et~al.} 2021, \apj, 912,
  39

\bibitem[{{Gold} {et~al.}(2020){Gold}, {Broderick}, {Younsi}, {Fromm},
  {Gammie}, {Mo{\'s}cibrodzka}, {Pu}, {Bronzwaer}, {Davelaar}, {Dexter},
  {Ball}, {Chan}, {Kawashima}, {Mizuno}, {Ripperda}, {Akiyama}, {Alberdi},
  {Alef}, {Asada}, {Azulay}, {Baczko}, {Balokovi{\'c}}, {Barrett}, {Bintley},
  {Blackburn}, {Boland}, {Bouman}, {Bower}, {Bremer}, {Brinkerink},
  {Brissenden}, {Britzen}, {Broguiere}, {Byun}, {Carlstrom}, {Chael},
  {Chatterjee}, {Chatterjee}, {Chen}, {Chen}, {Cho}, {Christian}, {Conway},
  {Cordes}, {Crew}, {Cui}, {De Laurentis}, {Deane}, {Dempsey}, {Desvignes},
  {Doeleman}, {Eatough}, {Falcke}, {Fish}, {Fomalont}, {Fraga-Encinas},
  {Freeman}, {Friberg}, {G{\'o}mez}, {Galison}, {Garc{\'\i}a}, {Gentaz},
  {Georgiev}, {Goddi}, {Gu}, {Gurwell}, {Hada}, {Hecht}, {Hesper}, {Ho}, {Ho},
  {Honma}, {Huang}, {Huang}, {Hughes}, {Inoue}, {Issaoun}, {James}, {Jannuzi},
  {Janssen}, {Jeter}, {Jiang}, {Jimenez-Rosales}, {Johnson}, {Jorstad}, {Jung},
  {Karami}, {Karuppusamy}, {Keating}, {Kettenis}, {Kim}, {Kim}, {Kim}, {Kino},
  {Koay}, {Koch}, {Koyama}, {Kramer}, {Kramer}, {Krichbaum}, {Kuo}, {Lauer},
  {Lee}, {Li}, {Li}, {Lico}, {Lindqvist}, {Liu}, {Liuzzo}, {Lo}, {Lobanov},
  {Loinard}, {Lonsdale}, {Lu}, {MacDonald}, {Markoff}, {Mao}, {Marrone},
  {Marscher}, {Mart{\'\i}-Vidal}, {Matsushita}, {Matthews}, {Medeiros},
  {Menten}, {Mizuno}, {Moran}, {Moriyama}, {M{\"u}ller}, {Nagai}, {Nakamura},
  {Nagar}, {Narayan}, {Narayanan}, {Natarajan}, {Neri}, {Ni}, {Noutsos},
  {Okino}, {Ortiz-Le{\'o}n}, {Oyama}, {{\"O}zel}, {Palumbo}, {Park}, {Patel},
  {Pen}, {Pesce}, {Plambeck}, {Pi{\'e}tu}, {PopStefanija}, {Porth},
  {Preciado-L{\'o}pez}, {Psaltis}, {Ramakrishnan}, {Rao}, {Rawlings},
  {Raymond}, {Rezzolla}, {Roelofs}, {Rogers}, {Ros}, {Rose}, {Roshanineshat},
  {Rottmann}, {Roy}, {Ruszczyk}, {Rygl}, {S{\'a}nchez},
  {S{\'a}nchez-Arguelles}, {Sasada}, {Savolainen}, {Schuster}, {Schloerb},
  {Shao}, {Shen}, {Small}, {Sohn}, {SooHoo}, {Tiede}, {Tazaki}, {Tilanus},
  {Titus}, {Toma}, {Torne}, {Trent}, {Traianou}, {Trippe}, {Tsuda}, {van
  Langevelde}, {van Bemmel}, {van Rossum}, {Wagner}, {Wardle}, {Wex},
  {Weintroub}, {Wharton}, {Wielgus}, {Wong}, {Wu}, {Yoon}, {Young}, {Young},
  {Yuan}, {Yuan}, {Zensus}, {Zhao}, {Zhao}, {Zhu}, \& {Event Horizon Telescope
  Collaboration}}]{Gold2020}
{Gold}, R., {Broderick}, A.~E., {Younsi}, Z., {et~al.} 2020, \apj, 897, 148

\bibitem[{{Gralla}(2020)}]{Gralla2020}
{Gralla}, S.~E. 2020, \prd, 102, 044017

\bibitem[{{Gralla} {et~al.}(2019){Gralla}, {Holz}, \& {Wald}}]{Gralla2019}
{Gralla}, S.~E., {Holz}, D.~E., \& {Wald}, R.~M. 2019, \prd, 100, 024018

\bibitem[{{Gralla} \& {Lupsasca}(2020{\natexlab{a}})}]{GrallaLupsasca2020a}
{Gralla}, S.~E. \& {Lupsasca}, A. 2020{\natexlab{a}}, \prd, 101, 044031

\bibitem[{{Gralla} \& {Lupsasca}(2020{\natexlab{b}})}]{GrallaLupsasca2020b}
{Gralla}, S.~E. \& {Lupsasca}, A. 2020{\natexlab{b}}, \prd, 101, 044032

\bibitem[{{Gralla} \& {Lupsasca}(2020{\natexlab{c}})}]{GrallaLupsasca2020c}
{Gralla}, S.~E. \& {Lupsasca}, A. 2020{\natexlab{c}}, \prd, 102, 124003

\bibitem[{{Gralla} {et~al.}(2020){Gralla}, {Lupsasca}, \& {Marrone}}]{GLM2020}
{Gralla}, S.~E., {Lupsasca}, A., \& {Marrone}, D.~P. 2020, \prd, 102, 124004

\bibitem[{{Gravity Collaboration}(2018)}]{GRAVITY}
{Gravity Collaboration}. 2018, \aap, 615, L15

\bibitem[{{Gurvits} {et~al.}(2022){Gurvits}, {Paragi}, {Amils}, {van Bemmel},
  {Boven}, {Casasola}, {Conway}, {Davelaar}, {D{\'\i}ez-Gonz{\'a}lez},
  {Falcke}, {Fender}, {Frey}, {Fromm}, {Gallego-Puyol}, {Garc{\'\i}a-Mir{\'o}},
  {Garrett}, {Giroletti}, {Goddi}, {G{\'o}mez}, {van der Gucht}, {Guirado},
  {Haiman}, {Helmich}, {Hudson}, {Humphreys}, {Impellizzeri}, {Janssen},
  {Johnson}, {Kovalev}, {Kramer}, {Lindqvist}, {Linz}, {Liuzzo}, {Lobanov},
  {L{\'o}pez-Fern{\'a}ndez}, {Malo-G{\'o}mez}, {Masania}, {Mizuno}, {Plavin},
  {Rajan}, {Rezzolla}, {Roelofs}, {Ros}, {Rygl}, {Savolainen}, {Schuster},
  {Venturi}, {Verkouter}, {de Vicente}, {Visser}, {Wiedner}, {Wielgus}, {Wiik},
  \& {Zensus}}]{Gurvits2021}
{Gurvits}, L.~I., {Paragi}, Z., {Amils}, R.~I., {et~al.} 2022, Acta
  Astronautica, 196, 314

\bibitem[{{Hadar} {et~al.}(2021){Hadar}, {Johnson}, {Lupsasca}, \&
  {Wong}}]{Hadar2021}
{Hadar}, S., {Johnson}, M.~D., {Lupsasca}, A., \& {Wong}, G.~N. 2021, \prd,
  103, 104038

\bibitem[{{Haworth} {et~al.}(2019){Haworth}, {Johnson}, {Pesce}, {Palumbo},
  {Blackburn}, {Akiyama}, {Boroson}, {Bouman}, {Farah}, {Fish}, {Honma},
  {Kawashima}, {Kino}, {Raymond}, {Silver}, {Weintroub}, {Wielgus}, {Doeleman},
  {Kauffmann}, {Keating}, {Krichbaum}, {Loinard}, {Narayanan}, {Doi}, {James},
  {Marrone}, {Mizuno}, \& {Nagai}}]{Haworth2019}
{Haworth}, K., {Johnson}, M., {Pesce}, D.~W., {et~al.} 2019, in Bulletin of the
  American Astronomical Society, Vol.~51, 235

\bibitem[{{Herdeiro} {et~al.}(2021){Herdeiro}, {Pombo}, {Radu}, {Cunha}, \&
  {Sanchis-Gual}}]{Herdeiro2021}
{Herdeiro}, C. A.~R., {Pombo}, A.~M., {Radu}, E., {Cunha}, P. V.~P., \&
  {Sanchis-Gual}, N. 2021, \jcap, 2021, 051

\bibitem[{{Himwich} {et~al.}(2020){Himwich}, {Johnson}, {Lupsasca}, \&
  {Strominger}}]{Himwich2020}
{Himwich}, E., {Johnson}, M.~D., {Lupsasca}, A., \& {Strominger}, A. 2020,
  \prd, 101, 084020

\bibitem[{{Igumenshchev} {et~al.}(2003){Igumenshchev}, {Narayan}, \&
  {Abramowicz}}]{Igumenshchev2003}
{Igumenshchev}, I.~V., {Narayan}, R., \& {Abramowicz}, M.~A. 2003, \apj, 592,
  1042

\bibitem[{{James} {et~al.}(2015){James}, {Tunzelmann}, {Franklin}, \&
  {Thorne}}]{James2015}
{James}, O., {Tunzelmann}, E.~v., {Franklin}, P., \& {Thorne}, K.~S. 2015,
  Classical and Quantum Gravity, 32, 065001

\bibitem[{{Janssen} {et~al.}(2021){Janssen}, {Falcke}, {Kadler}, {Ros},
  {Wielgus}, {Akiyama}, {Balokovi{\'c}}, {Blackburn}, {Bouman}, {Chael},
  {Chan}, {Chatterjee}, {Davelaar}, {Edwards}, {Fromm}, {G{\'o}mez}, {Goddi},
  {Issaoun}, {Johnson}, {Kim}, {Koay}, {Krichbaum}, {Liu}, {Liuzzo}, {Markoff},
  {Markowitz}, {Marrone}, {Mizuno}, {M{\"u}ller}, {Ni}, {Pesce},
  {Ramakrishnan}, {Roelofs}, {Rygl}, {van Bemmel}, {Event Horizon Telescope
  Collaboration}, {Alberdi}, {Alef}, {Algaba}, {Anantua}, {Asada}, {Azulay},
  {Baczko}, {Ball}, {Ball}, {Barrett}, {Benson}, {Bintley}, {Bintley},
  {Blundell}, {Boland}, {Boland}, {Bower}, {Boyce}, {Bremer}, {Brinkerink},
  {Brissenden}, {Britzen}, {Broderick}, {Broguiere}, {Bronzwaer}, {Byun},
  {Carlstrom}, {Chatterjee}, {Chen}, {Chen}, {Chesler}, {Cho}, {Christian},
  {Conway}, {Cordes}, {Crawford}, {Crew}, {Cruz-Osorio}, {Cui}, {Cui}, {De
  Laurentis}, {Deane}, {Dempsey}, {Desvignes}, {Dexter}, {Doeleman}, {Eatough},
  {Farah}, {Farah}, {Fish}, {Fomalont}, {Ford}, {Fraga-Encinas}, {Friberg},
  {Friberg}, {Fuentes}, {Galison}, {Gammie}, {Garc{\'\i}a}, {Gelles}, {Gentaz},
  {Georgiev}, {Georgiev}, {Gold}, {Gold}, {G{\'o}mez-Ruiz}, {Gu}, {Gurwell},
  {Hada}, {Haggard}, {Hecht}, {Hesper}, {Himwich}, {Ho}, {Ho}, {Honma},
  {Huang}, {Huang}, {Hughes}, {Ikeda}, {Inoue}, {Inoue}, {James}, {Jannuzi},
  {Jannuzi}, {Jeter}, {Jiang}, {Jimenez-Rosales}, {Jimenez-Rosales}, {Jorstad},
  {Jung}, {Karami}, {Karuppusamy}, {Kawashima}, {Keating}, {Kettenis}, {Kim},
  {Kim}, {Kim}, {Kim}, {Kino}, {Kino}, {Kofuji}, {Koyama}, {Kramer}, {Kramer},
  {Kramer}, {Kuo}, {Lauer}, {Lee}, {Levis}, {Li}, {Li}, {Lindqvist}, {Lico},
  {Lindahl}, {Lindahl}, {Liu}, {Liu}, {Lo}, {Lobanov}, {Loinard}, {Lonsdale},
  {Lu}, {MacDonald}, {Mao}, {Marchili}, {Marchili}, {Marchili}, {Marscher},
  {Mart{\'\i}-Vidal}, {Matsushita}, {Matthews}, {Medeiros}, {Menten}, {Mizuno},
  {Mizuno}, {Moran}, {Moriyama}, {Moscibrodzka}, {Moscibrodzka}, {Musoke},
  {Mej{\'\i}as}, {Nagai}, {Nagar}, {Nakamura}, {Narayan}, {Narayanan},
  {Natarajan}, {Nathanail}, {Neilsen}, {Neri}, {Neri}, {Noutsos}, {Nowak},
  {Okino}, {Olivares}, {Ortiz-Le{\'o}n}, {Oyama}, {{\"O}zel}, {Palumbo},
  {Park}, {Patel}, {Pen}, {Pen}, {Pi{\'e}tu}, {Plambeck}, {PopStefanija},
  {Porth}, {P{\"o}tzl}, {Prather}, {Preciado-L{\'o}pez}, {Psaltis}, {Pu}, {Pu},
  {Rao}, {Rawlings}, {Raymond}, {Rezzolla}, {Ricarte}, {Ripperda}, {Ripperda},
  {Rogers}, {Rogers}, {Rose}, {Roshanineshat}, {Rottmann}, {Roy}, {Ruszczyk},
  {Ruszczyk}, {S{\'a}nchez}, {S{\'a}nchez-Arguelles}, {Sasada}, {Savolainen},
  {Schloerb}, {Schuster}, {Shao}, {Shen}, {Small}, {Sohn}, {SooHoo}, {Sun},
  {Tazaki}, {Tetarenko}, {Tiede}, {Tilanus}, {Titus}, {Torne}, {Trent},
  {Traianou}, {Trippe}, {van Bemmel}, {van Langevelde}, {van Rossum}, {Wagner},
  {Ward-Thompson}, {Wardle}, {Weintroub}, {Wex}, {Wharton}, {Wharton}, {Wong},
  {Wu}, {Yoon}, {Young}, {Young}, {Younsi}, {Yuan}, {Yuan}, {Zensus}, {Zhao},
  \& {Zhao}}]{Janssen2021}
{Janssen}, M., {Falcke}, H., {Kadler}, M., {et~al.} 2021, Nature Astronomy, 5,
  1017

\bibitem[{{Johannsen}(2013)}]{Johannsen2013}
{Johannsen}, T. 2013, \apj, 777, 170

\bibitem[{{Johannsen} \& {Psaltis}(2010)}]{Johannsen2010}
{Johannsen}, T. \& {Psaltis}, D. 2010, \apj, 718, 446

\bibitem[{{Johnson} {et~al.}(2020){Johnson}, {Lupsasca}, {Strominger}, {Wong},
  {Hadar}, {Kapec}, {Narayan}, {Chael}, {Gammie}, {Galison}, {Palumbo},
  {Doeleman}, {Blackburn}, {Wielgus}, {Pesce}, {Farah}, \&
  {Moran}}]{Johnson2020}
{Johnson}, M.~D., {Lupsasca}, A., {Strominger}, A., {et~al.} 2020, Science
  Advances, 6, eaaz1310

\bibitem[{{Lamy} {et~al.}(2018){Lamy}, {Gourgoulhon}, {Paumard}, \&
  {Vincent}}]{Lamy2018}
{Lamy}, F., {Gourgoulhon}, E., {Paumard}, T., \& {Vincent}, F.~H. 2018,
  Classical and Quantum Gravity, 35, 115009

\bibitem[{{LIGO Scientific Collaboration} \& {Virgo
  Collaboration}(2016)}]{LIGO2016}
{LIGO Scientific Collaboration} \& {Virgo Collaboration}. 2016, \prl, 116,
  061102

\bibitem[{{Luminet}(1979)}]{Luminet1979}
{Luminet}, J.~P. 1979, \aap, 75, 228

\bibitem[{{Medeiros} {et~al.}(2022){Medeiros}, {Chan}, {Narayan}, {{\"O}zel},
  \& {Psaltis}}]{Medeiros2021}
{Medeiros}, L., {Chan}, C.-K., {Narayan}, R., {{\"O}zel}, F., \& {Psaltis}, D.
  2022, \apj, 924, 46

\bibitem[{{Medeiros} {et~al.}(2020){Medeiros}, {Psaltis}, \&
  {{\"O}zel}}]{Medeiros2020}
{Medeiros}, L., {Psaltis}, D., \& {{\"O}zel}, F. 2020, \apj, 896, 7

\bibitem[{{Narayan} {et~al.}(2003){Narayan}, {Igumenshchev}, \&
  {Abramowicz}}]{Narayan2003}
{Narayan}, R., {Igumenshchev}, I.~V., \& {Abramowicz}, M.~A. 2003, \pasj, 55,
  L69

\bibitem[{{Narayan} {et~al.}(2019){Narayan}, {Johnson}, \&
  {Gammie}}]{Narayan2019}
{Narayan}, R., {Johnson}, M.~D., \& {Gammie}, C.~F. 2019, \apjl, 885, L33

\bibitem[{{Okounkova} {et~al.}(2019){Okounkova}, {Scheel}, \&
  {Teukolsky}}]{Okounkova2019}
{Okounkova}, M., {Scheel}, M.~A., \& {Teukolsky}, S.~A. 2019, Classical and
  Quantum Gravity, 36, 054001

\bibitem[{{Pesce} {et~al.}(2019){Pesce}, {Haworth}, {Melnick}, {Blackburn},
  {Wielgus}, {Johnson}, {Raymond}, {Weintroub}, {Palumbo}, {Doeleman}, \&
  {James}}]{Pesce2019}
{Pesce}, D., {Haworth}, K., {Melnick}, G.~J., {et~al.} 2019, in Bulletin of the
  American Astronomical Society, Vol.~51, 176

\bibitem[{{Psaltis}(2019)}]{Psaltis2019}
{Psaltis}, D. 2019, General Relativity and Gravitation, 51, 137

\bibitem[{{Teo}(2003)}]{Teo2003}
{Teo}, E. 2003, General Relativity and Gravitation, 35, 1909

\bibitem[{{Vincent} {et~al.}(2022){Vincent}, {Gralla}, {Lupsasca}, \&
  {Wielgus}}]{Vincent2022}
{Vincent}, F.~H., {Gralla}, S.~E., {Lupsasca}, A., \& {Wielgus}, M. 2022, A\&A,
  667, A170

\bibitem[{{Vincent} {et~al.}(2016){Vincent}, {Meliani}, {Grandcl{\'e}ment},
  {Gourgoulhon}, \& {Straub}}]{Vincent2016}
{Vincent}, F.~H., {Meliani}, Z., {Grandcl{\'e}ment}, P., {Gourgoulhon}, E., \&
  {Straub}, O. 2016, Classical and Quantum Gravity, 33, 105015

\bibitem[{{Vincent} {et~al.}(2011){Vincent}, {Paumard}, {Gourgoulhon}, \&
  {Perrin}}]{Vincent2011}
{Vincent}, F.~H., {Paumard}, T., {Gourgoulhon}, E., \& {Perrin}, G. 2011,
  Classical and Quantum Gravity, 28, 225011

\bibitem[{{Vincent} {et~al.}(2021){Vincent}, {Wielgus}, {Abramowicz},
  {Gourgoulhon}, {Lasota}, {Paumard}, \& {Perrin}}]{Vincent2021}
{Vincent}, F.~H., {Wielgus}, M., {Abramowicz}, M.~A., {et~al.} 2021, \aap, 646,
  A37

\bibitem[{{Walker} {et~al.}(2018){Walker}, {Hardee}, {Davies}, {Ly}, \&
  {Junor}}]{Walker2018}
{Walker}, R.~C., {Hardee}, P.~E., {Davies}, F.~B., {Ly}, C., \& {Junor}, W.
  2018, \apj, 855, 128

\bibitem[{{Wielgus}(2021)}]{Wielgus2021}
{Wielgus}, M. 2021, \prd, 104, 124058

\bibitem[{{Wielgus} {et~al.}(2020){Wielgus}, {Hor{\'a}k}, {Vincent}, \&
  {Abramowicz}}]{Wielgus2020}
{Wielgus}, M., {Hor{\'a}k}, J., {Vincent}, F., \& {Abramowicz}, M. 2020, \prd,
  102, 084044

\bibitem[{{Wong}(2021)}]{Wong2021}
{Wong}, G.~N. 2021, \apj, 909, 217

\end{thebibliography}
\bibliographystyle{aa}


\begin{appendix}

\addtolength{\parskip}{2mm}

\section{Computing Kerr lensing bands}
\label{app:LensingBands}

We consider a null geodesic connecting a source at $(t_s,r_s,\theta_s,\varphi_s)$ to an observer at $(t_o,r_o,\theta_o,\varphi_o)$.
Its behavior can be characterized by its conserved quantities $(\lambda,\eta)$ [Eq.~\eqref{eq:ConservedQuantities} above] via the radial and angular potentials \citep[e.g.,][]{GrallaLupsasca2020b}
\begin{subequations}
\begin{align}
	\mathcal{R}(r)&=\pa{r^2+a_*^2-a_*\lambda}^2-\Delta(r)\br{\eta+\pa{\lambda-a_*}^2},\\
	\Theta(\theta)&=\eta+a_*^2\cos^2{\theta}-\lambda^2\cot^2{\theta},
\end{align}
\end{subequations}
whose zeros define the radial and angular turning points of the ray.
Integrating the geodesic equation along the trajectory gives
\begin{align}
    \label{eq:MinoTime}
	I_r\equiv\fint_{r_s}^{r_o}\frac{\ed r}{\pm_r\sqrt{\mathcal{R}(r)}}
	=\fint_{\theta_s}^{\theta_o}\frac{\ed\theta}{\pm_\theta \sqrt{\Theta(\theta)}}
	\equiv G_\theta,
\end{align}
where the signs $\pm_r$ and $\pm_\theta$
switch at radial and angular turning points, respectively.
These expressions keep track of the total Mino time $\tau=I_r=G_\theta$ elapsed along the trajectory.

We consider the case of a distant observer ($r_o=D\gg M$) at an inclination $\theta_o=i$ from the BH spin axis who shoots photons backwards into the geometry from a polar position $(\rho,\varphi_\rho)$ in the sky.
Such a photon has conserved quantities
\begin{align}
	\lambda=-D\rho\cos{\varphi_\rho}\sin{i},\qquad
	\eta=(D\rho)^2-\lambda^2-a_*^2\cos^2{i},
\end{align}
in terms of which the four roots $r_j$ of the quartic radial potential $\mathcal{R}(r)$ are given by [Eqs.~(79)--(95) of \citet{GrallaLupsasca2020b}]:
\begin{gather*}
	r_j=-z-\epsilon_j\sqrt{-\frac{\mathcal{A}}{2}-z^2+\tilde{\epsilon}_j\frac{\mathcal{B}}{4z}},\\
	\epsilon_j=\left\{
		\begin{array}{ll}
		    1 & \mbox{if }j\in\cu{1,3},\\
			-1 & \mbox{if }j\in\cu{2,4},
		\end{array}
	\right.\qquad
	\tilde{\epsilon}_j=\left\{
		\begin{array}{ll}
		    1 & \mbox{if }j\in\cu{1,2},\\
			-1 & \mbox{if }j\in\cu{3,4},
		\end{array}
	\right.\\
	\mathcal{A}=a_*^2-\eta-\lambda^2,\quad
	\mathcal{B}=2M\br{\eta+\pa{\lambda-a_*}^2},\quad
	\mathcal{C}=-a_*^2\eta,\\
	\mathcal{P}=-\frac{\mathcal{A}^2}{12}-\mathcal{C},\quad
	\mathcal{Q}=-\frac{\mathcal{A}}{3}\br{\pa{\frac{\mathcal{A}}{6}}^2-\mathcal{C}}-\frac{\mathcal{B}^2}{8},\\
	\omega_\pm=\pa{-\frac{\mathcal{Q}}{2}\pm\sqrt{\frac{\mathcal{P}^3}{27}+\frac{\mathcal{Q}^2}{4}}}^{1/3},\quad
	z=\sqrt{\frac{\omega_++\omega_--\mathcal{A}/3}{2}}.
\end{gather*}
Letting $r_{ij}=r_i-r_j$, we define $A=\sqrt{r_{32}r_{42}}$, $B=\sqrt{r_{31}r_{41}}$, and introduce the antiderivatives of $I_r$ \citep{GrallaLupsasca2020a,GrallaLupsasca2020b}:
\begin{align}
	\mathcal{I}_r^+(r)&=\frac{2}{\sqrt{r_{31}r_{42}}}F\pa{\left.\arcsin{\sqrt{\frac{r-r_4}{r-r_3}\frac{r_{31}}{r_{41}}}}\right|\frac{r_{32}r_{41}}{r_{31}r_{42}}},\\
	\mathcal{I}_r^-(r)&=\frac{1}{\sqrt{AB}}F\pa{\left.\arccos\pa{\frac{1-\frac{r-r_2}{r-r_1}\frac{B}{A}}{1+\frac{r-r_2}{r-r_1}\frac{B}{A}}}\right|\frac{(A+B)^2-r_{21}^2}{4AB}}.
\end{align}

\begin{itemize}
	\item[\textbullet] If $\rho>\tilde{\rho}(\varphi_\rho)$, so that the photon is shot from a position outside the critical curve, then all its roots are real and ordered as $r_4>r_3>r_+>r_->r_2>r_1$.
As such, it must encounter a radial turning point at $r=r_4$ and eventually be deflected back to infinity.
The Mino time $\tau(r)$ elapsed along its trajectory is
\begin{align}
	\label{eq:RadialIntegralOutside}
	I_r^{\mathcal{C}^+}(r)=\mathcal{I}_r^+(r_o)\mp\mathcal{I}_r^+(r),
\end{align}
with upper/lower sign before/after reaching the turning point.
The total Mino time elapsed along the full light ray is
\begin{align}
	I_r^{\rm total}=\frac{4}{\sqrt{r_{31}r_{42}}}F\pa{\left.\arcsin{\sqrt{\frac{r_{31}}{r_{41}}}}\right|\frac{r_{32}r_{41}}{r_{31}r_{42}}}.
\end{align}
	\item[\textbullet] If $\rho<\tilde{\rho}(\varphi_\rho)$, so that the photon is shot from a position inside the critical curve, then it cannot encounter a radial turning point and must therefore asymptotically reach the horizon at $r=r_+$.
The Mino time $\tau(r)$ elapsed along its trajectory is
\begin{align}
	\label{eq:RadialIntegralInside}
	I_r^{\mathcal{C}^-}(r)=\mathcal{I}_r^\pm(r_o)-\mathcal{I}_r^\pm(r),
\end{align}
with the upper sign when all roots are real (in which case $r_+>r_->r_4>r_3>r_2>r_1$) and the lower sign when $r_3=\bar{r}_4$ are complex conjugate roots with $r_+>r_->r_2>r_1$.\footnote{There also exist vortical ($\eta<0$) rays with both $r_1=\bar{r}_2$ and $r_3=\bar{r}_4$, but we may ignore them here as they cannot reach the equatorial plane.
}
The total Mino time elapsed along the full light ray is
\begin{align}
	I_r^{\rm total}=I_r^{\mathcal{C}^-}(r_+).
\end{align}
\end{itemize}

Regardless of where the photon is shot back from, it crosses the equatorial plane a maximum number $N+1$ of times given by
\begin{gather}
	\label{eq:MaxCrossing}
	N=\left\lfloor\frac{I_r^{\rm total}\sqrt{-u_-a_*^2}+\sign(\sin{\varphi_\rho})F_o}{2K}\right\rfloor-H(\sin{\varphi_\rho}),\\
	u_\pm=\Delta_\theta\pm\sqrt{\Delta_\theta^2+\frac{\eta}{a_*^2}},\qquad
	\Delta_\theta=\frac{1}{2}\pa{1-\frac{\eta+\lambda ^2}{a_*^2}},\notag\\
	K=K\pa{\frac{u_+}{u_-}},\qquad
	F_o=F\pa{\left.\arcsin\pa{\frac{\cos(i)}{\sqrt{u_+}}}\right|\frac{u_+}{u_-}},\notag
\end{gather}
where $H$, $F$, and $K$ respectively denote the Heaviside function, the incomplete elliptic integral of the first kind $F(u|k)$, and its completion $K(k)\equiv F(\pi/2|k)$.
$u_\pm$ are zeros $u=\cos^2{\theta}$ of the angular potential $\Theta(\theta)$; when evaluated on critical geodesics with $(\lambda,\eta)=(\tilde{\lambda},\tilde{\eta})$ [Eq.~\eqref{eq:AngMomentum}], they reproduce $\tilde{u}_\pm$ in Eq.~\eqref{eq:uCritical}.

The photon intersects the equatorial plane for the $(n+1)^\text{th}$ time (for $n\leq N$) when the Mino time $\tau(\theta)$ elapsed along its trajectory is [Eq.~(81) of \citet{GrallaLupsasca2020a}]
\begin{align}
    \label{eq:Gtheta}
	G_\theta(n)=\frac{2mK-\sign(\sin{\varphi_\rho})F_o}{\sqrt{-u_-a_*^2}},
\end{align}
where $m=n+H(\sin{\varphi_\rho})$ is the number of angular turning points encountered along the trajectory.
We note that $N$ is by definition the non-negative integer such that $G_\theta(N)\leq I_r^{\rm total}\leq G_\theta(N+1)$.

The $n^\text{th}$ image of an equatorial source ring of constant radius $r_s$ is the set of points $(\rho,\varphi_\rho)$ in the observer sky that satisfy
\begin{align}
	\label{eq:Contours}
	\boxed{I_r^{\mathcal{C}^\pm}(r_s)=G_\theta(n)}
\end{align}
with $G_\theta(n)$ given in Eq.~\eqref{eq:Gtheta} and $I_r^{\mathcal{C}^\pm}(r_s)$ given in Eqs.~\eqref{eq:RadialIntegralOutside} and \eqref{eq:RadialIntegralInside} for points outside and inside the critical curve.
These points always trace a closed, continuous curve $\mathcal{C}_n(r_s)$.

The contours $\mathcal{C}_0(r_s)$ and $\mathcal{C}_1(r_s)$ corresponding to a selection of radii $r_s$ are shown for various BH spins and inclinations in Fig.~6 of \citet{GrallaLupsasca2020a}.
While all these curves are continuous, some of them may cross the critical curve, in which case they are stitched together from segments solving Eq.~\eqref{eq:Contours} with opposite signs.
Aside from this minor inconvenience, it is straightforward to numerically solve Eq.~\eqref{eq:Contours} for any $\mathcal{C}_n(r_s)$.

In particular, as explained in Sec.~\ref{subsec:LensingBands}, the $n^\text{th}$ lensing band is the annular region in the sky bounded within the contours $\mathcal{C}_n(r_+)$ and $\mathcal{C}_n(\infty)$, which respectively trace its inner and outer edges.
These contours are relatively simple to compute because they can never cross the critical curve: trajectories that are deflected back to infinity must obey Eq.~\eqref{eq:RadialIntegralOutside}, so the outer edge $\mathcal{C}_n(\infty)$ is obtained by solving Eq.~\eqref{eq:Contours} with a $+$ sign only; likewise, trajectories that reach the BH must obey Eq.~\eqref{eq:RadialIntegralInside}, so the inner edge $\mathcal{C}_n(r_+)$ is obtained by solving Eq.~\eqref{eq:Contours} with a $-$ sign only.

We note that when crossing the inner and outer edges $\mathcal{C}_n(r_+)$ and $\mathcal{C}_n(\infty)$ of the $n^\text{th}$ lensing band, the maximal number $N+1$ of equatorial crossings \eqref{eq:MaxCrossing} must by definition change by 1.
This provides a means of checking that $\mathcal{C}_n(r_+)$ and $\mathcal{C}_n(\infty)$ have been correctly computed (or an alternative way to compute them).

Finally, we note that Eq.~\eqref{eq:MinoTime} may also be directly solved for $r_s(I_r)$ using a single ``unified inversion formula'' given in \citet{GrallaLupsasca2020b}.
Then $r_{\rm eq}^{(n)}(\rho,\varphi_\rho)\equiv r_s(G_\theta(n))$ are transfer functions that give the radii at which light rays shot back from position $(\rho,\varphi_\rho)$ cross the equatorial plane for the $(n+1)^\text{th}$ time.
Contours of these functions are precisely the $\mathcal{C}_n(r_s)$ defined above.
Hence, lensing bands may also be computed using these transfer functions, as described in App.~A of \citet{Chael2021}.

\section{Computing the range of allowed phovals within a lensing band}
\label{app:PhovalRegions}

Here, we provide a detailed solution to the following problem.
Given a lensing band $\mathcal{B}\subsetneq\mathbb{R}^2$, how do we compute the subregion $\mathcal{A}$ of the $(d_+,d_-)$ plane spanned by the phovals that fit within $\mathcal{B}$?

We glossed over this question after introducing it in Sec.~\ref{subsec:LensingBandsPhovals}, but its answer is crucial for obtaining the blue regions in Figs.~\ref{fig:IntermediateSpinSurvey}, \ref{fig:SurveyM87}, and \ref{fig:BoundingBox}.
At first glance, this problem seems straightforward: we could scan the 6D space of phovals \eqref{eq:PhovalProjectedPosition} and ask, for each $\vec{R}=\cu{R_0,R_1,R_2,X,\chi,\varphi_0}$ in this parameter space $\mathcal{P}_6$, whether the associated phoval fits inside the lensing band $\mathcal{B}$.
The phovals that fit inside the lensing band would form a 6D region $\mathcal{A}_6\subsetneq\mathcal{P}_6$ whose 2D projection $\Pi:\mathcal{P}_6\to\mathbb{R}^2$ via the map \eqref{eq:Projection},
\begin{align}
	\Pi:\vec{R}\mapsto(d_+,d_-)=2(R_0+R_1,R_0+R_2),
\end{align}
would give precisely the sought-after region in the $(d_+,d_-)$ plane:
\begin{align}
	\mathcal{A}=\Pi(\mathcal{A}_6).
\end{align}
In practice, however, the computational cost of this ``brute force'' approach is too high to achieve sufficient precision in reasonable time: the high-dimensionality of the parameter space $\mathcal{P}_6$ makes any detailed scan too slow.
Hence, another approach is needed.

The tractable method we came up with is quite general and consists of two parts.
First, we recast the question of whether a given point $(d_+,d_-)$ belongs in $\mathcal{A}$ into a minimization problem, which is easier to solve.
Second, we devised an efficient way to scan across regions of the $(d_+,d_-)$ plane while solving this minimization problem.
We now describe each part separately.

\subsection*{Part 1: Minimization method}

Let $\mathcal{B}\subsetneq\mathbb{R}^2$ denote an annular subregion of the plane with inner edge $\mathcal{I}$ and outer edge $\mathcal{O}$.
We can define a separation function characterizing the distance from a closed curve $\mathcal{C}$ to $\mathcal{B}$ as
\begin{align}
	\delta_\mathcal{B}(\mathcal{C})=\max_{z\in\mathcal{C}}d(z,\mathcal{B}),
\end{align}
where $d$ is the usual distance between a point and a set, namely:
\begin{align}
	d(z,\mathcal{B})=
	\begin{cases}
		\min_{a\in\mathcal{I}}|z-a|&\mbox{if $z$ lies inside of }\mathcal{I},\\
		\min_{a\in\mathcal{O}}|z-a|&\mbox{if $z$ lies outside of }\mathcal{O},\\
			0&\mbox{if }z\in\mathcal{B}.
	\end{cases}
\end{align}
For us, $\mathcal{C}$ will be a phoval shape [Eq.~\eqref{eq:PhovalProjectedPosition}] and $\mathcal{B}$ a Kerr lensing band with inner edge $\mathcal{I}=\mathcal{C}_n(r_+)$ and outer edge $\mathcal{O}=\mathcal{C}_n(\infty)$, as in App.~\ref{app:LensingBands}.

With this definition, $\delta_\mathcal{B}(\mathcal{C})=0$ if and only if $\mathcal{C}$ lies within the lensing band $\mathcal{B}$.
Thus, to check if a point $(d_+,d_-)$ corresponds to a phoval $\mathcal{C}$ in the lensing band $\mathcal{B}$, we can minimize the function
\begin{align*}
	\Delta:(d_+,d_-,\vec{r})\mapsto\delta_\mathcal{B}\pa{\mathcal{C}\pa{R_0,\frac{d_+}{2}-R_0,\frac{d_-}{2}-R_0,X,\chi,\varphi_0}},
\end{align*}
where $\mathcal{C}(\vec{R})$ is the phoval with projected position \eqref{eq:PhovalProjectedPosition}, over the 4D space $\mathcal{P}_4$ of parameters $\vec{r}=(R_0,X,\chi,\varphi_0)$.
In other words, we compute $m:\mathbb{R}^2\to\mathbb{R}$ defined as
\begin{align}
	m:(d_+,d_-)\mapsto\min_{\vec{r}\in\mathcal{P}_4}\Delta(d_+,d_-,\vec{r}).
\end{align}
If $m(d_+,d_-)$ vanishes (or stays below a tolerance threshold that accounts for numerical artifacts), then there exists a phoval with diameters $(d_+,d_-)$ that lies within the lensing band, and thus,
\begin{align}
	\mathcal{A}=\cu{(d_+,d_-)\in\mathbb{R}^2:m(d_+,d_-)=0}.
\end{align}
We have thus recast the problem of determining whether a given $(d_+,d_-)$ belongs to $\mathcal{A}$ to a problem of minimization over $\mathcal{P}_4$, which is only a 4D space.
This completes part 1.

\subsection*{Part 2: Random walk}

To compute $\mathcal{A}$, it now suffices to define a grid of values $(d_+,d_-)$ (typically, between the values found for $\mathcal{I}$ and $\mathcal{O}$) and compute $m(d_+,d_-)$ for each point on the grid.
Since the computational cost of the minimization method in part 1 is still quite high, we must still devise an efficient method to scan across a large grid, ideally without having to evaluate $m$ for each point.
This can be done using a random walk with a reflecting barrier:
\begin{enumerate}
	\item We start from the parameters $\vec{R}\in\mathcal{A}_6\subset\mathcal{P}_6$ of some phoval that we know to lie in the lensing band: for instance, some intermediate phoval between those approximating $\mathcal{I}$ and $\mathcal{O}$.
	\item We randomly pick one of the radii $\cu{R_0,R_1,R_2}$ and add to this parameter a predefined increment $\pm b$, with the sign picked at random (we recall our convention that $R_1\ge R_2$; if this changes, then we must interchange $R_1\leftrightarrow R_2$ to maintain it).
	\item After the increment, we test if the resulting phoval (with the same values of $X$, $\chi$, and $\varphi_0$) still lies within the lensing band.
	If that is the case, then we store $(d_+,d_-)=2(R_0+R_1, R_0+R_2)$ as an accepted value in the region $\mathcal{A}$ and then proceed with the random walk by repeating step 2.
	If not, then we compute $m(d_+,d_-)$, with two possible outcomes:
	\begin{itemize}
		\item[\textbullet] $m(d_+,d_-)=0$ (or the minimum is under threshold), in which case we store $(d_+,d_-)=2(R_0+R_1, R_0+R_2)$ as an accepted value in the region $\mathcal{A}$ and return to step 2 to continue the random walk; otherwise,
		\item[\textbullet] $m(d_+,d_-)>0$ (or the minimum is above threshold), in which case we store $(d_+,d_-)=2(R_0+R_1, R_0+R_2)$ as a rejected value, then restore the values of $\cu{R_0,R_1,R_2}$ to their previous state and return to step 2 to take another step of the random walk in a new direction.
	\end{itemize}
\end{enumerate}
Steps 2 and 3 can be repeated any number of times, resulting in a set of rejected values $\cu{(d_+,d_-)}_{\rm rejected}$ and a set of accepted values
\begin{align}
	\mathcal{A}=\cu{(d_+,d_-)}_{\rm accepted}.
\end{align}

One advantage is that the minimization routine $m$ is only evaluated in the rejected region of the $(d_+,d_-)$ plane, or inside the allowed region but very near its boundary.
This boundary acts as a reflecting barrier: each time it is crossed, the random walk returns to a previous position in $\mathcal{A}$.
Thus, it cannot escape far in the rejected region, where the computational cost is higher.

The boundaries of the region $\mathcal{A}$ (which corresponds to the blue boxes in Figs.~\ref{fig:IntermediateSpinSurvey}, \ref{fig:SurveyM87}, and \ref{fig:BoundingBox}) can be computed to arbitrary precision by tuning the value of the increment $b$ that controls the step size in the random walk, and also by running several instances of the random walk starting from different positions.

\section{Shape of the \texorpdfstring{$n=1$}{n=1} photon ring}
\label{app:FirstRing}

\begin{figure*}[hbtp]
	\centering
	\includegraphics[width=\textwidth]{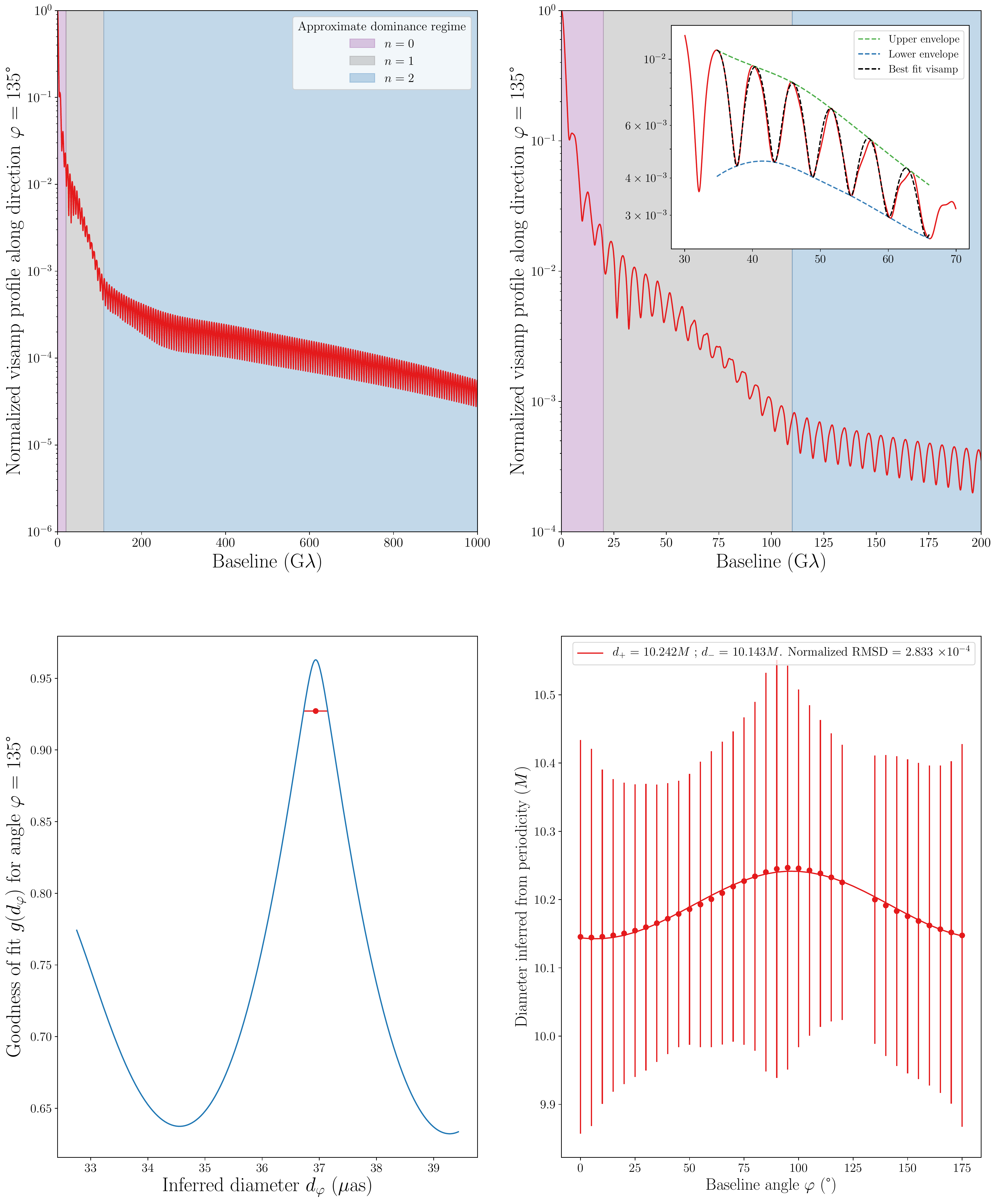}
	\caption{
	Kerr hypothesis test using the GLM method applied to the $n=1$ subring.
	An image was simulated for a BH of spin $a=0.94$ and inclination $i=17^\circ$, with equatorial emission profile parameters $\mu=r_-$, $\gamma=-\frac{3}{2}$, $\sigma=0.5$.
	(Upper left) Amplitude $|V(u,\varphi)|$ of the visibility at $\varphi=135^\circ$, with baselines colored according to which subring dominates the signal (Sec.~\ref{subsec:Subrings}).
	(Upper Right) Visibility amplitude fit (with envelope) in the $n=1$-dominated baseline window $u_w\in[30,70]\,$G$\lambda$ (Sec.~\ref{sec:Implementation}).
	(Lower left) Goodness of fit \eqref{eq:GoodnessOfFit} as a function of fitting diameter $d_\varphi^{(1)}$.
	The dot corresponds to the diameter with maximal goodness of fit $g_{\rm max}=e^{-\rmsd_{u,\rm  min}}$.
	The error bar includes nearby diameters such that $\rmsd_u(d_\varphi)\leq2\rmsd_{u,\rm min}$, or equivalently $g(d_\varphi)\geq g_{\rm max}^2$.
	(Lower Right) Fitting $d_\varphi^{(1)}$ to a circlipse using the multi-fit method (Sec.~\ref{subsec:MultiFit}), with  the same prescription for the error bars.
	Angles $\varphi=125^\circ$ and $130^\circ$ were removed because the visamp fit in the chosen baseline window was too poor.
	}
	\label{fig:FirstRing}
\end{figure*}

The $n=1$ ring of M87* will almost certainly be the first photon subring to be measured (since its signal dominates the visibility on the shortest baselines to space), and it is therefore interesting to predict its geometric shape and interferometric signature. Does the $n=1$ ring also follow the circlipse shape \eqref{eq:CirclipseShape}?
If so, to what precision?
Can its projected diameter $d_\varphi^{(1)}$ be recovered from the periodicity of the visibility amplitude $|V(u,\varphi)|$?

Answering these questions in detail would be the subject of a whole other paper.
Here, we merely record some observations that arose from our study and should inform future analyses:
\begin{enumerate}
	\item Only narrow rings with a very small width-to-diameter ratio $w/d\ll1$ admit a universal regime \eqref{eq:UniversalRegime}.
For thicker rings, there is not a large enough separation between the scales $1/d$ and $1/w$ for universality to set in.
The $n=1$ ring tends to be rather thick in most models (though not all, since there can be substantial variation in its width with the choice of astrophysical profile, as discussed below Eq.~\eqref{eq:BaselineTest} and in Sec.~\ref{sec:EmissionSurvey}), with a typical width of $w_1\approx1M$ and diameter $d\approx10M$.
Hence $w_1/d\approx10\%$, implying that subleading corrections to the universal visamp profile \eqref{eq:UniversalAmplitude} may be substantial.
Indeed, in most of our $n=1$ visamps, we observed a much faster decay than the predicted $u^{-1/2}$ universal power-law fall-off.
	\item Geometrically, this also means that the periodicity $d_\varphi^{(1)}$ of the $n=1$ visamp no longer admits a simple interpretation as the projected diameter of the $n=1$ ring in the image, as its diameter is not sharply defined to better than $w_1/d\approx10\%$.
	\item Even in the absence of a universal regime $1/d\ll u\ll1/w$ with simple interferometric signature \eqref{eq:UniversalAmplitude}, the $n=1$ ring can still dominate the visamp on some baselines $b_1\ll u\ll b_2$ where we may hope to measure
the periodicity of $|V(u,\varphi)|$, and thus $d_\varphi^{(1)}$,
but only so long as the visamp displays enough periods within this baseline domain of width $\Delta u=b_2-b_1$.
\end{enumerate}
While these issues may complicate a measurement of the $n=1$ ring and its interpretation, none of them seems insurmountable (see Fig.~\ref{fig:FirstRing}), and we briefly address them in reverse order.

Regarding point 3: a baseline region of width $\Delta u$ contains a number of visamp hops $N=\Delta u/T$, where $T$ is the period of a hop, which for a ring is $T=1/d_\varphi$.
Hence, the number of visamp hops in the $n=1$-dominated regime $b_1\ll u\ll b_2$ is
\begin{align}
	N_1(\varphi)\approx\br{b_2(\varphi)-b_1(\varphi)}d_\varphi.
\end{align}
Likewise, for higher-order rings $n\gtrsim2$, using Eq.~\eqref{eq:ThresholdScaling}, we have
\begin{align}
	N_n(\varphi)&\approx\br{b_{n+1}(\varphi)-b_n(\varphi)}d_\varphi
	\approx e^{\gamma(\varphi)}N_{n-1}(\varphi),
\end{align}
so that the number of visamp hops within the regime dominated by the $n^\text{th}$ ring grows exponentially with $n$ at a rate controlled by the image demagnification factor $e^\gamma$.
Thus, the numbers of hops $N_n(\varphi)$ for higher-order rings are predicted by Kerr lensing once the number of hops $N_1(\varphi)$ for the first ring is known, but this latter number depends on astrophysical details.
For instance, for the model in Fig.~\ref{fig:FirstRing}, we have $\Delta u=b_2-b_1\approx80\,$G$\lambda$ at $\varphi=135^\circ$.
Since the photon ring of M87* has an expected diameter of
\begin{align}
	d_\varphi\approx 40\,\mu\rm{as}
	\approx\frac{1}{5\,\rm{G}\lambda},
\end{align}
where we manipulated units as in Eq.~\eqref{eq:Units}, it follows that
\begin{align}
	N_1(135^\circ)\approx16,
\end{align}
and indeed, we count 16 visamp nulls within the gray region in Fig.~\ref{fig:FirstRing}.
This is in principle a sufficient number of hops to obtain a good estimate of the periodicity, and hence of $d_\varphi^{(1)}$, which in the example of Fig.~\ref{fig:FirstRing} follows a circlipse shape, albeit with larger error bars than for the $n=2$ ring.

Regarding point 2: in the limit $w/d\to0$, a ring's projected diameter $d_\varphi$ becomes sharply defined in its image.
Otherwise, it is only defined up to ambiguities of order $w/d$.
If the $n=1$ ring displays a visamp periodicity $1/d_\varphi^{(1)}$, which feature of the ring image does that correspond to?
A natural guess would be the distance $d_\varphi^{\rm peak}$ between peaks of the ring's intensity profile, but in simple models (such as the Gaussian ring below) one can prove that $d_\varphi^{(1)}<d_\varphi^{\rm peak}$, with the difference vanishing as $w/d\to0$.
This effect can be attributed to the everywhere-positive curvature of a ring, which implies that its Radon transform has its center of mass shifted inward relative to its peaks, with the corresponding 1D Fourier transform $|V(u,\varphi)|$ displaying a slightly smaller $d_\varphi^{(1)}$.

Regarding point 1: it is helpful to consider an axisymmetric ring with purely radial intensity profile $I(\rho,\varphi)=I(\rho)$, such as
\begin{align}
	I(\rho)=\frac{1}{2\pi w^2}e^{-\frac{d^2}{8w^2}}I_0\pa{\frac{d\rho}{2w^2}}e^{-\frac{\rho^2}{2w^2}},
\end{align}
with $I_0$ denoting the $0^\text{th}$ modified Bessel function of the first kind.
This is known as the Gaussian ring because $I_0(x)\stackrel{x\to\infty}{\approx}(2\pi x)^{-\frac{1}{2}}e^x$, so that as $d\to\infty$ the radial profile approaches a Gaussian of width $w$ centered as $\rho=d/2$, with normalization chosen such that the total flux is unity, $V(0)=1$.
The visibility is exactly
\begin{align}
	V(u,\varphi)=V(u)
	=J_0(\pi du)e^{-2\pi^2w^2u^2}.
\end{align}
The key point is the following.
If the ring is very thin, then there is a wide range of baselines $u\ll1/w$ in which the exponential is negligible and the visibility is simply $V(u)\approx J_0(\pi du)$, which matches the universal prediction \eqref{eq:UniversalAmplitude} as soon as $du\gtrsim1\gg wu$.
If the ring is thick, then we may never ignore that exponential term, yet we still expect to see the universal visamp \eqref{eq:UniversalAmplitude}, only multiplied by that exponential; in general, we expect $|V(u,\varphi)|$ to take the form \eqref{eq:UniversalAmplitude} times $e^{-c(\varphi)u^p}$ (with $p=2$ for a Gaussian ring).

\end{appendix}

\end{document}